\documentclass{aa}  

\usepackage{graphicx}
\usepackage[varg]{txfonts}
\usepackage[compatibility=false]{caption}

\usepackage{xcolor}  
\usepackage{hyperref}

\begin{document}

   \title{Lithium and the evolution of intermediate-mass T Tauri and Herbig stars. Rotation, accretion, and planets}
 \titlerunning{Lithium and the evolution of intermediate-mass T Tauri and Herbig stars}

   \author{I. Mendigutía\inst{1}
        \and
        J. Campbell-White\inst{2}
        \and
        B. Montesinos\inst{1}
        \and
        J. Maldonado\inst{3}
        \and
        L. Fullana-García\inst{1}
        \and
        G.M. Mirouh\inst{4}
        \and
        G. Meeus\inst{5}
        \and
        M. Vioque\inst{2}
        \and
        A. Sicilia-Aguilar\inst{6}
        \and
        M.R. Zapatero-Osorio\inst{1}
        \and
        E. Villaver\inst{7,8}
        \and
        R. Kahar\inst{6}
        }

   \institute{Centro de Astrobiología (CAB), CSIC-INTA, Camino Bajo del Castillo s/n, 28692, Villanueva de la Cañada, Madrid, Spain.\\
             \email{imendigutia@cab.inta-csic.es}
            \and European Southern Observatory, Karl-Schwarzschild-Str. 2, 85748 Garching bei München, Germany \and INAF – Osservatorio Astronomico di Palermo, Piazza del Parlamento 1, 90134 Palermo, Italy \and Dep. de fisica teórica y del cosmos, Universidad de Granada, Campus de Fuentenueva s/n, 18071 Granada, Spain \and Departamento Física Teórica, Centro de Investigación Avanzada en física Fundamental (CIAFF), Facultad de Ciencias, Universidad Autónoma de Madrid, Campus de Cantoblanco, Carretera Colmenar s/n - km 15, 28049, Madrid, Spain \and SUPA, School of Science and Engineering, University of Dundee, Nethergate, DD1 4HN, Dundee, UK \and Instituto de Astrofísica de Canarias, Vía Láctea s/n, 38200 La Laguna, Tenerife, Spain \and Universidad de La Laguna (ULL), Astrophysics Department, 38206 La Laguna, Tenerife, Spain
 }

   \date{Received 4 February 2026. Accepted 26 March 2026}

 
  \abstract
   {Interior models predict that stellar envelopes change from convective to radiative during the pre-main-sequence (pre-MS) evolution of intermediate-mass stars. Although the amount of surface lithium (Li) is a direct probe of mixing in stellar interiors, analyses focused on this type of source are practically absent.}
   {We contribute to our understanding of the evolution of young intermediate-mass stars by providing a comprehensive analysis of their Li content.} 
   {A sample of 71 intermediate-mass T Tauri (IMTT) and Herbig stars within the mass range 1.5 -- 3.5 M$_{\odot}$ was carefully selected for the analysis. Metallicities, rotational velocities, and accretion rates were obtained from spectra. The curves of growth for stars hotter than 8000 K were built to infer the Li abundances, which were interpreted considering standard models of stellar interiors and non-standard processes affecting Li depletion.}
   {Li is generally less strongly depleted in intermediate-mass stars than in their lower-mass counterparts, as expected from standard evolution models. However, Li abundances significantly below the cosmic value are observed in 25 -- 30$\%$ of intermediate-mass stars. It is also unexpected that the results show no significant difference between the 1.5 --2.5 M$_{\odot}$ and 2.5 -- 3.5 M$_{\odot}$ subsamples. Evidence is provided showing that disk-locking works in young intermediate-mass stars. This constitutes independent support for the hypothesis that magnetospheric accretion scenario operates in these sources. We found that disk-locking is effective for a timescale that is about twice shorter than for lower-mass stars, before magnetospheres reduce their sizes during the transition from the IMTT to the Herbig regime. This contraction of the magnetosphere can explain the increase in rotation by a factor of about 3 and in accretion by a factor of about 4 that is observed during this transition. We propose a complex scenario linking rotation, accretion, and the surface Li abundance. Finally, we tentatively suggest that the known relation between the presence of planets and Li depletion might also be present in intermediate-mass MS stars and might originate in the pre-MS.}
   {We provide the most complete Li analysis and database focused on IMTT and Herbig stars to date. However, we emphasize the need of additional observations and non-standard models such as those available for their lower-mass analogs.}

   \keywords{Stars: evolution -- Stars: abundances -- Stars: interiors -- Stars: pre-main sequence -- Stars: variables: T Tauri, Herbig Ae/Be}

   \maketitle
\nolinenumbers

\section{Introduction}
\label{Sect:intro}
Lithium (Li) is the heaviest element produced in significant amounts immediately after the Big Bang, and it is one of the most frequently studied elements in astrophysics, mainly through the \ion{Li}{i} optical transitions at 6707.856 {\AA} \citep{Campbell23}. Studies of Li range from cosmological nucleosynthesis analyses to evolutionary investigations based on the internal processes of stars, brown dwarfs, and planets \citep[e.g., the review by][]{Martin23}. 

It is well known that Li is progressively depleted during the early evolution of T Tauri stars \citep[TTs hereafter; with stellar masses M$_*$ < 1.5 M$_{\odot}$; e.g.,][]{MartinRebolo94}, for which the Li abundance\footnote{This is defined as A(Li) = log[N(Li)/N(H)] + 12, where N is the atomic number density} is an age indicator for low-mass stars and solar analogs \citep[e.g.,][]{Bildsten97,Jeffries23,Gutierrez24}.  A(Li) typically ranges from the so-called cosmic abundance (typical of the youngest sources in our Galaxy) to lower values during the main sequence (MS). This evolution is explained by the deep convective sub-photospheric layer that characterizes TTs. This layer transports Li from the surfaces to the interiors until Li is burned at temperatures $\gtrsim$ 2.5 x 10$^6$ K \citep[e.g., the review by][]{Jeffries06}. The observed spread among stars with similar ages and masses is mainly attributed to differences in accretion and magnetic activity, rotation, or the presence of planets \citep[e.g.,][]{Piau02,Baraffe17,Israelian01,Israelian09,Jackson25}.

In contrast, Li measurements of young intermediate-mass Herbig stars \citep[e.g., the review by][]{Brittain23} are merely anecdotal and limited to a few sources \citep[e.g.,][]{King93,Cowley13}. Moreover, some of these Li detections do not refer to the central star, but serve as a probe of unresolved, less massive companions \citep[e.g.,][]{Martin94,Corporon99}. The main reason for the relatively small number of studies devoted to Herbig stars is that Li ionizes at $\sim$ 5.39 eV. Thus, the probability of observing the \ion{Li}{i} optical transition is virtually zero for sources hotter than $\sim$ 10000 K (which roughly corresponds to stars more massive than 3.5 M$_{\odot}$ during the pre-MS). No curves of growth (COG) that transform \ion{Li}{i} equivalent widths (EWs) into abundances have been systematically inferred for stellar temperatures above $\sim$ 8000 K \citep[but see, e.g.,][and references therein]{North05,Takeda12}. In addition, the initial mass function implies that intermediate-mass stars are intrinsically less numerous than low-mass stars, for which in most nearby star-forming regions the number of well-characterized Herbig stars remains small. Moreover, in distant massive clusters, observational limitations and crowding often hamper the homogeneous determination of stellar and circumstellar properties. Therefore, comparative evolutionary studies across regions remain challenging for intermediate-mass stars \citep[but see, e.g.,][]{Hernandez05,Ribas15}.

Nevertheless, measuring A(Li) of young stars with intermediate masses between $\sim$ 1.5 and 3.5 M$_{\odot}$ is essential to probe the critical change predicted by stellar interior models. Notably, during the pre-MS evolution of such stars, the stellar envelopes are expected to change from convective to radiative \citep[e.g.,][]{Palla93}. In turn, the appearance of radiative envelopes is expected to lead to the disappearance of magnetic fields \citep{Villebrun19}, and thus, to a change in the disk-to-star accretion paradigm \citep{Mendi20}. Although magnetic field measurements of young intermediate-mass stars are still affected by strong uncertainties \citep[see, e.g., the discussions in][]{Bagnulo12,Villebrun19,Mendi20}, many lines of evidence suggest that the magnetospheric accretion scenario currently accepted for TTs is also valid for late-type Herbig stars, but hardly for sources more massive than $\sim $ 3-4 M$_{\odot}$ \citep[e.g.,][]{Mendi20,Wichittanakom20,Vioque22,Grant22,Rogers25}. On the other hand, the transition to radiative envelopes is also expected to lead to a lower convective mixing efficiency with the stellar interiors, which could be directly probed by the Li content measured at the surfaces of the stars. Such measurements may help us to better understand the evolution of Herbig stars based on their precursors, the so-called intermediate-mass T Tauri stars \citep[IMTTs; e.g.,][]{Calvet04,Villebrun19,Valegard21}. It is particularly relevant to compare the evolution of stellar interiors, based on Li measurements, with recent results indicating that mass accretion rates do not show the expected decline from the IMTT to the Herbig regime \citep{Brittain25}.

We contribute to our understanding of the evolution of young intermediate-mass stars by focusing on A(Li). Section \ref{Sect:sample_obs} describes the sample and observations. In addition to the Li equivalent widths and abundances based on extended COG, our results include new determinations of stellar metallicities, rotational velocities and periods, and accretion rates, which are presented in Sect. \ref{Sect:results}. Section \ref{Sect:analysis_discussion} analyses the evolution of A(Li) in young intermediate-mass stars and its potential relation with nonstandard processes such as stellar rotation, accretion, or the presence of planets. Finally, our main results and conclusions are summarized in Sect. \ref{Sect:summary_conclusions}.

\section{Sample selection and observations}
\label{Sect:sample_obs}
The sample was mainly composed of young intermediate-mass stars analyzed in \citet{GuzmanDiaz23}, where the surface temperature (T$_*$) and gravity (log g), stellar mass (M$_*$), luminosity (L$_*$), age (t$_*$), metallicity ([Fe/H]), and projected rotational velocity ($v \sin i$) were derived for each star based on ESO optical spectra. We selected sources with 1.5 $\leq$ M$_*$/M${_\odot}$ $\leq$ 3.5 and 5000 $\leq$ T$_*$ (K) $<$ 10000. Stars showing the \ion{Li}{i} optical feature but having confirmed close companions (separated by $\leq$ 1 $\arcsec$) were removed, making a total of 43 sources. This restriction avoids spectral contamination and prevented us from including sources whose Li evolution is affected by interactions with close companions. The sample was complemented with 14 additional sources from \citet{Valegard21} and with one source from \citet{GuzmanDiaz21} with the same previous properties and appropriate ESO optical spectra (see below). Finally, 13 young intermediate-mass stars without close companions and with \ion{Li}{i} detections from \citet{King93} were added to the sample. 

Table \ref{table:sample} lists the main stellar properties of the 71 stars we selected in this way. The use of different pre-MS stellar evolution models to infer masses, ages, and surface gravities does not significantly affect our results because it was shown that differences between these models become relevant mainly for colder sources \citep{Hillenbrand04,Stassun14}. Table \ref{table:sample} also lists the spectral energy distribution (SED) groups based on the classification by \citet{Meeus01}, which are available in \citet{GuzmanDiaz21} and \citet{Valegard21} for most stars in the sample and were also necessary for our analysis. 

\begin{figure}
   \centering
   \includegraphics[width=9cm]{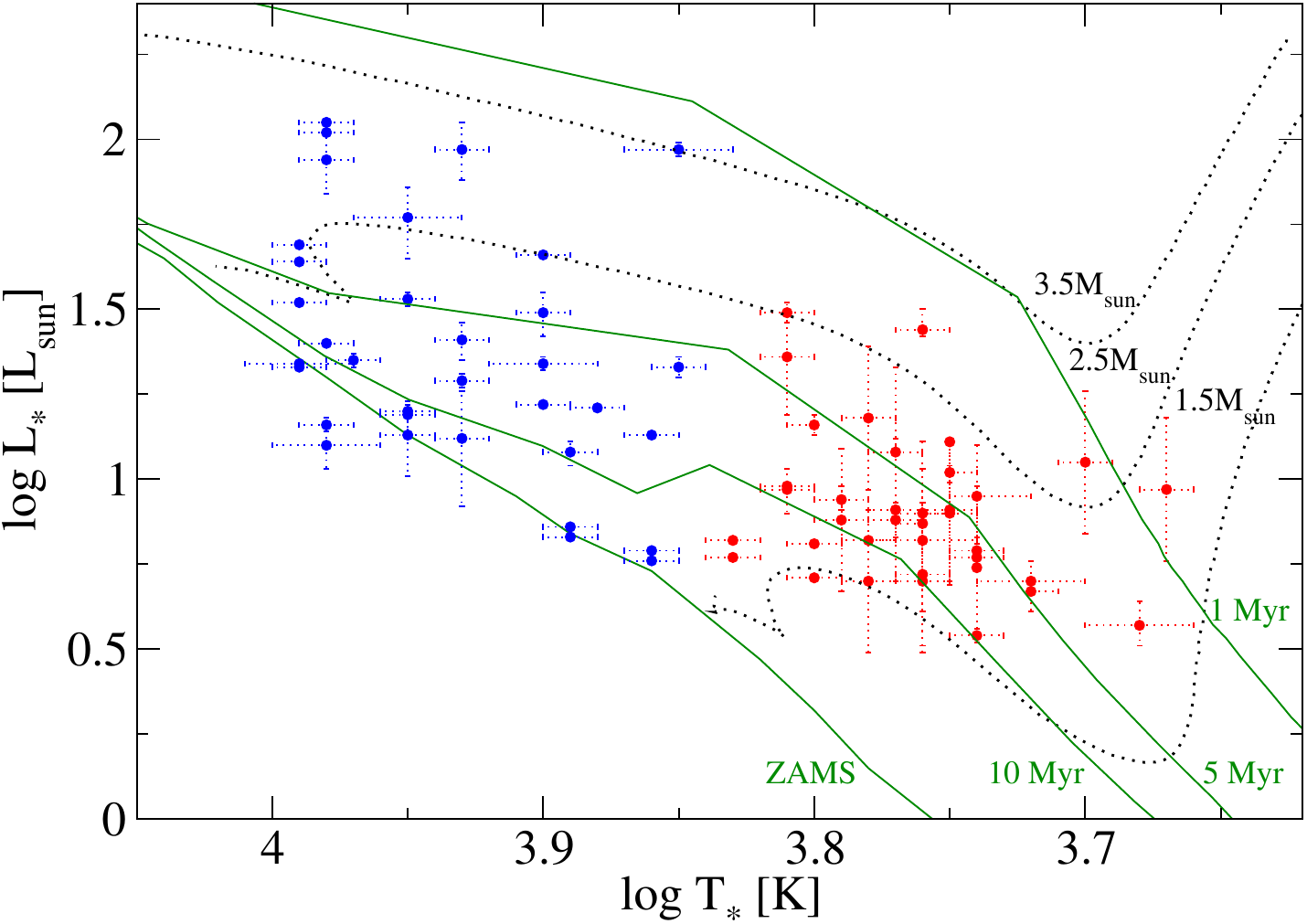}
      \caption{HR diagram for the 34 Herbig (blue) and 37 IMTT (red) sample stars. Representative pre-MS tracks (black) and isochrones (green) from \citet{Siess00} are overplotted.} 
         \label{fig:HRdiagram}
   \end{figure}

While all stars in the sample share the same stellar mass range, the term "Herbig" is reserved in this work for sources with T$_*$ > 7000 K (i.e., spectral types earlier than F), the rest are their IMTT precursors. Figure \ref{fig:HRdiagram} shows the Hertzsprung-Russell (HR) diagram for the 37 IMTTs and 34 Herbigs in our sample, illustrating the evolutionary pattern that links the two regimes and that all sources will finally be mostly A-type stars when they reach the zero-age main sequence (ZAMS). Our sample is the most complete to date for a Li-based evolution analysis of intermediate-mass pre-MS stars with protoplanetary disks. For instance, we cross-matched the almost half million sources with A(Li) measurements from the Large Sky Area Multi-Object Fiber Spectroscopic Telescope (LAMOST) DR9 survey \citep{Ding24} with the 20 young intermediate-mass stars identified from LAMOST DR8 \citep{Zhang22} and with the 145975 young accreting stars recently identified based on \textit{Gaia} \citep{Delfini25}. This led to only 5 additional sources with Li spectra with a signal-to-noise ratio (S/N) that was high enough ($>$ 20) and potentially M$_*$ $\geq$ 1.5 M${_\odot}$. Because their stellar and multiplicity properties are comparatively less well known, and also for consistency reasons, we decided to keep the analysis focused on our 71 sample stars.

Table \ref{table:observations} lists the properties of the spectra we used to perform the analysis. All spectra have a high spectral resolution (R = $\lambda$/$\delta$$\lambda$ $\geq$ 30000), S/N $\geq$ 30, and most were taken with the UVES or X-Shooter spectrographs. In all cases, the \ion{Li}{i} and H$\alpha$ features were covered simultaneously. For stars with several optical spectra, we selected the spectrum with values of R and S/N that minimized the uncertainty of the Li EW (Sect. \ref{Sect:abundances}). All processed spectra are available at the ESO Science Portal\footnote{https://archive.eso.org/scienceportal/home} and were reduced following standard procedures by the corresponding pipelines. The spectral information for the stars selected from \citet{King93} (see Table \ref{table:sample}) is not included in Table \ref{table:observations} because their Li and H$\alpha$ measurements are adopted here. In short, those spectra were taken at the Lick Observatory 3 m telescope, and their spectral resolution and S/N are similar to the spectra used here.

\begin{table*}
\centering
\scriptsize
\caption{Sample properties.}
\label{table:sample}
\centering
\begin{tabular}{lllllllll}
\hline\hline
Name & T$_*$ & log g & M$_*$ & log L$_*$ & t$_*$ & d & SED & Ref \\
... & K & [cm s$^{-2}$] & M$_{\odot}$ & [L$_{\odot}$] & Myr & pc & ... & ... \\ 
\hline\hline
PDS 2 & 6750$^{+125}_{-125}$ & 4.23 $\pm$ 0.05 & 1.46$^{+0.04}_{-0.01}$ & 0.77$^{+0.01}_{-0.01}$ & 15.04$^{+2.67}_{-2.50}$ & 399.0$^{+2.7}_{-2.7}$ & I & GD+23 \\
HD 9672 & 9000$^{+125}_{-125}$ & 3.97 $\pm$ 0.07 & 1.93$^{+0.02}_{-0.03}$ & 1.20$^{+0.02}_{-0.02}$ & 11.02$^{+8.96}_{-1.15}$ & 57.1$^{+0.2}_{-0.2}$ & II & GD+23 \\
LkH$\alpha$ 330 & 6240$^{+100}_{-70}$ & 3.70 $\pm$ 0.04 & 1.93$^{+0.07}_{-0.07}$ & 1.16$^{+0.03}_{-0.03}$ & 4.66$^{+0.65}_{-0.26}$ & 308.4$^{+7.7}_{-7.4}$ & I & V+21 \\
RY Tau & 5945$^{+143}_{-143}$ & 3.70 $\pm$ 0.21 & 1.95$^{+0.10}_{-0.17}$ & 1.08$^{+0.25}_{-0.15}$ & 4.29$^{+1.43}_{-0.89}$ & 133.5$^{+54.8}_{-30.1}$ & II & V+21 \\
UX Tau A & 5490$^{+130}_{-210}$ & 3.77 $\pm$ 0.17 & 2.34$^{+0.29}_{-0.43}$ & 0.95$^{+0.15}_{-0.14}$ & 1.26$^{+1.03}_{-0.63}$ & 139.4$^{+2.0}_{-2.0}$ & I & V+21 \\
HBC 415 & 5770$^{+110}_{-50}$ & 3.81 $\pm$ 0.03 & 1.78$^{+0.07}_{-0.08}$ & 0.87$^{+0.02}_{-0.01}$ & 5.19$^{+0.96}_{-0.61}$ & 165.2$^{+1.3}_{-1.3}$ & Debris & V+21 \\
SU Aur & 5680$^{+40}_{-20}$ & 3.65 $\pm$ 0.01 & 2.22$^{+0.02}_{-0.05}$ & 1.11$^{+0.01}_{-0.01}$ & 2.92$^{+0.19}_{-0.41}$ & 157.7$^{+1.5}_{-1.5}$ & I & V+21 \\
HD 31648 & 8000$^{+125}_{-125}$ & 4.13 $\pm$ 0.05 & 1.85$^{+0.04}_{-0.01}$ & 1.22$^{+0.01}_{-0.01}$ & 7.71$^{+0.23}_{-0.38}$ & 155.2$^{+1.3}_{-1.2}$ & II & GD+23 \\
UX Ori & 8500$^{+250}_{-250}$ & 3.75 $\pm$ 0.12 & 1.91$^{+0.04}_{-0.00}$ & 1.12$^{+0.14}_{-0.20}$ & 9.84$^{+0.17}_{-0.00}$ & 319.8$^{+2.9}_{-2.9}$ & II & GD+23 \\
HD 34282 & 9500$^{+250}_{-250}$ & 4.43 $\pm$ 0.10 & $<$1.90 & 1.16$^{+0.02}_{-0.02}$ & $<$19.87 & 306.5$^{+2.2}_{-2.2}$ & I & GD+23 \\
HD 290380 & 6250$^{+125}_{-125}$ & 3.98 $\pm$ 0.05 & 1.59$^{+0.06}_{-0.06}$ & 0.81$^{+0.01}_{-0.01}$ & 9.32$^{+0.75}_{-1.31}$ & 343.3$^{+2.5}_{-2.5}$ & II & GD+23 \\
V346 Ori & 7750$^{+250}_{-250}$ & 4.50 $\pm$ 0.05 & 1.65$^{+0.04}_{-0.04}$ & 0.86$^{+0.01}_{-0.01}$ & 16.21$^{+3.37}_{-5.51}$ & 336.2$^{+2.2}_{-2.2}$ & I & GD+23 \\
CO Ori & 6500$^{+215}_{-215}$ & 3.76 $\pm$ 0.05 & 2.30$^{+0.30}_{-0.35}$ & 1.36$^{+0.12}_{-0.17}$ & 3.92$^{+2.07}_{-1.19}$ & 394.7$^{+3.2}_{-3.2}$ & II & GD+23 \\
HD 35929 & 7000$^{+250}_{-250}$ & 3.50 $\pm$ 0.05 & 3.53$^{+0.08}_{-0.13}$ & 1.97$^{+0.02}_{-0.02}$ & 1.20$^{+0.30}_{-0.16}$ & 376.7$^{+3.4}_{-3.4}$ & II & GD+23 \\
HD 290500 & 9500$^{+500}_{-500}$ & 3.79 $\pm$ 0.25 & 1.85$^{+0.05}_{-0.00}$ & 1.10$^{+0.06}_{-0.07}$ & $<$19.92 & 402.5$^{+5.5}_{-5.3}$ & I & GD+23 \\
HD 244314 & 8500$^{+250}_{-250}$ & 4.08 $\pm$ 0.05 & 2.12$^{+0.04}_{-0.07}$ & 1.29$^{+0.02}_{-0.02}$ & 6.99$^{+0.63}_{-0.00}$ & 398.1$^{+3.2}_{-3.1}$ & II & GD+23 \\
HD 244604 & 9000$^{+250}_{-250}$ & 3.95 $\pm$ 0.18 & 2.16$^{+0.04}_{-0.01}$ & 1.53$^{+0.02}_{-0.02}$ & 5.08$^{+0.16}_{-0.08}$ & 398.4$^{+3.2}_{-3.2}$ & II & GD+23 \\
V1044 Ori & 5500$^{+140}_{-140}$ & 3.84 $\pm$ 0.06 & 1.89$^{+0.12}_{-0.17}$ & 0.79$^{+0.04}_{-0.03}$ & 4.00$^{+1.38}_{-1.12}$ & 388.0$^{+5.9}_{-5.8}$ & I & V+21 \\
EZ Ori & 5830$^{+88}_{-87}$ & 3.82 $\pm$ 0.06 & 1.75$^{+0.09}_{-0.09}$ & 0.88$^{+0.05}_{-0.05}$ & 5.51$^{+1.06}_{-0.81}$ & 399.0$^{+26.5}_{-23.4}$ & I/II & V+21 \\
Brun 225 & 5781$^{+93}_{-93}$ & 3.94 $\pm$ 0.20 & 1.60$^{+0.16}_{-0.16}$ & 0.70$^{+0.21}_{-0.21}$ & 6.03$^{+4.17}_{-4.17}$ & 395.1$^{+3.8}_{-3.8}$ & ... & K+93 \\
Brun 381 & 6040$^{+83}_{-83}$ & 3.68 $\pm$ 0.20 & 2.20$^{+0.22}_{-0.22}$ & 1.18$^{+0.21}_{-0.21}$ & 2.00$^{+1.38}_{-1.38}$ & 135.8$^{+0.3}_{-0.3}$ & ... & K+93 \\
V2087 Ori & 5768$^{+213}_{-213}$ & 3.84 $\pm$ 0.20 & 1.70$^{+0.17}_{-0.17}$ & 0.82$^{+0.21}_{-0.21}$ & 3.47$^{+2.40}_{-2.40}$ & 378.9$^{+3.4}_{-3.4}$ & ... & K+93 \\
Brun 555 & 4699$^{+151}_{-151}$ & 3.45 $\pm$ 0.20 & 2.20$^{+0.22}_{-0.22}$ & 0.97$^{+0.21}_{-0.21}$ & 0.60$^{+0.42}_{-0.42}$ & 407.2$^{+8.4}_{-8.4}$ & ... & K+93 \\
Brun 656 & 5770$^{+150}_{-90}$ & 3.43 $\pm$ 0.06 & 2.72$^{+0.21}_{-0.15}$ & 1.44$^{+0.06}_{-0.02}$ & 1.53$^{+0.48}_{-0.18}$ & 466.8$^{+10.2}_{-9.8}$ & Debris & V+21 \\
Brun 684 & 5623$^{+78}_{-78}$ & 3.76 $\pm$ 0.20 & 1.85$^{+0.19}_{-0.19}$ & 0.90$^{+0.21}_{-0.21}$ & 3.02$^{+2.09}_{-2.09}$ & 333.7$^{+74.6}_{-74.6}$ & ... & K+93 \\
NV Ori & 7000$^{+125}_{-125}$ & 3.77 $\pm$ 0.05 & 2.09$^{+0.06}_{-0.09}$ & 1.33$^{+0.03}_{-0.03}$ & 5.04$^{+0.85}_{-0.25}$ & 383.9$^{+3.0}_{-3.0}$ & I & GD+23 \\
BD-07 1129 & 4966$^{+114}_{-114}$ & 3.49 $\pm$ 0.20 & 2.30$^{+0.23}_{-0.23}$ & 1.05$^{+0.21}_{-0.21}$ & 0.89$^{+0.62}_{-0.62}$ & 364.9$^{+1.9}_{-1.9}$ & ... & K+93 \\
T Ori & 9000$^{+500}_{-500}$ & 3.65 $\pm$ 0.17 & 2.52$^{+0.19}_{-0.21}$ & 1.77$^{+0.09}_{-0.12}$ & 3.82$^{+0.59}_{-0.82}$ & 398.6$^{+4.5}_{-4.4}$ & ... & GD+23 \\
V815 Ori & 5530$^{+60}_{-40}$ & 3.87 $\pm$ 0.02 & 1.76$^{+0.06}_{-0.05}$ & 0.74$^{+0.01}_{-0.01}$ & 4.82$^{+0.45}_{-0.50}$ & 397.5$^{+4.9}_{-4.8}$ & I & V+21 \\
CQ Tau & 6750$^{+125}_{-125}$ & 4.13 $\pm$ 0.05 & 1.50$^{+0.01}_{-0.01}$ & 0.82$^{+0.01}_{-0.01}$ & 12.42$^{+2.40}_{-1.43}$ & 148.6$^{+1.3}_{-1.3}$ & I & GD+23 \\
HD 294258 & 6194$^{+185}_{-185}$ & 3.92 $\pm$ 0.20 & 1.75$^{+0.18}_{-0.18}$ & 0.88$^{+0.21}_{-0.21}$ & 4.47$^{+3.09}_{-3.09}$ & 393.8$^{+2.5}_{-2.5}$ & ... & K+93 \\
Parenago 2374 & 5848$^{+67}_{-67}$ & 3.78 $\pm$ 0.20 & 1.70$^{+0.17}_{-0.17}$ & 0.91$^{+0.21}_{-0.21}$ & 3.09$^{+2.13}_{-2.13}$ & 244.6$^{+0.8}_{-0.8}$ & ... & K+93 \\
Brun 973 & 5821$^{+94}_{-94}$ & 3.78 $\pm$ 0.20 & 1.70$^{+0.17}_{-0.17}$ & 0.90$^{+0.21}_{-0.21}$ & 3.39$^{+2.34}_{-2.34}$ & 385.7$^{+3.0}_{-3.0}$ & ... & K+93 \\
Brun 1004 & 5546$^{+64}_{-64}$ & 3.81 $\pm$ 0.20 & 1.65$^{+0.17}_{-0.17}$ & 0.77$^{+0.21}_{-0.21}$ & 2.95$^{+2.04}_{-2.04}$ & 385.2$^{+2.5}_{-2.5}$ & ... & K+93 \\
Brun 1025 & 6095$^{+112}_{-112}$ & 4.01 $\pm$ 0.20 & 1.50$^{+0.15}_{-0.15}$ & 0.70$^{+0.21}_{-0.21}$ & 6.46$^{+4.46}_{-4.46}$ & ... & ... & K+93 \\
HD 294260 & 6115$^{+168}_{-168}$ & 3.82 $\pm$ 0.06 & 1.68$^{+0.13}_{-0.07}$ & 0.94$^{+0.04}_{-0.03}$ & 6.72$^{+0.84}_{-1.54}$ & 406.7$^{+7.0}_{-6.7}$ & II & V+21 \\
Brun 1060 & 5768$^{+93}_{-93}$ & 3.92 $\pm$ 0.20 & 1.60$^{+0.16}_{-0.16}$ & 0.72$^{+0.21}_{-0.21}$ & 4.79$^{+3.31}_{-3.31}$ & 317.6$^{+2.3}_{-2.3}$ & ... & K+93 \\
BF Ori & 9000$^{+250}_{-250}$ & 3.87 $\pm$ 0.34 & 1.85$^{+0.15}_{-0.00}$ & 1.13$^{+0.09}_{-0.12}$ & 17.14$^{+2.80}_{-0.00}$ & 377.9$^{+4.0}_{-3.9}$ & II & GD+23 \\
HD 37357 & 9500$^{+250}_{-250}$ & 4.19 $\pm$ 0.10 & 2.80$^{+0.20}_{-0.20}$ & 1.94$^{+0.08}_{-0.10}$ & 2.96$^{+0.44}_{-0.74}$ & 465.1$^{+53.2}_{-41.3}$ & II & GD+23 \\
HD 290764 & 7875$^{+375}_{-375}$ & 3.94 $\pm$ 0.04 & 1.99$^{+0.03}_{-0.04}$ & 1.34$^{+0.02}_{-0.02}$ & 6.10$^{+0.62}_{-0.12}$ & 397.1$^{+3.1}_{-3.1}$ & I & GD+23 \\
HD 294290 & 5598$^{+129}_{-129}$ & 3.75 $\pm$ 0.20 & 1.90$^{+0.19}_{-0.19}$ & 0.91$^{+0.21}_{-0.21}$ & 2.69$^{+1.86}_{-1.86}$ & 323.6$^{+2.6}_{-2.6}$ & ... & K+93 \\
V599 Ori & 8000$^{+250}_{-250}$ & 3.87 $\pm$ 0.05 & 2.15$^{+0.11}_{-0.10}$ & 1.49$^{+0.06}_{-0.07}$ & 5.09$^{+0.91}_{-0.51}$ & 401.3$^{+3.2}_{-3.2}$ & I & GD+23 \\
HD 39014 & 8000$^{+125}_{-125}$ & 3.80 $\pm$ 0.05 & 2.48$^{+0.07}_{-0.06}$ & 1.66$^{+0.01}_{-0.01}$ & 3.52$^{+0.28}_{-0.31}$ & 45.7$^{+0.2}_{-0.2}$ & II & GD+23 \\
HBC 217 & 6000$^{+125}_{-125}$ & 3.95 $\pm$ 0.05 & 1.75$^{+0.10}_{-0.10}$ & 0.82$^{+0.01}_{-0.01}$ & 6.89$^{+1.13}_{-1.55}$ & 704.6$^{+6.9}_{-6.7}$ & I & GD+23 \\
HBC 222 & 6500$^{+174}_{-174}$ & 3.94 $\pm$ 0.05 & 1.70$^{+0.12}_{-0.11}$ & 0.97$^{+0.06}_{-0.07}$ & 8.24$^{+1.77}_{-0.00}$ & 701.1$^{+7.0}_{-6.8}$ & ... & GD+23 \\
HD 68695 & 9250$^{+250}_{-250}$ & 4.35 $\pm$ 0.08 & 2.08$^{+0.06}_{-0.03}$ & 1.35$^{+0.02}_{-0.02}$ & 8.02$^{+0.91}_{-0.74}$ & 374.6$^{+2.1}_{-2.1}$ & I & GD+23 \\
PDS 277 & 6500$^{+125}_{-125}$ & 3.89 $\pm$ 0.04 & 1.70$^{+0.05}_{-0.05}$ & 0.98$^{+0.01}_{-0.01}$ & 8.15$^{+1.18}_{-0.98}$ & 340.7$^{+1.5}_{-1.5}$ & I & GD+21 \\
GSC 8581-2002 & 9750$^{+250}_{-250}$ & 3.90 $\pm$ 0.05 & 2.40$^{+0.00}_{-0.09}$ & 1.52$^{+0.01}_{-0.01}$ & 5.12$^{+0.86}_{-0.12}$ & 534.1$^{+3.6}_{-3.6}$ & I & GD+23 \\
PDS 33 & 9750$^{+250}_{-250}$ & 4.41 $\pm$ 0.08 & $<$2.10 & 1.33$^{+0.01}_{-0.01}$ & $<$9.19 & 880.2$^{+8.7}_{-8.5}$ & I & GD+23 \\
CR Cha & 4800$^{+230}_{-230}$ & 3.75 $\pm$ 0.14 & 1.62$^{+0.28}_{-0.39}$ & 0.57$^{+0.07}_{-0.06}$ & 1.51$^{+1.59}_{-0.80}$ & 186.5$^{+0.8}_{-0.8}$ & II & V+21 \\
Ass Cha T 2-21 & 5660$^{+20}_{-70}$ & 3.71 $\pm$ 0.03 & 2.09$^{+0.09}_{-0.06}$ & 1.02$^{+0.02}_{-0.03}$ & 3.23$^{+0.17}_{-0.36}$ & 164.8$^{+4.0}_{-3.8}$ & Debris & V+21 \\
CV Cha & 5280$^{+60}_{-110}$ & 3.88 $\pm$ 0.03 & 1.85$^{+0.05}_{-0.05}$ & 0.67$^{+0.02}_{-0.01}$ & 3.44$^{+0.51}_{-0.87}$ & 192.2$^{+1.0}_{-1.0}$ & II & V+21 \\
Ass ChaT2-54 & 5260$^{+200}_{-200}$ & 3.86 $\pm$ 0.11 & 1.92$^{+0.08}_{-0.24}$ & 0.70$^{+0.06}_{-0.09}$ & 3.06$^{+2.24}_{-1.16}$ & 202.9$^{+19.2}_{-16.2}$ & Debris & V+21 \\
HD 100453 & 7250$^{+250}_{-250}$ & 4.47 $\pm$ 0.05 & 1.60$^{+0.05}_{-0.04}$ & 0.79$^{+0.01}_{-0.01}$ & 19.28$^{+0.70}_{-0.68}$ & 103.6$^{+0.2}_{-0.2}$ & I & GD+23 \\
HD 100546 & 9750$^{+500}_{-500}$ & 4.01 $\pm$ 0.03 & 2.10$^{+0.05}_{-0.03}$ & 1.34$^{+0.01}_{-0.01}$ & 7.67$^{+0.36}_{-0.67}$ & 108.0$^{+0.4}_{-0.4}$ & I & GD+23 \\
HD 101412 & 9750$^{+250}_{-250}$ & 4.32 $\pm$ 0.19 & 2.39$^{+0.01}_{-0.02}$ & 1.69$^{+0.01}_{-0.01}$ & 4.06$^{+0.08}_{-0.01}$ & 409.9$^{+2.5}_{-2.5}$ & II & GD+23 \\
HD 135344B & 6375$^{+125}_{-125}$ & 4.15 $\pm$ 0.05 & 1.46$^{+0.07}_{-0.04}$ & 0.71$^{+0.01}_{-0.01}$ & 10.48$^{+0.97}_{-0.49}$ & 134.4$^{+0.4}_{-0.4}$ & I & GD+23 \\
HD 139614 & 7750$^{+250}_{-250}$ & 4.45 $\pm$ 0.05 & 1.60$^{+0.02}_{-0.00}$ & 0.83$^{+0.01}_{-0.01}$ & 19.35$^{+0.64}_{-0.00}$ & 133.1$^{+0.5}_{-0.5}$ & I & GD+23 \\
HD 141569 & 9500$^{+250}_{-250}$ & 4.09 $\pm$ 0.08 & 2.12$^{+0.04}_{-0.01}$ & 1.40$^{+0.01}_{-0.01}$ & 7.97$^{+0.03}_{-0.03}$ & 111.1$^{+0.4}_{-0.4}$ & II & GD+23 \\
HD 142666 & 7250$^{+250}_{-250}$ & 3.99 $\pm$ 0.05 & 1.75$^{+0.03}_{-0.00}$ & 1.13$^{+0.01}_{-0.01}$ & 8.73$^{+0.08}_{-0.74}$ & 145.5$^{+0.5}_{-0.5}$ & II & GD+23 \\
HD 143006 & 5500$^{+125}_{-125}$ & 4.10 $\pm$ 0.05 & 1.74$^{+0.10}_{-0.13}$ & 0.54$^{+0.02}_{-0.02}$ & 5.10$^{+1.89}_{-0.00}$ & 166.4$^{+0.5}_{-0.5}$ & I & GD+23 \\
HD 144432 & 7500$^{+250}_{-250}$ & 4.00 $\pm$ 0.05 & 1.82$^{+0.03}_{-0.01}$ & 1.21$^{+0.01}_{-0.01}$ & 7.98$^{+0.02}_{-0.21}$ & 154.0$^{+0.6}_{-0.6}$ & II & GD+23 \\
HR 5999 & 8500$^{+250}_{-250}$ & 3.50 $\pm$ 0.11 & 3.19$^{+0.21}_{-0.29}$ & 1.97$^{+0.08}_{-0.09}$ & 1.98$^{+0.33}_{-0.41}$ & 157.8$^{+0.9}_{-0.8}$ & II & GD+23 \\
V718 Sco & 7750$^{+250}_{-250}$ & 4.27 $\pm$ 0.05 & 1.71$^{+0.04}_{-0.02}$ & 1.08$^{+0.03}_{-0.04}$ & 9.02$^{+0.56}_{-0.11}$ & 153.9$^{+0.5}_{-0.5}$ & II & GD+23 \\
HD 149914 & 9500$^{+125}_{-125}$ & 3.50 $\pm$ 0.11 & 3.07$^{+0.07}_{-0.05}$ & 2.05$^{+0.01}_{-0.01}$ & 2.02$^{+0.05}_{-0.02}$ & 153.5$^{+0.6}_{-0.6}$ & II & GD+23 \\
HD 163296 & 9000$^{+250}_{-250}$ & 4.06 $\pm$ 0.20 & 1.91$^{+0.12}_{-0.00}$ & 1.19$^{+0.04}_{-0.05}$ & 10.00$^{+9.50}_{-2.00}$ & 100.6$^{+0.4}_{-0.4}$ & II & GD+23 \\
HD 169142 & 7250$^{+125}_{-125}$ & 4.34 $\pm$ 0.05 & 1.55$^{+0.03}_{-0.00}$ & 0.76$^{+0.01}_{-0.01}$ & $<$20.00 & 114.4$^{+0.4}_{-0.4}$ & I & GD+23 \\
HD 176386 & 9750$^{+125}_{-125}$ & 3.70 $\pm$ 0.07 & 2.63$^{+0.00}_{-0.25}$ & 1.64$^{+0.01}_{-0.01}$ & 3.88$^{+0.57}_{-0.01}$ & 154.3$^{+0.7}_{-0.7}$ & ... & GD+23 \\
HD 179218 & 9500$^{+250}_{-250}$ & 3.90 $\pm$ 0.04 & 2.99$^{+0.01}_{-0.04}$ & 2.02$^{+0.01}_{-0.01}$ & 2.35$^{+0.19}_{-0.16}$ & 258.0$^{+2.2}_{-2.2}$ & I & GD+23 \\
WW Vul & 8500$^{+125}_{-125}$ & 3.65 $\pm$ 0.01 & 2.04$^{+0.06}_{-0.05}$ & 1.41$^{+0.05}_{-0.06}$ & 6.02$^{+0.52}_{-0.20}$ & 480.1$^{+4.3}_{-4.2}$ & II & GD+23 \\
PX Vul & 6500$^{+125}_{-125}$ & 3.63 $\pm$ 0.05 & 2.59$^{+0.13}_{-0.18}$ & 1.49$^{+0.03}_{-0.03}$ & 2.95$^{+0.20}_{-0.64}$ & 621.7$^{+20.5}_{-19.3}$ & II & GD+23 \\
\hline
\hline
\end{tabular}
\begin{minipage}{18cm}
\textbf{Notes:} For each star in Col. 1, the surface temperature and gravity, stellar mass, luminosity, and age are listed on Cols. 2 -- 6. \textit{Gaia} distances and SED group according with the classification by \citet{Meeus01} are listed in Cols. 7 and 8. The last column indicates the literature reference. The full table is available electronically at the CDS. \\
\textbf{References:} GD+23: \citet{GuzmanDiaz23}, V+21: \citet{Valegard21}, K+93: \citet{King93}, GD+21: \citet{GuzmanDiaz21}\\
\end{minipage}
\end{table*}

\begin{table*}
\centering
\caption{Observations.}
\label{table:observations}
\centering
\begin{tabular}{llllll}
\hline\hline
Name & Instrument & Date & S/N$_{Li}$ & S/N$_{H\alpha}$ &  $\delta$$\lambda$\\
... & ... & year-month-day hour:min:sec & ... & ...& {\AA} pixel$^{-1}$\\ 
\hline\hline
PDS 2 & XSHOOTER & 2010-08-01 09:50:39 & 365 & 390 & 0.2 \\
HD 9672 & UVES & 2012-09-23 08:15:31 & 750 & 750 & 0.02 \\
LkH$\alpha$ 330 & UVES & 2009-08-27 09:08:07 & 100 & 150 & 0.02 \\
RY Tau & UVES & 2018-12-11 05:32:16 & 65 & 80 & 0.02 \\
UX Tau A & XSHOOTER & 2012-11-15 06:32:37 & 140 & 200 & 0.2 \\
HBC 415 & XSHOOTER & 2014-12-10 01:56:09 & 240 & 250 & 0.2 \\
SU Aur & UVES & 2008-12-05 02:23:18 & 35 & 45 & 0.04 \\
HD 31648 & XSHOOTER & 2014-12-02 03:41:51 & 690 & 740 & 0.2 \\
UX Ori & XSHOOTER & 2009-10-05 07:27:31 & 220 & 225 & 0.2 \\
HD 34282 & UVES & 2007-12-29 04:54:13 & 135 & 135 & 0.02 \\
HD 290380 & UVES & 2017-09-24 08:37:03 & 90 & 140 & 0.02 \\
V346 Ori & XSHOOTER & 2009-12-06 05:24:11 & 360 & 380 & 0.2 \\
CO Ori & UVES & 2010-01-09 03:22:14 & 75 & 80 & 0.02 \\
HD 35929 & UVES & 2009-10-04 08:56:05 & 160 & 165 & 0.02 \\
HD 290500 & XSHOOTER & 2009-12-17 03:23:29 & 310 & 340 & 0.2 \\
HD 244314 & XSHOOTER & 2010-01-02 00:55:30 & 175 & 175 & 0.2 \\
HD 244604 & XSHOOTER & 2009-12-17 04:45:34 & 270 & 270 & 0.2 \\
V1044 Ori & XSHOOTER & 2014-12-10 00:54:05 & 200 & 275 & 0.2 \\
EZ Ori & UVES & 2009-01-04 04:01:19 & 30 & 50 & 0.04 \\
Brun 656 & UVES & 2001-12-30 01:12:18 & 275 & 250 & 0.02 \\
NV Ori & XSHOOTER & 2012-12-19 06:28:09 & 235 & 250 & 0.2 \\
T Ori & UVES & 2009-12-29 05:13:02 & 125 & 125 & 0.02 \\
V815 Ori & UVES & 2009-02-13 02:34:09 & 120 & 150 & 0.03 \\
CQ Tau & XSHOOTER & 2012-12-12 05:42:30 & 180 & 195 & 0.2 \\
HD 294260 & XSHOOTER & 2014-12-05 01:58:12 & 180 & 200 & 0.2 \\
BF Ori & XSHOOTER & 2014-12-02 03:13:58 & 452 & 480 & 0.2 \\
HD 37357 & UVES & 2010-01-01 04:24:31 & 115 & 115 & 0.02 \\
HD 290764 & XSHOOTER & 2009-12-26 06:58:55 & 295 & 315 & 0.2 \\
V599 Ori & XSHOOTER & 2010-01-06 01:54:43 & 205 & 210 & 0.2 \\
HD 39014 & UVES & 2007-10-30 04:08:34 & 350 & 350 & 0.02 \\
HBC 217 & GIRAFFE & 2012-10-25 08:21:25 & 40 & 40 & 0.05 \\
HBC 222 & GIRAFFE & 2012-10-26 08:05:10 & 85 & 75 & 0.05 \\
HD 68695 & XSHOOTER & 2009-12-21 06:22:47 & 245 & 260 & 0.2 \\
PDS 277 & XSHOOTER & 2021-12-03 07:55:50 & 100 & 125 & 0.2 \\
GSC 8581-2002 & XSHOOTER & 2009-12-21 07:15:17 & 235 & 255 & 0.2 \\
PDS 33 & XSHOOTER & 2009-12-21 06:44:05 & 195 & 205 & 0.2 \\
CR Cha & EXPRESSO & 2019-06-08 03:34:48 & 90 & 120 & 0.01 \\
Ass Cha T 2-21 & XSHOOTER & 2010-01-20 07:07:57 & 140 & 160 & 0.2 \\
CV Cha & UVES & 2016-05-20 00:53:46 & 150 & 200 & 0.02 \\
Ass ChaT2-54 & UVES & 2012-03-16 02:13:20 & 80 & 75 & 0.02 \\
HD 100453 & XSHOOTER & 2010-03-29 03:59:48 & 435 & 475 & 0.2 \\
HD 100546 & XSHOOTER & 2010-06-05 00:31:59 & 315 & 340 & 0.2 \\
HD 101412 & UVES & 2009-04-07 02:38:15 & 150 & 150 & 0.02 \\
HD 135344B & UVES & 2006-05-17 04:19:43 & 125 & 125 & 0.02 \\
HD 139614 & XSHOOTER & 2010-03-28 08:35:20 & 355 & 385 & 0.2 \\
HD 141569 & XSHOOTER & 2010-03-28 08:54:21 & 175 & 185 & 0.2 \\
HD 142666 & UVES & 2010-02-01 08:26:11 & 200 & 200 & 0.02 \\
HD 143006 & XSHOOTER & 2016-07-25 03:37:58 & 270 & 295 & 0.2 \\
HD 144432 & UVES & 2010-02-25 09:23:28 & 225 & 225 & 0.02 \\
HR 5999 & UVES & 2010-02-26 08:15:19 & 175 & 175 & 0.02 \\
V718 Sco & XSHOOTER & 2010-03-29 06:27:49 & 370 & 400 & 0.2 \\
HD 149914 & UVES & 2023-09-17 01:38:08 & 265 & 245 & 0.02 \\
HD 163296 & UVES & 2023-09-14 02:24:30 & 215 & 190 & 0.02 \\
HD 169142 & UVES & 2009-04-24 09:17:27 & 115 & 115 & 0.02 \\
HD 176386 & XSHOOTER & 2010-09-26 03:18:23 & 355 & 390 & 0.2 \\
HD 179218 & UVES & 2006-05-17 06:50:00 & 140 & 140 & 0.02 \\
WW Vul & UVES & 2009-04-15 08:50:07 & 90 & 90 & 0.02 \\
PX Vul & UVES & 2009-04-16 09:57:17 & 60 & 65 & 0.02 \\
\hline
\hline
\end{tabular}
\begin{minipage}{18cm}
\textbf{Notes:} For each star in Col. 1, the instrument, date of observation, S/N at the \ion{Li}{i}@6707.9 {\AA} and H$\alpha$@6562.8 transitions, and the spectral dispersion of the spectra we used are listed. Details of the spectra for the sources taken from \citet{King93} (see Table \ref{table:sample}) are provided in that work. The full table is available electronically at the CDS.\\  
\end{minipage}
\end{table*}

\section{Results}
\label{Sect:results}
\subsection{Stellar metallicity, rotation, and accretion}
\label{Sect:stellar_parameters}
We derived [Fe/H] and $v \sin i$ values for the sample stars from \citet{Valegard21} based on the corresponding ESO spectra indicated in Table \ref{table:observations}. Iron abundances were computed through spectral fitting within a Bayesian framework. Following \cite{2008JKAS...41...83T},  up to three spectral regions (537.5 -- 539.0 nm, 614.0 -- 617.0 nm, and 776.5 -- 778.5 nm) were used. In addition to [Fe/H], v$\sin i$ was varied in the synthetic spectra until it fit the observed one. Synthetic spectra were computed using the 2019 version of the MOOG\footnote{\url{https://www.as.utexas.edu/~chris/moog.html}} code \citep{1973PhDT.......180S}, together with the ATLAS9 atmospheric models \citep{1993KurCD..13.....K}. Further details will be provided in a forthcoming paper. [Fe/H] and $v \sin i$ values inferred by \citet{GuzmanDiaz23} for the corresponding sample stars were adopted. The $v \sin i$ values from \citet{King93} were also used for the corresponding 13 sample stars, but their metallicities are unknown, and no appropriate spectra are available. For this subsample we assumed a solar value, that is, the mean and median metallicity inferred for the remaining stars, with a conservative uncertainty of $\pm$ 0.25 dex. Columns 2 and 3 of Table \ref{table:results} list the [Fe/H] and $v \sin i$ values for the whole sample.

Inclinations to the line of sight were found in the literature for 26 sample stars, based on high-resolution imaging or interferometry of the surrounding disks (see below for the references). For sources with a significant misalignment between the inner and outer disks, the inclination of the former was selected because this likely reflects that of the central stars better \citep[e.g.,][]{Barber24}. We assumed i = $\pi$/4 for the rest; this is the typical value expected from random orientations \citep{Jackson10}. Stellar rotation periods (P) were then derived by combining the values of $v \sin i$, i, and the corresponding stellar radius (R$_*$). The error bars were obtained by propagating the uncertainties for the previous parameters, and we assumed an uncertainty of 10 $\%$ for the inclination errors not available in the literature. Inclinations with the corresponding references and periods are listed in Cols. 4 and 5 of Table \ref{table:results}. 

Accretion luminosities (L$_{\rm acc}$) and mass accretion rates ($\dot{\rm M}_{\rm acc}$) were also necessary for our study. Although accretion estimates are available for many sources in our sample, they were derived based on different methods and epochs \citep[e.g.,][]{GuzmanDiaz21,Brittain25}. Because our spectra cover the \ion{Li}{i} and H$\alpha$ features simultaneously, we inferred accretion rates based on the latter spectral line following the most homogeneous procedure possible. The H$\alpha$ EWs were measured from the observed spectra, and we corrected for photospheric absorption based on Kurucz model atmospheres with the same T$_*$ and log g as our stars. A conservative error bar of $\sim$ 0.01 {\AA} was assumed, which is significantly higher than that inferred from the spectral S/N and R. Optical extinctions and \textit{Gaia} distances were used to infer de-reddened line luminosities using the R or V magnitudes listed in SIMBAD, the corresponding zero-magnitude fluxes from \citet{Bessell79}, and the extinction law in \citet{Cardelli89}. The optical extinction A$_V$ was directly taken from \citet{GuzmanDiaz21} for 44 stars in our sample. The E(B-V) values listed in \citet{Valegard21} were converted into A$_V$ values using a standard total-to-selective extinction ratio R$_V$ = 3.1 for 12 sample stars included in that work. The error bars for the line luminosities consider 10$\%$ errors in continuum fluxes and the errors for the EWs, distances and A$_V$ values. We derived the de-reddened H$\alpha$ luminosities for 10 stars in our sample from the corresponding EWs (and errors) listed in \citet{King93}, the E(B-V) values inferred from SIMBAD, and intrinsic colors from \citet{Kenyon95}. The H$\alpha$ line luminosities were converted into accretion luminosities using the empirical calibration by \citet{Fairlamb17}, which is best suited for our sources (see Sect. \ref{Sect:Li_accretion}). The error bars for the accretion luminosities consider those of the calibrations and the uncertainties of the H$\alpha$ luminosities. Finally, mass accretion rates were derived from the expression $\dot{\rm M}_{\rm acc}$ = L$_{\rm acc}$R$_*$/GM$_*$, and their uncertainties were inferred from the propagation of errors in L$_{\rm acc}$, R$_*$, and M$_*$. Columns 6 to 8 of Table \ref{table:results} list the derived H$\alpha$ EWs, accretion luminosities, and mass accretion rates.   

\begin{table*}
\centering
\scriptsize
\caption{Additional properties and results.}
\label{table:results}
\centering
\begin{tabular}{lllllllllll}
\hline\hline
Name & [Fe/H]  & $v \sin i$ & i & P & EW$_{H\alpha}$ & log L$_{\rm acc}$ & log $\dot{\rm M}_{\rm acc}$ & EW$_{Li}$ & A(Li) & $\Delta$\\
... & ... & km s$^{-1}$ & deg & days & \AA & [L$_{\odot}$] & M$_{\odot}$ yr$^{-1}$ & nm & ... & ...\\ 
\hline\hline
PDS 2 & -0.10 & 15$\pm$1 & ... & 4.08$^{+0.47}_{-0.46}$ & -6.77 & -0.12$\pm$0.13 & -7.60$\pm$0.13 & <0.62 & <1.2 & ... \\
HD 9672 & 0.20$\pm$0.10 & 200$\pm$10 & 80.6$\pm$0.4$^1$ & 0.60$^{+0.06}_{-0.06}$ & -0.69 & -0.75$\pm$0.16 & -8.15$\pm$0.16 & <0.03 & <1.8 & ... \\
LkH$\alpha$ 330 & -0.12$\pm$0.20 & 32$\pm$1 & 27.5$\pm$0.2$^2$ & 2.36$^{+0.14}_{-0.12}$ & -14.71 & 1.12$\pm$0.10 & -6.16$\pm$0.10 & 108$^{+2}_{-0}$ & 3.1$^{+0.1}_{-0.1}$ & -0.09 \\
RY Tau & 0.32$\pm$0.28 & 50$\pm$2 & 65.0$\pm$0.1$^3$ & 2.97$^{+0.88}_{-0.55}$ & -13.88 & 0.23$\pm$0.30 & -7.05$\pm$0.32 & 243$^{+5}_{-13}$ & 3.6$^{+0.1}_{-0.4}$ & -0.31 \\
UX Tau A & 0.02$\pm$0.02 & 24$\pm$1 & 46$\pm$2$^4$ & 4.97$^{+0.94}_{-0.93}$ & -10.75 & ... & ... & 330$^{+0}_{-5}$ & >3.6 & -0.49 \\
HBC 415 & 0.02$\pm$0.56 & 128$\pm$12 & ... & 0.84$^{+0.11}_{-0.10}$ & -2.04 & -0.55$\pm$0.15 & -7.86$\pm$0.15 & 487$^{+6}_{-15}$ & >4.0 & ... \\
SU Aur & 0.48$\pm$0.06 & 58$\pm$3 & 56.9$\pm$0.4$^5$ & 2.68$^{+0.15}_{-0.14}$ & -7.27 & -0.06$\pm$0.13 & -7.33$\pm$0.13 & 234$^{+1}_{-1}$ & 3.4$^{+0.1}_{-0.1}$ & -0.17 \\
HD 31648 & -0.25$\pm$0.13 & 102$\pm$5 & 36$\pm$1$^6$ & 0.57$^{+0.05}_{-0.04}$ & -12.76 & 0.32$\pm$0.11 & -7.15$\pm$0.12 & <0.33 & <1.9 & ... \\
UX Ori & 0.00$\pm$0.10 & 225$\pm$11 & 70$\pm$5$^7$ & 0.65$^{+0.10}_{-0.10}$ & -11.44 & 0.46$\pm$0.13 & -6.83$\pm$0.15 & <1.03 & <2.9 & ... \\
HD 34282 & <0.00 & 105$\pm$5 & 56$\pm$1$^8$ & 0.56$^{+0.04}_{-0.04}$ & -8.07 & 0.06$\pm$0.12 & >-7.57 & <0.15 & <2.6 & ... \\
HD 290380 & 0.00$\pm$0.10 & 75$\pm$4 & 21$\pm$3$^9$ & 0.52$^{+0.08}_{-0.08}$ & -8.42 & -0.17$\pm$0.13 & -7.53$\pm$0.14 & 106$^{+1}_{-2}$ & 3.1$^{+0.2}_{-0.2}$ & -0.09 \\
V346 Ori & 0.00$\pm$0.13 & 115$\pm$6 & ... & 0.41$^{+0.04}_{-0.04}$ & -3.11 & -0.51$\pm$0.15 & -8.15$\pm$0.15 & <0.63 & <1.9 & ... \\
CO Ori & 0.15$\pm$0.10 & 55$\pm$3 & 30.2$\pm$2.2$^{10}$ & 1.54$^{+0.19}_{-0.20}$ & -11.89 & 0.67$\pm$0.16 & -6.66$\pm$0.18 & 149$^{+4}_{-2}$ & 3.5$^{+0.2}_{-0.2}$ & -0.13 \\
HD 35929 & 0.12$\pm$0.10 & 60$\pm$3 & ... & 3.67$^{+0.39}_{-0.39}$ & -8.10 & 1.16$\pm$0.09 & -6.14$\pm$0.09 & 60$^{+0}_{-3}$ & 3.2$^{+0.2}_{-0.1}$ & -0.05 \\
HD 290500 & 0.00$\pm$0.10 & 80$\pm$4 & ... & 1.43$^{+0.43}_{-0.43}$ & -13.71 & 0.47$\pm$0.12 & -6.83$\pm$0.18 & <0.73 & <3.7 & ... \\
HD 244314 & 0.00$\pm$0.10 & 55$\pm$3 & ... & 1.59$^{+0.17}_{-0.17}$ & -30.98 & 0.96$\pm$0.09 & -6.52$\pm$0.10 & <1.29 & <2.9 & ... \\
HD 244604 & 0.50$\pm$0.10 & 100$\pm$5 & ... & 1.03$^{+0.23}_{-0.23}$ & -12.78 & 0.88$\pm$0.09 & -6.54$\pm$0.13 & <0.84 & <3.1 & ... \\
V1044 Ori & 0.02$\pm$0.01 & 24$\pm$1 & ... & 4.48$^{+0.47}_{-0.46}$ & -12.95 & 0.10$\pm$0.13 & -7.24$\pm$0.14 & 266$^{+3}_{-4}$ & 3.3$^{+0.3}_{-0.3}$ & -0.33 \\
EZ Ori & 0.13$\pm$0.04 & 45$\pm$1 & ... & 2.36$^{+0.23}_{-0.23}$ & -20.09 & 0.33$\pm$0.13 & -6.98$\pm$0.14 & 287$^{+2}_{-6}$ & >3.6 & -0.38 \\
Brun 225 & ... & 14$\pm$5 & ... & 6$^{+3}_{-3}$ & -6.30 & -0.46$\pm$0.15 & -7.80$\pm$0.18 & 224$^{+22}_{-22}$ & 3.3$^{+0.2}_{-0.2}$ & -0.23 \\
Brun 381 & ... & 6$\pm$5 & ... & 24$^{+20}_{-20}$ & -7.00 & -0.85$\pm$0.17 & -8.14$\pm$0.20 & 62$^{+7}_{-7}$ & 2.6$^{+0.2}_{-0.2}$ & -0.01 \\
V2087 Ori & ... & 68$\pm$5 & ... & 1.52$^{+0.39}_{-0.39}$ & -4.50 & -0.58$\pm$0.15 & -7.90$\pm$0.19 & 318$^{+33}_{-33}$ & >3.7 & -0.45 \\
Brun 555 & ... & 35$\pm$5 & ... & 5$^{+1}_{-1}$ & -0.20 & ... & ... & 564$^{+59}_{-59}$ & >3.9 & -0.15 \\
Brun 656 & 0.95$\pm$0.04 & 101$\pm$1 & ... & 2.05$^{+0.23}_{-0.17}$ & -1.86 & -0.12$\pm$0.14 & -7.33$\pm$0.15 & 256$^{+6}_{-0}$ & 3.5$^{+0.3}_{-0.0}$ & -0.19 \\
Brun 684 & ... & 51$\pm$5 & ... & 2.32$^{+0.61}_{-0.61}$ & -3.70 & ... & ... & 294$^{+31}_{-31}$ & 3.4$^{+0.2}_{-0.2}$ & -0.37 \\
NV Ori & 0.14$\pm$0.10 & 75$\pm$4 & ... & 1.66$^{+0.18}_{-0.18}$ & -7.93 & 0.36$\pm$0.11 & -6.96$\pm$0.12 & 62$^{+2}_{-1}$ & 3.3$^{+0.1}_{-0.2}$ & -0.08 \\
BD-07 1129 & ... & 8$\pm$5 & ... & 22$^{+15}_{-15}$ & -2.90 & -0.56$\pm$0.15 & -7.77$\pm$0.19 & 24$^{+5}_{-5}$ & 1.3$^{+0.2}_{-0.2}$ & 0.15 \\
T Ori & 0.10$\pm$0.10 & 150$\pm$8 & ... & 1.04$^{+0.23}_{-0.23}$ & -16.18 & 0.91$\pm$0.12 & -6.39$\pm$0.15 & <0.16 & <2.8 & ... \\
V815 Ori & 0.20$\pm$0.04 & 35$\pm$1 & ... & 2.87$^{+0.23}_{-0.23}$ & -12.33 & -0.13$\pm$0.13 & -7.47$\pm$0.13 & 279.9$^{+0.3}_{-0.3}$ & 3.4$^{+0.1}_{-0.1}$ & -0.32 \\
CQ Tau & <0.00 & 98$\pm$5 & ... & 0.71$^{+0.07}_{-0.07}$ & -18.82 & -0.04$\pm$0.13 & -7.47$\pm$0.13 & <1.26 & <1.5 & ... \\
HD 294258 & ... & 88$\pm$5 & ... & 1.09$^{+0.27}_{-0.27}$ & -7.90 & -0.33$\pm$0.14 & -7.69$\pm$0.18 & 166$^{+17}_{-17}$ & 3.4$^{+0.2}_{-0.2}$ & -0.20 \\
Parenago 2374 & ... & 11$\pm$5 & ... & 10$^{+5}_{-5}$ & -6.50 & -0.70$\pm$0.15 & -7.98$\pm$0.19 & 39$^{+5}_{-5}$ & 2.2$^{+0.2}_{-0.2}$ & 0.03 \\
Brun 973 & ... & 58$\pm$5 & ... & 1.91$^{+0.50}_{-0.50}$ & -6.10 & -0.46$\pm$0.15 & -7.74$\pm$0.18 & 184$^{+19}_{-19}$ & 3.1$^{+0.2}_{-0.2}$ & -0.12 \\
Brun 1004 & ... & 22$\pm$5 & ... & 5$^{+2}_{-2}$ & -5.00 & -0.50$\pm$0.15 & -7.79$\pm$0.19 & 246$^{+26}_{-26}$ & 3.2$^{+0.2}_{-0.2}$ & -0.23 \\
Brun 1025 & ... & 7$\pm$5 & ... & 11$^{+9}_{-9}$ & -7.90 & ... & ... & 125$^{+13}_{-13}$ & 3.0$^{+0.2}_{-0.2}$ & -0.08 \\
HD 294260 & 0.27$\pm$0.27 & 12$\pm$2 & 22$\pm$3$^{9}$ & 4.14$^{+0.92}_{-0.91}$ & -11.83 & ... & ... & 174$^{+6}_{-7}$ & 3.3$^{+0.2}_{-0.1}$ & -0.11 \\
Brun 1060 & ... & 119$\pm$5 & ... & 0.77$^{+0.19}_{-0.19}$ & -6.80 & -0.56$\pm$0.15 & -7.90$\pm$0.19 & 107$^{+12}_{-12}$ & 2.7$^{+0.2}_{-0.2}$ & -0.01 \\
BF Ori & 0.20$\pm$0.13 & 37$\pm$2 & ... & 3$^{+1}_{-1}$ & -14.16 & 0.37$\pm$0.14 & -6.98$\pm$0.22 & <0.5 & <3.0 & ... \\
HD 37357 & 0.50$\pm$0.20 & 140$\pm$7 & ... & 0.63$^{+0.09}_{-0.09}$ & -11.15 & 1.18$\pm$0.12 & -6.41$\pm$0.14 & <0.18 & <2.9 & ... \\
HD 290764 & -0.15$\pm$0.10 & 55$\pm$10 & ... & 1.81$^{+0.36}_{-0.36}$ & -14.43 & 0.77$\pm$0.10 & -6.62$\pm$0.10 & <0.77 & <2.1 & ... \\
HD 294290 & ... & 7$\pm$5 & ... & 17$^{+13}_{-13}$ & -4.80 & -0.56$\pm$0.15 & -7.85$\pm$0.19 & 78$^{+9}_{-9}$ & 2.3$^{+0.2}_{-0.2}$ & 0.04 \\
V599 Ori & -0.35$\pm$0.10 & 47$\pm$2 & 57$\pm$4$^{9}$ & 2.55$^{+0.23}_{-0.22}$ & -11.80 & 0.52$\pm$0.11 & -6.85$\pm$0.12 & 7$^{+1}_{-1}$ & 2.8$^{+0.1}_{-0.2}$ & -0.07 \\
HD 39014 & 0.20$\pm$0.20 & 195$\pm$10 & 54$\pm$5.4$^{11}$ & 0.69$^{+0.07}_{-0.07}$ & -2.43 & 0.15$\pm$0.12 & -7.22$\pm$0.12 & <0.06 & <1.3 & ... \\
HBC 217 & -0.25$\pm$0.10 & 36$\pm$1 & ... & 2.57$^{+0.26}_{-0.26}$ & -13.72 & 0.22$\pm$0.12 & -7.15$\pm$0.12 & 143$^{+1}_{-1}$ & 3.1$^{+0.2}_{-0.2}$ & -0.13 \\
HBC 222 & 0.00$\pm$0.10 & 52$\pm$2 & ... & 1.77$^{+0.19}_{-0.19}$ & -6.62 & -0.06$\pm$0.14 & -7.43$\pm$0.15 & 127$^{+1}_{-1}$ & 3.4$^{+0.2}_{-0.2}$ & -0.14 \\
HD 68695 & -0.50$\pm$0.10 & 45$\pm$2 & ... & 1.41$^{+0.18}_{-0.18}$ & -17.17 & 0.78$\pm$0.10 & -6.83$\pm$0.11 & <0.92 & <3.2 & ... \\
PDS 277 & 0.93$\pm$0.04 & 38$\pm$1 & ... & 0.63$^{+0.03}_{-0.03}$ & -3.69 & -0.20$\pm$0.14 & -7.53$\pm$0.14 & 65$^{+3}_{-3}$ & 3.0$^{+0.2}_{-0.2}$ & 0.00 \\
GSC 8581-2002 & 0.00$\pm$0.10 & 155$\pm$8 & ... & 0.74$^{+0.08}_{-0.08}$ & -0.59 & -0.60$\pm$0.15 & -8.01$\pm$0.15 & <0.96 & <3.7 & ... \\
PDS 33 & -0.50$\pm$0.17 & 140$\pm$7 & ... & 0.46$^{+0.05}_{-0.05}$ & -18.62 & 0.77$\pm$0.10 & >-6.85 & <1.16 & <3.6 & ... \\
CR Cha & 0.50$\pm$0.03 & 34$\pm$1 & ... & 3.24$^{+0.47}_{-0.45}$ & -15.14 & -0.09$\pm$0.18 & -7.36$\pm$0.20 & 309$^{+5}_{-2}$ & 2.9$^{+0.4}_{-0.3}$ & -0.05 \\
Ass Cha T 2-21 & 0.72$\pm$0.08 & 86$\pm$2 & ... & 1.54$^{+0.12}_{-0.13}$ & -3.09 & -0.06$\pm$0.13 & -7.35$\pm$0.13 & 150$^{+0}_{-8}$ & 3.0$^{+0.0}_{-0.3}$ & 0.02 \\
CV Cha & 0.14$\pm$0.08 & 25$\pm$1 & 43$\pm$5$^{12}$ & 3.53$^{+0.37}_{-0.39}$ & -58.58 & 0.50$\pm$0.11 & -6.86$\pm$0.11 & 295$^{+9}_{-0}$ & 3.2$^{+0.1}_{-0.0}$ & -0.27 \\
Ass ChaT2-54 & 0.32$\pm$0.15 & 8$\pm$1 & ... & 13$^{+2}_{-3}$ & -0.97 & -1.02$\pm$0.20 & -8.37$\pm$0.21 & 336$^{+5}_{-23}$ & 3.5$^{+0.2}_{-0.6}$ & -0.32 \\
HD 100453 & -0.10$\pm$0.10 & 50$\pm$3 & 48$\pm$4.8$^{13}$ & 0.92$^{+0.10}_{-0.10}$ & -3.24 & -0.43$\pm$0.15 & -8.05$\pm$0.15 & <0.52 & <1.4 & ... \\
HD 100546 & <1.00 & 58$\pm$3 & 47$\pm$1$^{14}$ & 1.52$^{+0.10}_{-0.10}$ & -31.49 & 1.19$\pm$0.09 & -6.25$\pm$0.09 & <0.72 & <3.6 & ... \\
HD 101412 & <0.00 & 3$\pm$1 & ... & 24$^{+10}_{-10}$ & -17.44 & 1.09$\pm$0.09 & -6.53$\pm$0.13 & <0.14 & <2.9 & ... \\
HD 135344B & 0.00$\pm$0.13 & 70$\pm$4 & 11.6$\pm$0.8$^{15}$ & 0.25$^{+0.03}_{-0.03}$ & -10.22 & -0.07$\pm$0.13 & -7.51$\pm$0.13 & 117$^{+2}_{-4}$ & 3.2$^{+0.1}_{-0.1}$ & -0.10 \\
HD 139614 & -0.25$\pm$0.10 & 32$\pm$2 & 21$\pm$1$^{16}$ & 0.71$^{+0.07}_{-0.07}$ & -11.65 & 0.16$\pm$0.12 & -7.44$\pm$0.12 & 10.8$^{+0.1}_{-0.2}$ & 2.8$^{+0.1}_{-0.2}$ & -0.07 \\
HD 141569 & -0.50$\pm$0.38 & 220$\pm$11 & 51$\pm$3$^{17}$ & 0.39$^{+0.04}_{-0.04}$ & -7.12 & 0.23$\pm$0.12 & -7.25$\pm$0.12 & <1.29 & <3.7 & ... \\
HD 142666 & 0.00$\pm$0.10 & 65$\pm$3 & 58$\pm$4$^{18}$ & 1.47$^{+0.13}_{-0.13}$ & -5.74 & 0.00$\pm$0.13 & -7.39$\pm$0.13 & 31.1$^{+0.4}_{-0.5}$ & 3.0$^{+0.2}_{-0.1}$ & -0.07 \\
HD 143006 & <0.15 & 15$\pm$1 & 27$\pm$9$^{19}$ & 2.99$^{+0.96}_{-0.96}$ & -9.18 & -0.31$\pm$0.14 & -7.75$\pm$0.15 & >221.6 & >2.9 & -0.10 \\
HD 144432 & 0.15$\pm$0.10 & 77$\pm$4 & <28$^{20}$ & <0.69 & -10.74 & 0.37$\pm$0.11 & -7.04$\pm$0.12 & 35$^{+1}_{-1}$ & 3.3$^{+0.1}_{-0.2}$ & -0.07 \\
HR 5999 & 0.20$\pm$0.10 & 180$\pm$9 & 45$\pm$5$^{21}$ & 1.05$^{+0.17}_{-0.18}$ & -17.57 & 1.11$\pm$0.11 & -6.17$\pm$0.13 & 9.7$^{+0.1}_{-0.1}$ & 3.6$^{+0.3}_{-0.3}$ & ... \\
V718 Sco & 0.00$\pm$0.10 & 120$\pm$6 & ... & 0.53$^{+0.06}_{-0.06}$ & -3.70 & -0.06$\pm$0.13 & -7.59$\pm$0.13 & <0.61 & <1.9 & ... \\
HD 149914 & 0.00$\pm$0.14 & 200$\pm$10 & ... & 1.03$^{+0.16}_{-0.16}$ & -1.03 & 0.23$\pm$0.12 & -7.04$\pm$0.13 & <0.08 & <2.6 & ... \\
HD 163296 & 0.20$\pm$0.10 & 126$\pm$6 & 45$\pm$1$^{22}$ & 0.61$^{+0.14}_{-0.14}$ & -24.61 & 0.59$\pm$0.10 & -6.86$\pm$0.15 & <0.09 & <2.2 & ... \\
HD 169142 & -0.50$\pm$0.33 & 48$\pm$2 & 13$\pm$1.3$^{23}$ & 0.33$^{+0.04}_{-0.04}$ & -12.28 & 0.08$\pm$0.12 & -7.46$\pm$0.13 & 13.4$^{+0.3}_{-0.8}$ & 2.4$^{+0.1}_{-0.2}$ & -0.06 \\
HD 176386 & 0.20$\pm$0.10 & 170$\pm$9 & ... & 0.89$^{+0.08}_{-0.09}$ & -0.51 & -1.07$\pm$0.17 & -8.41$\pm$0.18 & <0.64 & <3.5 & ... \\
HD 179218 & -0.50$\pm$0.17 & 75$\pm$4 & 54$\pm$5.4$^{24}$ & 1.76$^{+0.17}_{-0.17}$ & -13.70 & 1.15$\pm$0.09 & -6.32$\pm$0.09 & <0.15 & <2.7 & ... \\
WW Vul & >0.20 & 190$\pm$10 & ... & 0.74$^{+0.07}_{-0.07}$ & -16.41 & 0.66$\pm$0.11 & -6.59$\pm$0.11 & <0.23 & <2.1 & ... \\
PX Vul & 0.00$\pm$0.10 & 78$\pm$4 & ... & 2.08$^{+0.22}_{-0.23}$ & -17.02 & 0.85$\pm$0.10 & -6.45$\pm$0.11 & 26$^{+1}_{-1}$ & 2.6$^{+0.2}_{-0.2}$ & -0.03 \\
\hline
\hline
\end{tabular}
\begin{minipage}{18cm}
\textbf{Notes:} For each star in Col. 1, the remaining columns list the metallicity, projected rotational velocity, inclination, rotational period, H$\alpha$ equivalent width (corrected for photospheric absorption; the nominal error bars are $\sim$ 10$\%$ for the \citet{King93} sources (see Table \ref{table:sample}) and lower than 0.01 $\AA$ for the rest), accretion luminosity, mass accretion rate, \ion{Li}{i} equivalent width (corrected for blending with the adjacent \ion{Fe}{i}6707.43 transition), Li abundance, and the deviation from local thermodynamical equilibrium ($\Delta$ = A(Li)$_{non-LTE}$ - A(Li)$_{LTE}$). The full table is available electronically at the CDS. \\
\textbf{References:} [Fe/H] and $v \sin i$ values were taken from \citet{GuzmanDiaz23} and \citet{King93} when available (see Table \ref{table:sample}). References for the inclinations are $^1$\citet{Hughes17}, $^2$\citet{Pinilla22}, $^3$\citet{Long19}, $^4$\citet{Tanii12}, $^5$\citet{Labdon23}, $^6$\citet{Pietu06}, $^7$\citet{Kreplin16}, $^8$\citet{deBoer21}, $^9$\citet{Valegard24}, $^{10}$\citet{Davies18}, $^{11}$\citet{Lazareff17}, $^{12}$\citet{Ginski24}, $^{13}$\citet{Benisty17}, $^{14}$\citet{Jamialahmadi18}, $^{15}$\citet{Cazzoletti18}, $^{16}$\citet{Muro-Arena20}, $^{17}$\cite{Weinberger99}, $^{18}$\citet{Davies18b}, $^{19}$\citet{Benisty18}, $^{20}$\citet{Chen12}, $^{21}$\citet{Benisty11}, $^{22}$\citet{Muro-Arena18}, $^{23}$\citet{Bertrang18}, $^{24}$\citet{Kokoulina21}.\\ 
\end{minipage}
\end{table*}

\subsection{Li equivalent widths, curves of growth, and abundances}
\label{Sect:abundances}
Figure \ref{Fig:lithium} shows the \ion{Li}{i} optical feature at 6707.856 {\AA} for all stars in which the line is apparent in our spectra. The representative Li features observed in the \citet{King93} subsample are shown in Fig. 1 of that work. Their Li EWs were adopted here, and for the remaining stars, they were measured using the STAR-MELT Python package, which is fully described in \citet{Campbell-White21}. Further details of the Li fitting procedure are described in \citet{Campbell23}. For the stars with T$_*$ < 8000 K, the measured EWs include the contribution of the \ion{Fe}{i} transition at 6707.43 {\AA}, which is blended with the \ion{Li}{i} line in most of our spectra. We subtracted the contribution of \ion{Fe}{i} using Table A.2 from \citet{Franciosini22a}. The error bars for the Li EWs were derived considering the different \ion{Fe}{i} contributions resulting from the uncertainties associated with the stellar parameters T$_*$, log g, and [Fe/H]. For the featureless stars, only upper limits for the Li EWs could be inferred and were derived from the expression 7590 $\times$ R/S/N \citep{Martin23}. The Li EWs corrected for contamination by \ion{Fe}{i} are listed in Col. 9 of Table \ref{table:results}. 

Some earlier work focused on low-mass stars corrected the Li EWs by a low percentage due to the contribution of veiling. However, this correction depends on the accretion rate and on a very accurate characterization of the stellar parameters, and it might introduce additional uncertainties that are difficult to calibrate. Furthermore, veiling tends to be significant mainly for stars colder than 5000 K \citep[e.g.,][]{Carini26}. Hotter sources such as those in our sample have intrinsically shallow photospheric lines, and the excess emission, characterized by a temperature very similar than that of the stellar surfaces, causes negligible line veiling that can hardly be associated with accretion \citep[e.g.,][]{Muzerolle04}. Taking all this into account, we decided not to correct the Li EWs for veiling. 

\begin{figure}
   \centering
   \includegraphics[width=9cm]{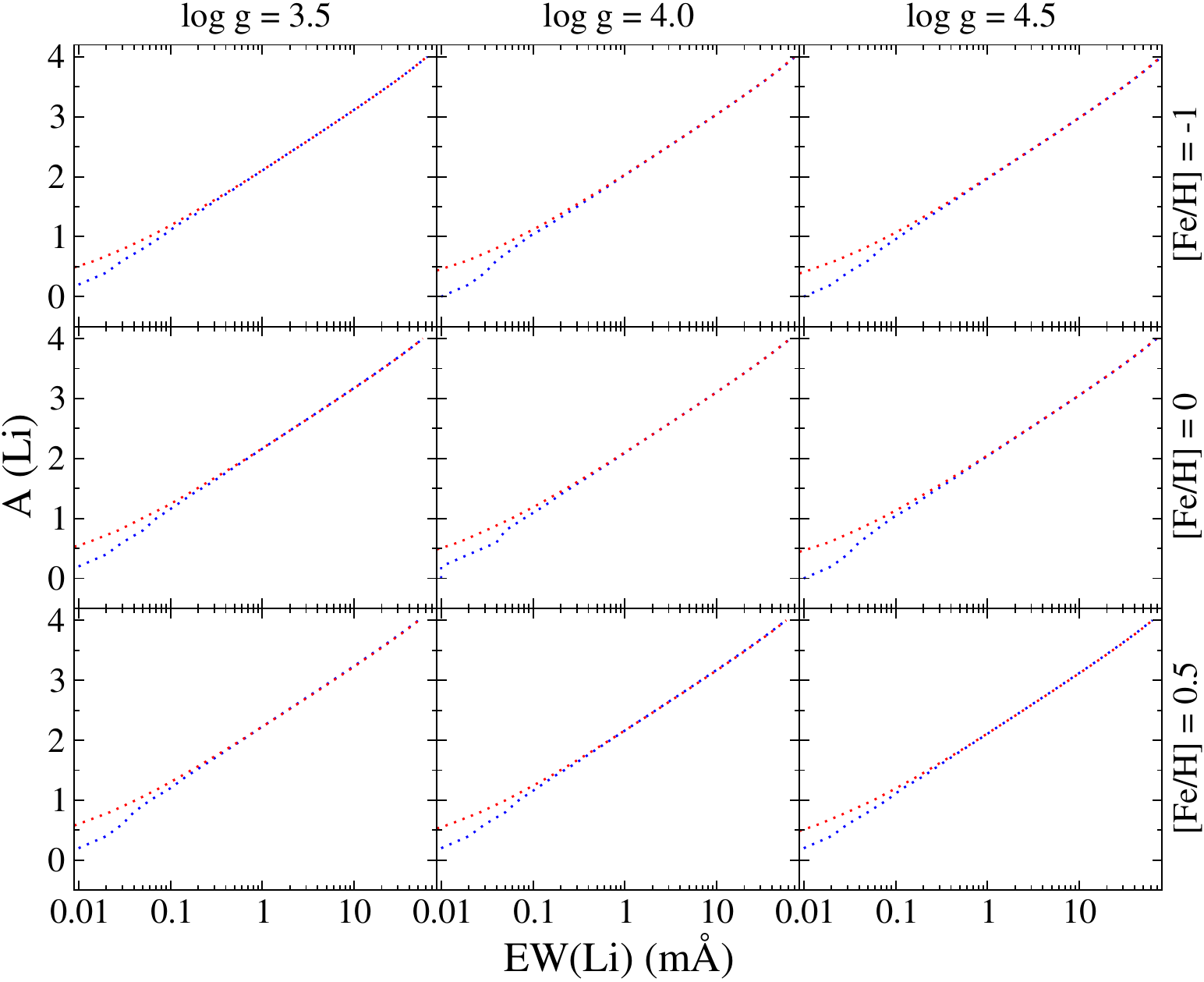}
      \caption{Comparison between COG from this work (red) and from \citet{Franciosini22a} (blue) for T$_*$ = 8000 K  and the indicated values of log g and [Fe/H].} 
         \label{fig:cog_compare}
   \end{figure}

The Li EWs were transformed into abundances using the COG of \citet{Franciosini22a} for stars with T$_*$ $\leq$ 8000 K. For hotter stars, we derived the COG based on synthetic Castelli-Kurucz spectra computed with the ATLAS9 suite of codes \citep{Castelli03}. High-resolution spectra (R = 150000) were synthesized in the range 6700-6715 {\AA} for the following stellar parameters: T$_*$ from 8000 to 12000 K in steps of 200 K, log g from 3.5 to 4.5 in steps of 0.5, and [M/H] from -1 to 0.5 in steps of 0.5. In each case, A(Li) was varied from -1 to 4 in steps of 0.2. The Li EWs were then measured in the synthetic spectra, and we tabulated these values versus A(Li) for the corresponding stellar parameters. Table \ref{table:cog} provides an excerpt of the COG derived in this way. Figure \ref{fig:cog_compare} compares our COG with those in \citet{Franciosini22a} for the only stellar temperature in common (8000 K). The COGs are consistent for large enough EWs ($\geq$ 0.1 m{\AA}). For smaller EWs, our A(Li) values tend to be slightly higher ($\leq$ 0.5 dex) than those in \citet{Franciosini22a}. This difference likely reflects the intrinsic difficulty of measuring and modeling extremely weak lines. Importantly, this mismatch does not affect the trends discussed in this work. 

\begin{table}
\centering
\caption{COG for hot stars (extract).}
\label{table:cog}
\centering
\begin{tabular}{l l l l l}
\hline\hline
T$_*$ & log g  & [M/H] & A(Li) & EW (Li) \\
K &  [cm s$^{-2}$] & ... & ... & m{\AA} \\ 
\hline\hline
9000&4.00&-0.50&-1.0&1.0000$\times$10$^{-3}$\\
9000&4.00&-0.50&-0.8&1.0000$\times$10$^{-3}$\\
...&...&...&...&...\\
\hline
\hline
\end{tabular}
\begin{minipage}{9cm}
\textbf{Notes:} The full table for all stellar parameters indicated in the text is available electronically at the CDS. 
\end{minipage}
\end{table}

The abundances from \ion{Li}{i} detections were corrected to include departures from local thermodynamical equilibrium (LTE) by using the INSPECT database\footnote{https://www.inspect-stars.com/}, based on the non-LTE models in \citet{Lind09}. Microturbulence velocities were varied within the allowed range of 1 -- 5 km s$^{-1}$, and we selected the velocity that provided the highest value of $\Delta$ = A(Li)$_{non-LTE}$ - A(Li)$_{LTE}$. This corresponds to 1 -- 2 km s$^{-1}$ in all cases, resulting in small deviations from LTE (|$\Delta$| $<$ 0.4 dex, the median being 0.1 dex). The error bars for A(Li) were estimated by propagating those for the Li EWs resulting from the uncertainties in T$_*$, log g, and [Fe/H]. Overall, A(Li) is most sensitive to T$_*$ variations, but is barely affected by uncertainties in log g and [Fe/H]. For the stars without detections, upper limits are provided based on those for the EWs and the highest possible values of A(Li), considering the uncertainties of stellar parameters and metallicities. Finally, lower limits for A(Li) are provided for the few stars with upper limits for [Fe/H] or with strong Li EWs that lead to A(Li) values higher than computed in the COG, A(Li) = 4.  The abundances and departures from LTE are listed in Cols. 10 and 11 of Table \ref{table:results}.

\section{Analysis and discussion}
\label{Sect:analysis_discussion}

\subsection{Preliminary considerations}
 \label{Sect:preliminary}
\begin{figure}
   \centering
   \includegraphics[width=9cm]{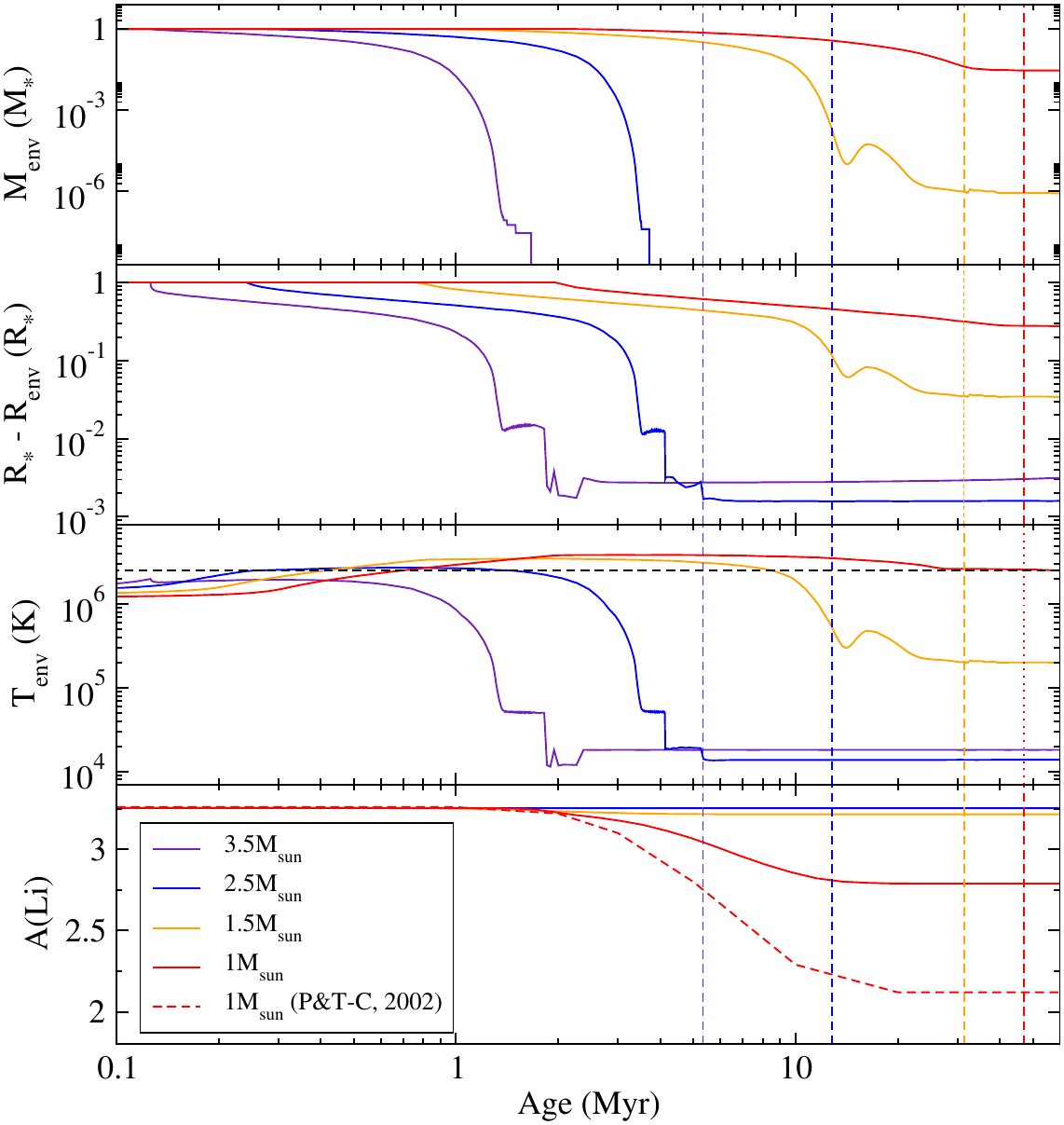}
      \caption{Evolution of the mass and thickness of the convective envelope (in units of stellar mass and radius; 1 means fully convective), the temperature of the envelope at its base (the horizontal dashed line shows the typical Li burning temperature at 2.5 $\times$ 10$^6$ K), and the resulting A(Li), according to models by \citet{Siess00}. The bottom panel includes the solar model from \citet{Piau02}. The vertical dashed lines mark the ZAMS ages for the different stellar masses and use the same color-code as indicated in the legend.}
         \label{fig:interiors}
   \end{figure}

In order to interpret the observations in the following sections, it is necessary to grasp the expected evolution of the Li content of young intermediate-mass stars based on standard models of stellar interiors. Figure \ref{fig:interiors} plots the age evolution of the mass fraction of the star contained in the convective envelope and its thickness (1 means that the star is fully convective), the maximum temperature of the convective envelope (at its base), and the resulting A(Li) based on the \citet{Siess00} models for 3.5, 2.5, 1.5, and 1.0 M$_{\odot}$ with solar metallicity. For reference, the A(Li) expected for a 1M$_{\odot}$ star from the models by \citet{Piau02} is included to illustrate differences from specific treatments of the stellar interiors (see the solid line in their Fig. 4 and the related discussion). Standard models show that while all the stars start their evolution as fully convective, the envelope reduces faster as the stellar mass increases, and the temperature necessary to burn lithium is only reached for low stellar masses. In particular, standard models predict that the stars in the 1.5 -- 3.5 M$_{\odot}$ range we studied are expected to generally show negligible Li depletion. This becomes larger and more similar to the 1M$_{\odot}$ case as the stellar mass decreases, and a thicker convective envelope and a hotter base temperature facilitate Li mixing and burning. 

\begin{figure}
   \centering
   \includegraphics[width=9cm]{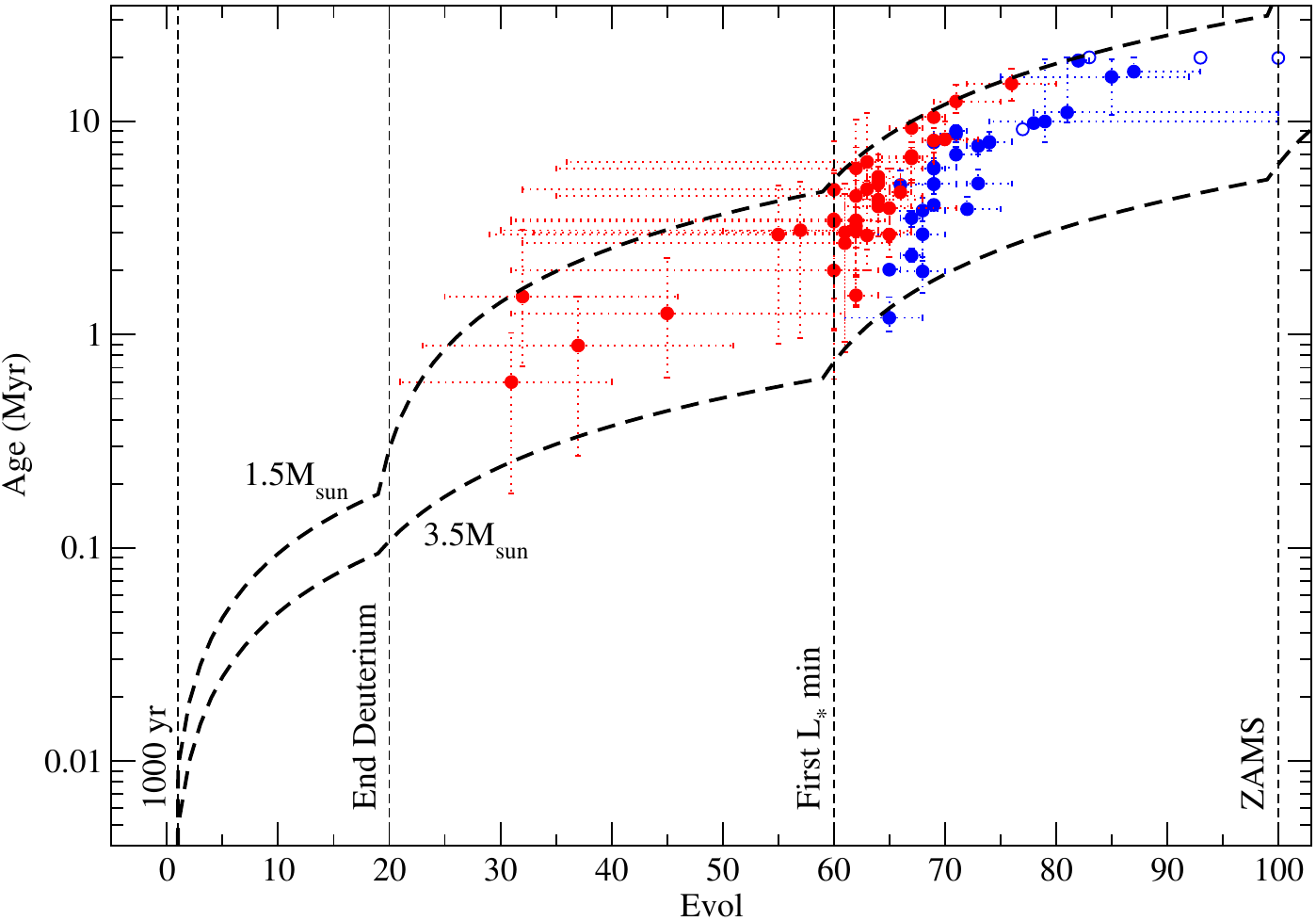}
      \caption{Relations between stellar age and evolutionary parameter based on the BaSTI stellar evolution models and prescriptions in \citet{Hidalgo18} for 1.5 and 3.5 M$_{\odot}$ (dashed lines). The IMTT and Herbig stars in our sample are indicated in red and blue, respectively. The open symbols indicate upper limits on both axes. The vertical dashed lines indicate the main stages during the pre-MS evolution (see text).}
         \label{fig:age_evol}
   \end{figure}

Figure \ref{fig:interiors} also illustrates that the stellar age is not best suited for comparative evolution analyses. Based on pre-MS tracks and isochrones, a given age can correspond to different evolutionary stages, especially for stars with different stellar masses. In the following, we use the evolutionary parameter Evol, which corresponds to the row number in the updated BaSTI stellar evolution models \citep[see details in][]{Hidalgo18}. A given value of Evol refers to a fixed location in the evolutionary tracks in the HR diagram, and thus, it is directly related to a specific phase during the stellar evolution. Figure \ref{fig:age_evol} shows the correspondence between age and Evol for the stellar mass range of the stars in our sample. The vertical lines indicate the main pre-MS stages as identified in \citet{Hidalgo18}: Evol = 1, 20, 60, and 100 correspond to the initial age of all evolutionary tracks (1000 yr), the end of the deuterium-burning phase, the first minimum of the surface luminosity in the HR diagram, and the ZAMS, respectively. The figure shows that while IMTTs and Herbigs share roughly the same age range, the latter tend to be more evolved. 

It is also necessary for the analysis to assess the meaning of the non-detections of the \ion{Li}{I} feature in our sample. These non-detections can be interpreted in three different ways. First, the amount of Li at the stellar surface may be negligible. Second, Li may be present, but a high surface temperature may cause it to be fully ionized. Third, spectroscopic line broadening due to stellar rotation may dilute the \ion{Li}{I} feature. In our sample, the \ion{Li}{I} feature is not detected in any star hotter than 9000 K or rotating faster than 200 km s$^{-1}$. In addition, stars rotating faster than 50 km s$^{-1}$ only show firm Li detections when the abundance is greater than 2.5. Thus, upper limits for A(Li) of stars with T$_*$ $\geq$ 9000 K, $v \sin i$ $\geq$ 200 km s$^{-1}$, or with $v \sin i$ $>$ 50 km s$^{-1}$ and A(Li) $<$ 2.5, may reflect not only that Li is absent, but also that Li is ionized, rotationally diluted, or a combination of the two. Our Li analysis only considered the upper limits of A(Li) for the remaining stars because these provide constraints on the maximum abundance that is compatible with the observations.

\subsection{Li evolution}
 \label{Sect:Li_evol}

\begin{figure}
   \centering
   \includegraphics[width=9cm]{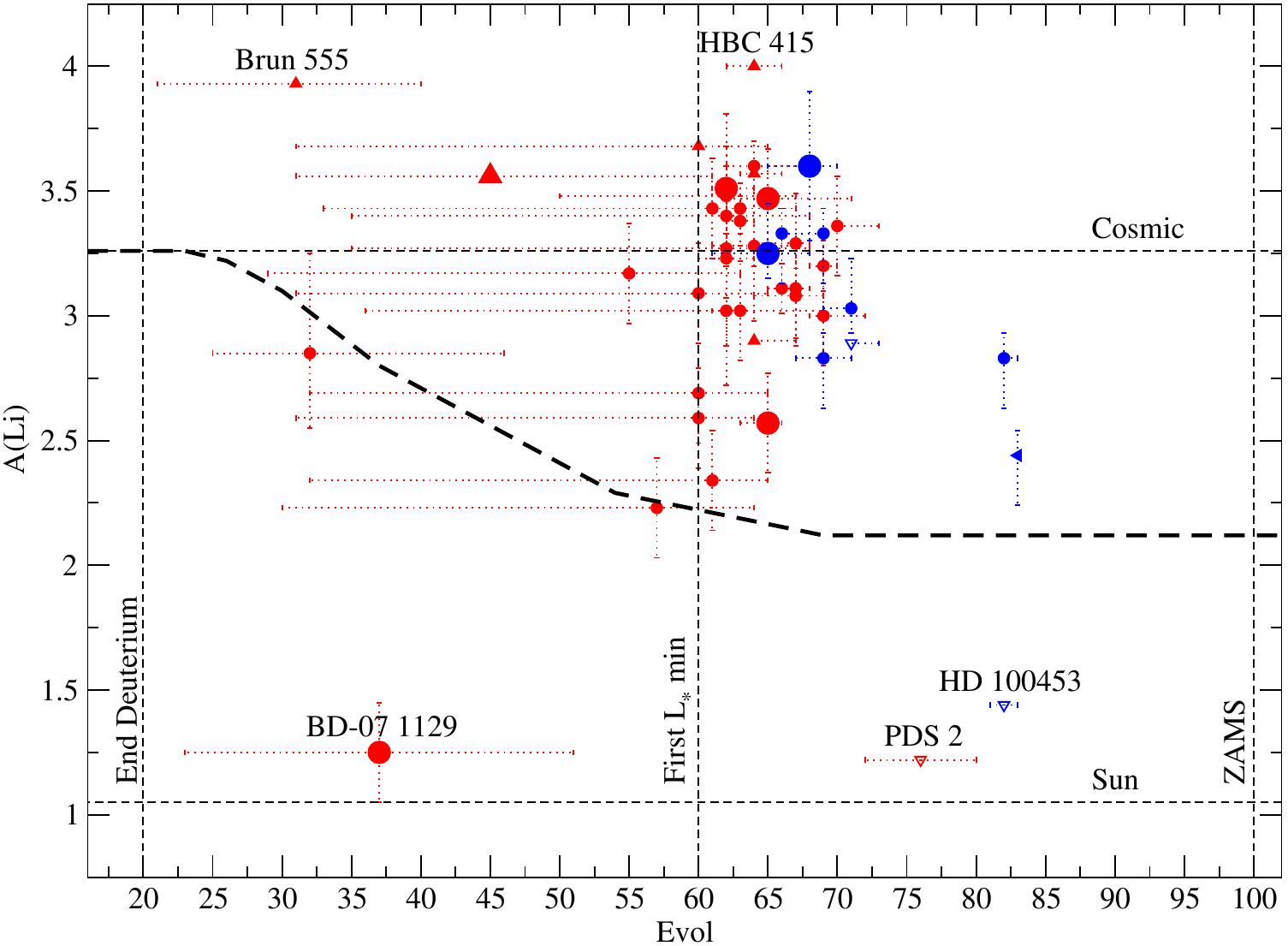}
      \caption{Evolution of A(Li) for the mass ranges 1.5 -- 2.5 M${_\odot}$ (small symbols) and 2.5 -- 3.5 M${_\odot}$ (large symbols) and for IMTTs (red) and Herbigs (blue). The solid symbols represent A(Li) detections, and triangles indicate lower or upper limits depending on their orientation. The open triangles show upper limits for A(Li). The lower boundary predicted from the solar model by \citet{Piau02} is overplotted with a dashed line. The horizontal dashed lines show the levels for the cosmic and solar abundances for reference, and the vertical dashed lines are the same as in Fig. \ref{fig:age_evol}}  
         \label{fig:Li_age}
   \end{figure}

Figure \ref{fig:Li_age} shows the evolution of A(Li) considering two different stellar mass ranges: the low-mass subsample including the 39 stars with 1.5 $\leq$ M$_*$/M${_\odot}$ $<$ 2.5, and the high-mass subsample comprising the 7 sources with 2.5 $\leq$ M$_*$/M${_\odot}$ $<$ 3.5. For reference, the cosmic A(Li) value as measured in primitive meteorites of the Solar System (3.26 $\pm$ 0.05, also representative of that of the interstellar medium) and the A(Li) for the Sun (1.05 $\pm$ 0.10) are indicated \citep[][]{Asplund09}. The lower boundary expected for our sample based on the \citet{Piau02} pre-MS model for solar-type stars is overplotted. A(Li) values $<$ 3 (i.e., $>$ 5$\sigma$ times below the cosmic abundance) are observed in $\sim$ 26 $\%$ of our sources and contrast with the negligible Li depletion expected from the standard models of stellar interiors discussed above. Moreover, these models predict that the high-mass subsample is less strongly depleted than the low-mass subsample, but this difference is not observed. In particular, in 2 out of 7 high-mass sources ($\sim$ 29 $\%$), A(Li) is significantly lower than the cosmic value. The names of the 3 stars that fall below the lower boundary indicated with the dashed line and those of the 2 super-Li-rich stars in our sample \citep[with A(Li) $>$ 3.8 considering error bars, following][] {Zhou25} are indicated in Fig. \ref{fig:Li_age}. Specific notes on these stars can be found in Appendix \ref{appendix:stars}. 

\begin{figure}
   \centering
   \includegraphics[width=9cm]{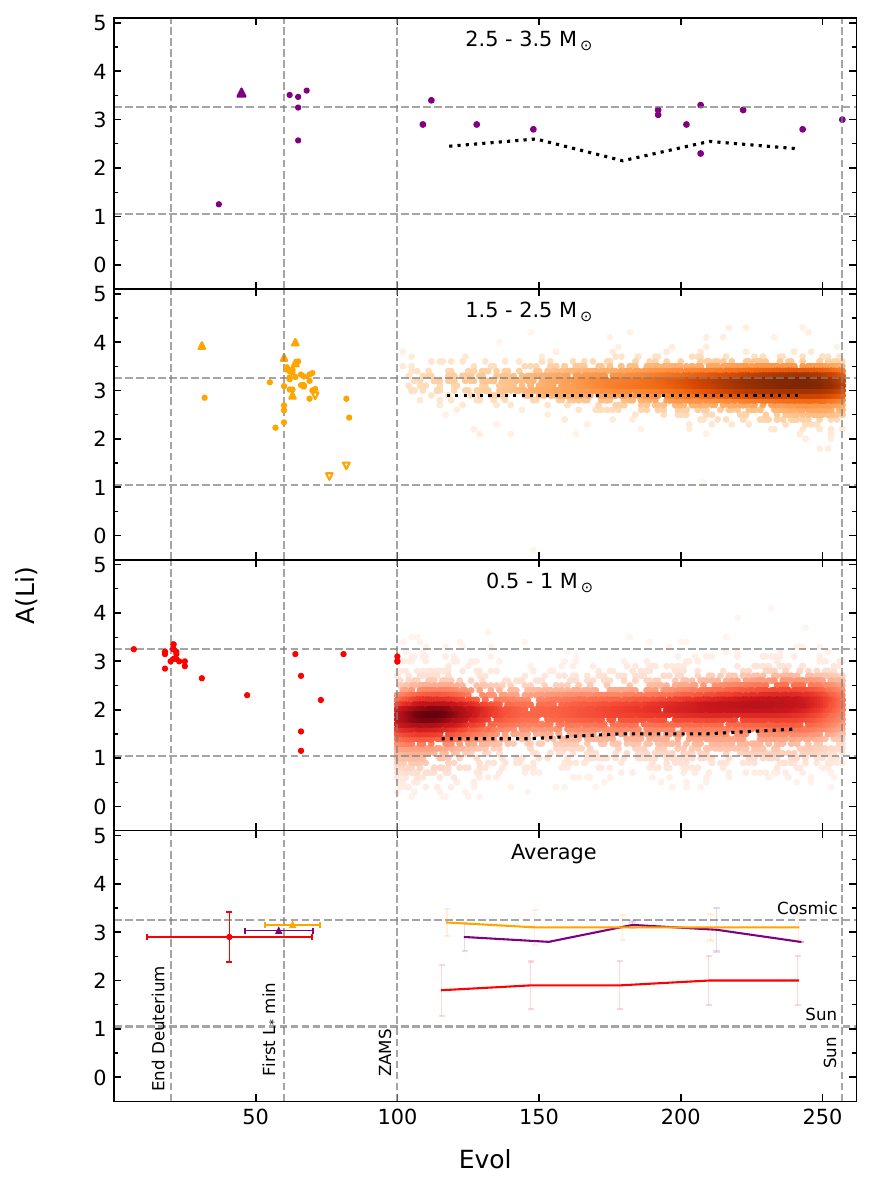}
      \caption{Evolution of A(Li) for the mass ranges indicated in the first three panels. The A(Li) uncertainties are typically $\sim$ 0.2 and 0.3 dex for the intermediate- and (sub-)solar mass pre-MS stars and $\sim$ 0.1 dex for MS stars. The medians of the non-detections in the MS are indicated with dotted lines. The bottom panel shows the mean values of the detections, with standard deviations as error bars. The solid and open symbols, triangles, and dashed lines are the same as in Fig. \ref{fig:Li_age} (see Fig. \ref{fig:Li_evol} for an expanded version). }   
         \label{fig:Li_evol_pre_MS}
   \end{figure}

In order to place the previous results in perspective, the first three panels of Fig. \ref{fig:Li_evol_pre_MS} show the evolution of A(Li) including nonbinary stars in the first half of the MS (up to the evolutionary stage of the Sun) for the 2.5 -- 3.5 M${_\odot}$, 1.5 -- 2.5 M${_\odot}$, and 0.5 -- 1 M${_\odot}$ ranges. The upper limits are represented by their median values. The bottom panel shows the mean A(Li) values for the pre-MS and MS phases. A complete version of Fig. \ref{fig:Li_evol_pre_MS} (including the post-MS) and details about the plotted data are given in Appendix \ref{appendix:whole_sample}. The typical A(Li) of intermediate-mass stars barely declines, in contrast to the evolution observed in (sub-)solar mass sources and in general agreement with standard models of stellar interiors. In particular, while the mean A(Li) for intermediate-mass stars remains at $\sim$ 3, it rapidly drops down to $\sim$ 2 for (sub-)solar mass stars (the Sun has less Li than average; see Sect. \ref{Sect:Li_SEDs}). Nevertheless, the fraction of intermediate-mass stars in the MS with A(Li) $<$ 3 is $\sim$ 25$\%$, which roughly matches the fraction for the pre-MS. This fraction for the MS might increase when non-detections are considered, which were not included in the calculations (see Appendix \ref{appendix:whole_sample}). In either case, the fraction of intermediate-mass stars showing significant Li depletion is non-negligible and differs from the expectations from standard models of stellar interiors. In addition, as we argued before for the pre-MS, it is intriguing that the low- and high-mass subsamples within the intermediate-mass regime show no significant difference during the MS. Moreover, the averages in the bottom panel of Fig. \ref{fig:Li_evol_pre_MS} suggest that the high-mass subsample undergoes slightly more Li depletion than the low-mass subsample. This agrees with \citet{Maldonado25}, who already indicated that the mixing with the interiors for the 2.5 -- 3.5 M${_\odot}$ range seems slightly more effective than for the 1.5 -- 2.5 M${_\odot}$ range. However, a larger number of stars within the 2.5 -- 3.5 M${_\odot}$ range, which is barely explored so far, is necessary before firmer conclusions can be made.

Figure \ref{fig:Li_evol_pre_MS} shows another difference between the low- and high-mass subsamples, which concerns the fraction of super-Li rich stars with A(Li) $>$ 3.8. Only the low-mass subsample shows this type of sources, which reduces from $\sim$ 5 $\%$ during the pre-MS to $\sim$ 0.3 $\%$ during the MS. The corresponding fractions for the remaining samples reduces to practically 0$\%$ for the pre-MS and the MS. The origin of the Li super-abundances remains a mystery, and we refer to the recent work by \citet{Zhou25}.

\begin{figure}
   \centering
   \includegraphics[width=9cm]{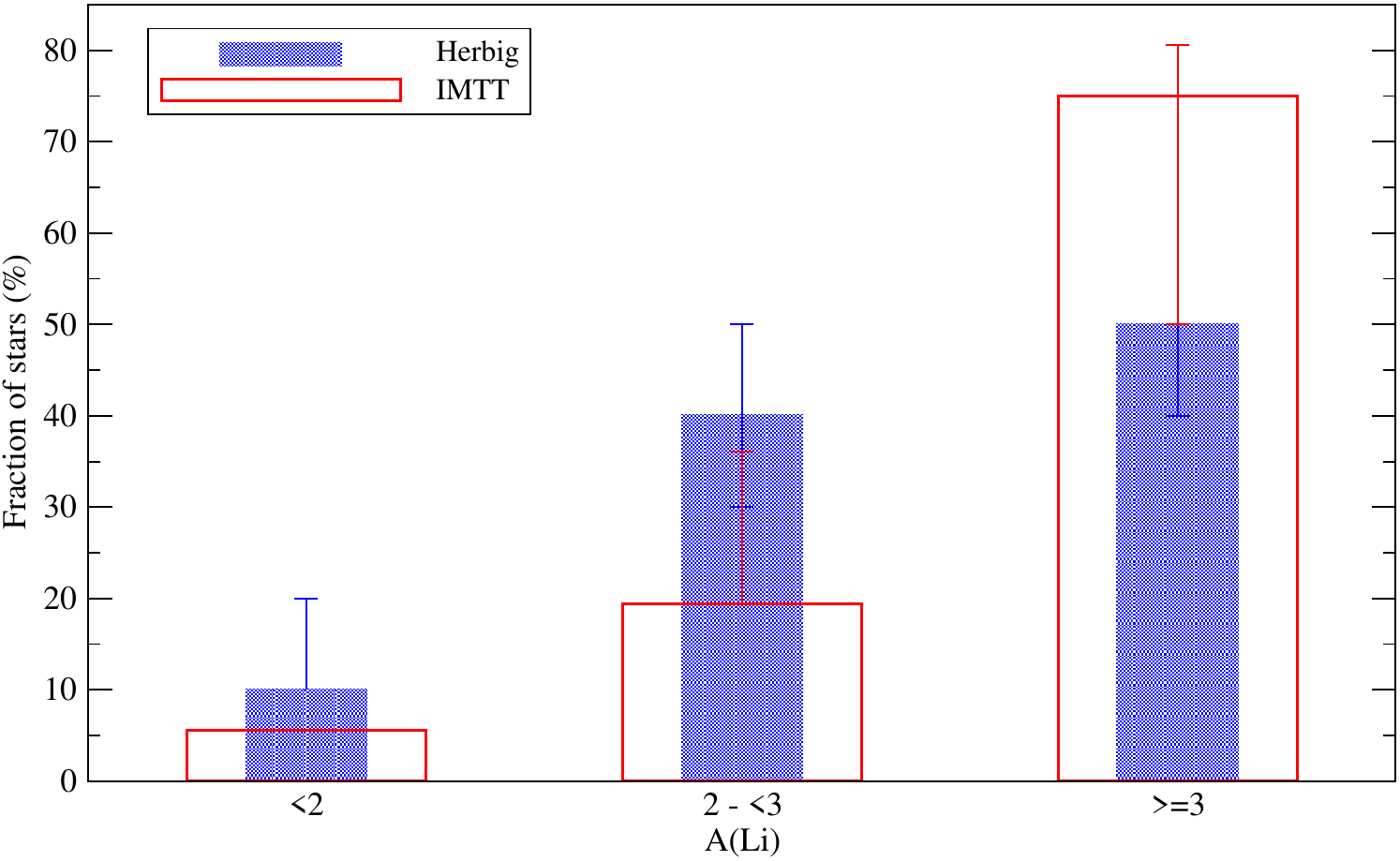}
      \caption{Distribution of A(Li) for 36 IMTTs (red) and 10 Herbigs (blue).} 
         \label{fig:Li_hist_IMTT_vs_Herbig}
   \end{figure}

On the other hand, when we consider the IMTT and Herbig regimes in Fig. \ref{fig:Li_age}, a weak evolutionary pattern seems to be present, especially for the latter sources, with the more evolved Herbigs showing lower values of A(Li). This is explored in more detail in Figure \ref{fig:Li_hist_IMTT_vs_Herbig}, which shows the A(Li) distributions for the 36 IMTT and 10 Herbig stars. The median A(Li) is 2.96 for the Herbigs and 3.22 for the IMTTs. Thus, in terms of typical number density, the lithium content of IMTTs is almost twice larger. However, the two-sample Kolmogorov-Smirnov (KS) and Anderson-Darling (AD) tests do not reject the null hypothesis that the Herbig and IMTT populations are drawn by the same parent distribution with a default significance level of 5$\%$. Additional Herbig stars are necessary to test the weak evidence suggesting that A(Li) decreases from the IMTT to the Herbig regime, which may be consistent with longer mixing timescales or with additional nonstandard processes that may affect the Li content during the early evolution of intermediate-mass stars. The potential effects of rotation, accretion, and the presence of planets are explored in the following sections. 

\subsection{Li abundance and rotation}
\label{Sect:Li_rotation}

Surveys of young clusters have revealed that solar-type and lower-mass stars that rotate fast tend to retain almost cosmic A(Li), while slow rotators are heavily depleted \citep{Soderblom93,Bouvier16,Bouvier18,Arancibia20}. Disk locking, that is, the coupling between the stellar surface and the inner protoplanetary disk through magnetospheric accretion flows, prolongs the slow rotation phase and might facilitate stronger Li depletion through different mechanisms \citep[e.g., shear-driven mixing at the base of the convection zone;][and references therein]{Bouvier08,Eggenberger12}. By contrast, analogous effects in intermediate-mass stars are barely constrained. \citet{King93} found that young intermediate-mass stars in Orion Ic showed a trend similar to that of their lower-mass analogs because the stars in their sample with the lowest $v \sin i$ values also had the lowest A(Li). Nevertheless, this early result was based on a small sample of 23 sources and needs additional testing. The few observational works that included stars more massive than the Sun suggested that disk locking might also apply to intermediate-mass stars \citep{Rebull06,Bu25}.

\begin{figure}
   \centering
   \includegraphics[width=9cm]{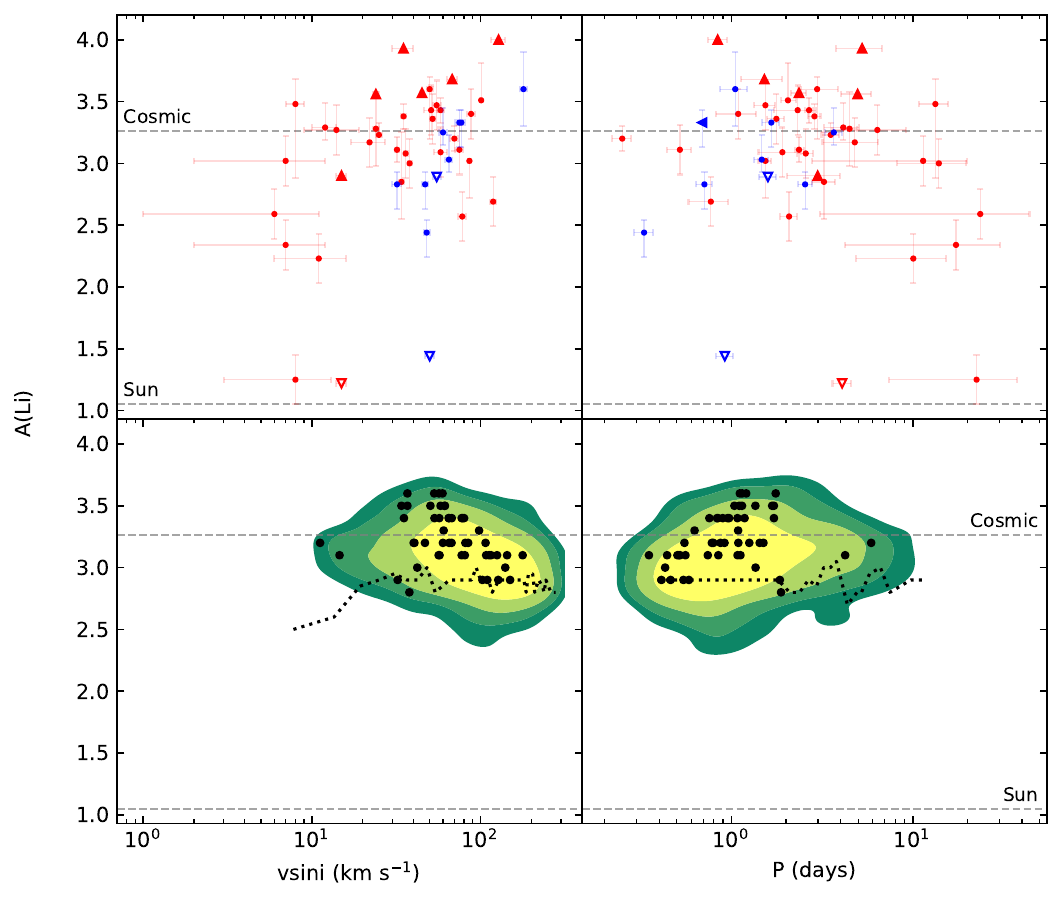}
      \caption{Li abundance vs. projected rotational velocity (left) and period (right) for pre-MS (top) and MS (bottom) intermediate-mass stars. The solid and open symbols, triangles, and horizontal dashed lines are the same as in Fig. \ref{fig:Li_age}. Top: Red and blue symbols refer to IMTTs and Herbigs, respectively. Bottom: Darker to lighter contours correspond to the 80th, 85th, 90th, and 95th percentiles of the kernel density estimate, and the stars on the early MS are plotted with black circles (the mean error bars are $\sim$ 1 km s$^{-1}$, 0.05 days, and 0.1 dex for $v \sin i$, P, and A(Li)). The medians of the non-detections are indicated with dotted lines for reference.}  
         \label{fig:Li_rot}
   \end{figure}

The top panels of Fig. \ref{fig:Li_rot} show A(Li) versus the rotational velocities and periods for the stars in our sample. The trend indicated by \citet{King93} is recovered: the lowest abundances correspond to the stars that rotate slower in terms of their projected rotational velocities, with a low Spearman probability of a false correlation (p $\sim$ 2 $\%$). However, when projection effects and stellar radii are considered by means of the rotational periods, no trend is observed (p $\sim$ 42 $\%$). This lack of a trend is mainly driven by the Herbig stars because the IMTTs alone still show a weak anticorrelation between rotational periods and abundances (p $\sim$ 7 $\%$) that is similar to their lower-mass TT counterparts \citep{Bouvier16}. In contrast, the Herbigs with the lowest A(Li) values ($<$ 2.5) are the fastest rotators (P $<$ 1 day). 

The bottom panels of Fig. \ref{fig:Li_rot} also show A(Li) versus rotational velocities and periods, but this time, for 1129 intermediate-mass stars in the MS. These are single stars from \textit{Gaia} with reliable LAMOST A(Li) measurements (see Appendix \ref{appendix:whole_sample}) and $v \sin i$ values in \citet{Sun21}. The periods in the bottom right panel were derived based on the stellar radii inferred from the stellar luminosities and temperatures also provided in \citet{Sun21}, assuming a typical inclination of $\pi$/4 (Sect. \ref{Sect:stellar_parameters}). The mean error bars are $\sim$ 1 km s$^{-1}$ for $v \sin i$ and $\sim$ 0.05 days for P (after propagation of the uncertainties in L$_*$ and T$_ *$). The data in \citet{Sun21} refer to A- and F-type MS stars within the same 1.5 -- 3.5 M$_{ \odot}$ mass range as we studied here, and they also come from the LAMOST survey. Thus, the stars plotted in the bottom panels of Fig. \ref{fig:Li_rot} constitute the largest possible sample of intermediate-mass MS stars with measurements of A(Li) and stellar rotation, to our knowledge. For these stars, A(Li) tends to be smaller for fast-rotating stars, which is confirmed based on rotational velocities and periods (p $<<$ 1$\%$). In order to avoid possible effects due to certain rotational evolution during the MS \citep[e.g.,][and references therein]{Sun21b,Sun24}, we repeated the exercise considering the 50 stars that spent less than 20$\%$ of their MS life (black symbols in the bottom panels of Fig. \ref{fig:Li_rot}). The same trend is statistically confirmed (p $<$ 1$\%$), in sharp contrast to previous results, which indicated that rapid rotators tend to retain almost cosmic Li abundances for low-mass stars in the early MS \citep{Soderblom93,Bouvier18,Arancibia20}.    

In summary, our results suggest that while IMTTs and low-mass young stars behave similarly concerning the relation between faster rotation and larger amounts of Li, the relation reverses during the Herbig phase. From this stage on, the opposite trend relating faster rotation with smaller amounts of Li characterizes intermediate-mass stars. We propose that this switch is related to the rotational and accretion evolution of young intermediate-mass stars. The remainder of this section shows that disk-locking also plays a major role in IMTTs, and Sect. \ref{Sect:Li_accretion} is devoted to analyzing the link with accretion.  

\begin{figure}
   \centering
   \includegraphics[width=9cm]{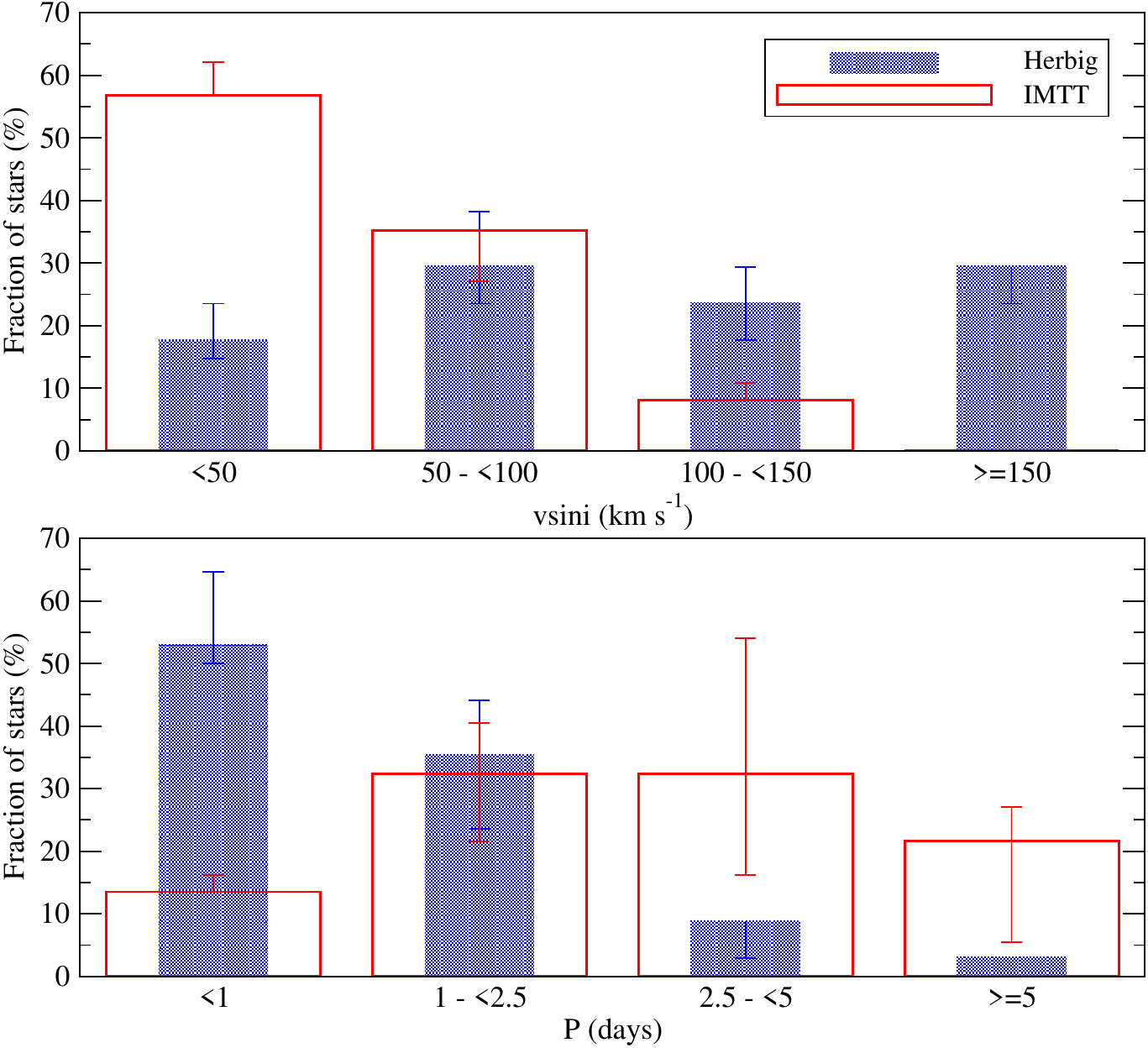}
      \caption{Distributions of rotational velocities and periods for 37 IMTTs (red) and 34 Herbigs (blue).} 
         \label{fig:hist_rot}
   \end{figure}

Figure \ref{fig:hist_rot} shows the distribution of rotational velocities and periods for all IMTTs and Herbigs in our sample. The KS and AD tests both reject the null hypothesis that the parent distribution is the same, with a default significance level of 5$\%$. The median rotational periods for the IMTT and Herbig stars is 2.7 and 0.90 days, respectively. Thus, the rotation of Herbig stars is faster by typically a factor of 3 than for IMTTs.

\begin{figure}
   \centering
   \includegraphics[width=9cm]{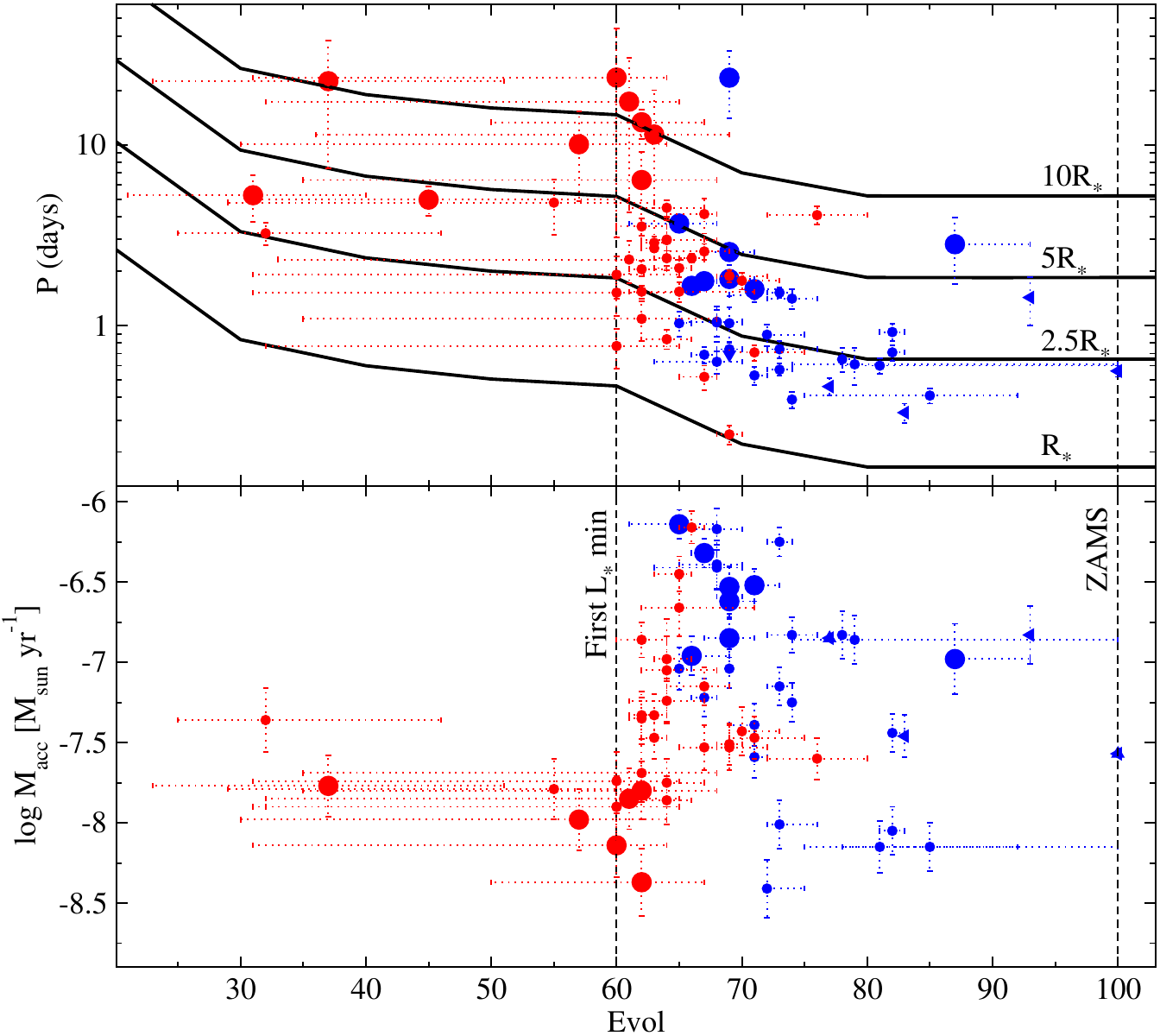}
      \caption{Rotation (top) and accretion (bottom) evolution for the IMTTs (red) and Herbigs (blue) in our sample. The slowest rotators (first quartile in each regime) are indicated with large symbols. The solid lines in the top panel show the rotational periods predicted by Keplerian rotation at 1, 2.5, 5, and 10R$_*$, as indicated (see text). The triangles and dashed lines are the same as in Fig. \ref{fig:Li_age}.}  
         \label{fig:disk_locking}
   \end{figure}

Figure \ref{fig:disk_locking} (top) shows the evolution of the rotational periods for the IMTTs and Herbigs. We computed the rotational periods expected from Keplerian rotation at a distance of 1, 2.5, 5, and 10R$_*$, with 5R$_*$ being the typical magnetospheric radius at which the star is anchored to the disk in low-mass TTs \citep[e.g.,][]{Bouvier20,Wojtczak23}. The updated BaSTI stellar evolution models \citep{Hidalgo18} were used to compute R$_*$ at different evolutionary stages for stellar masses 1.5, 2.5, and 3.5 M$_ {\odot}$. Since the resulting Keplerian periods are similar, Fig. \ref{fig:disk_locking} (top) shows the averaged curves based on the previous models (solid lines). The rotational periods of all stars are compatible with disk Keplerian rotation at distances beyond the stellar surface. However, the rotational periods of the Herbigs are consistent with Keplerian rotation at typically smaller disk distances than for the IMTTs. In particular, when we consider the error bars, all Herbigs are consistent with Keplerian rotation at small magnetospheres ($\leq$ 5R$_*$, except for the strongly magnetic star HD 101412; see Appendix \ref{appendix:stars}), but nearly half of the IMTTs are consistent with Keplerian rotation at larger magnetospheres. 

The compatibility of stellar rotation with Keplerian rotation at distances beyond the stellar surface is consistent with the hypothesis that disk-locking also operates in young intermediate-mass stars. However, the region in which the disk and the star rotation is coupled decreases during the pre-MS evolution: the size is similar to that of classical TTs during the IMTT phase, and it is then significantly smaller during the Herbig phase. This contraction is expected because the convective envelopes and magnetic fields become smaller during the pre-MS evolution of intermediate-mass stars \citep[Fig. \ref{fig:interiors} and, e.g.,][]{Villebrun19}, which is expected to lead to smaller magnetospheric disk truncation radii \citep{Koenigl91,Shu94}. The presence of disk-locking constitutes independent support for the current evidence that indicates that magnetospheric accretion can be extended from the low-mass to the intermediate-mass regime \citep[][and references therein]{Mendi20}. Nevertheless, the disk lifetimes tend to be shorter for intermediate-mass stars \citep{Hernandez05,Ribas15}, which as we have shown experience substantial braking by the disk (at distances $\geq$ 5R$_*$) only for a short period of time that is limited to the IMTT phase. Pre-MS tracks show that for a 2.5 M$_ {\odot}$ star, this phase takes less than $\sim$ 3 Myr, or $<$ 25$\%$ of the pre-MS lifetime. In contrast, disk-locking has been reported to operate for a timescale twice longer in 0.4 -- 1.2 M$_ {\odot}$ TTs \citep{Serna21}.        

\subsection{Li abundance and accretion}
\label{Sect:Li_accretion}
In addition to the effect that accretion bursts during the earliest stages of evolution might have on the Li content \citep[e.g.,][]{Baraffe10,Baraffe17}, this can be altered by current accretion onto TTs delivering interstellar medium material with pristine A(Li) to the stellar surface \citep[e.g.,][]{Piau02}. Observational studies have indeed revealed that accreting classical TTs tend to have higher A(Li) than non-accreting weak TTs \citep[e.g.,][]{Sestito08}. Although any straightforward A(Li)-$\dot{\rm M}_{\rm acc}$ relation can be diluted by evolutionary effects and additional factors such as observational biases, rotation (Sect. \ref{Sect:Li_rotation}), or the presence of planets (Sect. \ref{Sect:Li_SEDs}), Li replenishment by accretion has been unambiguously observed in the TT RW Aur at least. \citet{Stout00} observed this source during low and high accretion states and showed that the high accretion state leads to super-meteoritic A(Li) levels. Because the timescale for Li burning in stellar interiors is longer as the stellar mass increases, Li-enhancement produced by accretion might persist longer and might thus be better observable in the intermediate-mass regime we studied.      

Nevertheless, previous observations were limited to solar-like stars or to less massive stars. For more massive sources, the situation might be more puzzling because it was reported recently that the median accretion rates of Herbig stars are higher by an order of magnitude than those of their IMTT precursors \citep{Brittain25}. In order to explain this contradiction with canonical models of accretion evolution, \citet{Brittain25} proposed that the sharp increase in the far-ultraviolet (FUV) radiation expected during the evolution from the IMTT to the Herbig regime drives the reported increase in the mass accretion rate. Before we addressed potential links between accretion and the Li content, we therefore revisited the accretion history of young intermediate-mass stars.  

\begin{figure}
   \centering
   \includegraphics[width=9cm]{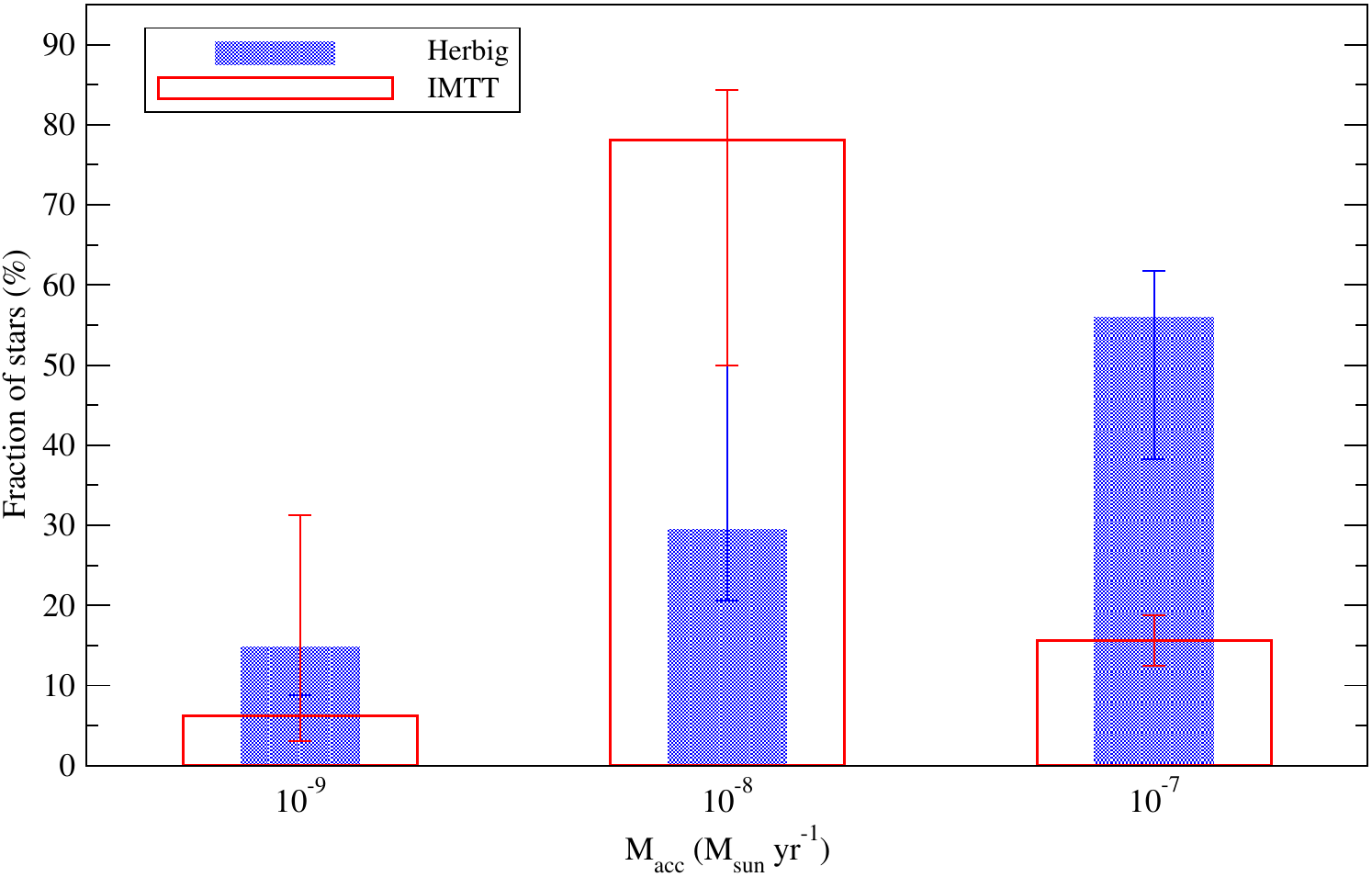}
      \caption{Distribution of the mass accretion rates for 32 IMTTs (red) and 34 Herbigs (blue).}  
         \label{fig:accretion_IMTTs_Herbigs}
   \end{figure}

Figure \ref{fig:accretion_IMTTs_Herbigs} shows the distributions of mass accretion rates for the IMTT and Herbig stars in our sample. The KS and AD tests reject the null hypothesis that they are drawn from the same parent distribution with a default significance level of 5$\%$. However, the only difference between the two distributions considering the error bars concerns the number of strong accretors ($>$10$^{-7}$ M$_{\odot}$ yr$^{-1}$), which are dominated by the Herbigs. At the typical median mass accretion rates, IMTTs show 3.02 $\times$ 10$^{-8}$ M$_{\odot}$ yr$^{-1}$, and Herbigs  1.25 $\times$ 10$^{-7}$ M$_{\odot}$ yr$^{-1}$, which is higher by only a factor of $\sim$ 4. 

The distributions of the mass accretion rates and the ratio of the median values for Herbigs and IMTTs differ from those reported by \citet{Brittain25}. Although we discarded several stars included in their sample because they have close companions, most sources are the same. The main reason for the discrepancy comes from the method adopted to infer the accretion rates of the IMTTs. For the majority of such sources, and in particular, for the weak accretors with log $\dot{\rm M}_{\rm acc}$ $<$ -8.5, \citet{Brittain25} used the calibration with the H$\alpha$ luminosities provided in \citet{Alcala17}. Based on the same H$\alpha$ luminosities, their calibration leads to accretion luminosities and mass accretion rates that are lower by 1 dex than those inferred from the \citet{Fairlamb17} calibration, which we used (Sect. \ref{Sect:results}). The \citet{Alcala17} calibration was inferred based on a sample of TTs with M$_*$ $<$ 1.5M$_{\odot}$, T$_*$ $<$ 5000 K, and log L$_*$/L$_{\odot}$ $<$ 0.67, but the stellar parameters of the sample analyzed in this work and by \citet{Brittain25} both lie above the previous values. Conversely, the H$\alpha$ empirical calibration we used is based on young stars with stellar properties that cover the IMTT and Herbig regimes \citep{Fairlamb17}. Because of the previous reason, and also based on the homogeneous procedure we applied to infer accretion rates on a sample of stars presumably unaffected by the presence of close stellar companions, our results probably reflect the difference in accretion properties between IMTTs and Herbigs better.

The increase in the mass accretion rate observed during the Herbig phase with respect to the IMTT phase might be explained as a natural consequence of the reduced size of the magnetosphere discussed in the previous section. The mass accretion rate in the innermost region of the disk (which we assumed to be equal to that onto the central star) is $\dot{\rm M}_{\rm acc}$ = $\rho$ $\times$ A $\times$ v$_{l}$, with $\rho$ the gas density at the disk truncation radius R$_t$, A the area over which matter is loaded onto the stellar magnetic field lines, and v$_{l}$ the launching velocity at the base of the magnetic funnels. v$_{l}$ scales with R$_t^{-1/2}$ (assuming Keplerian velocity), and A scales with the area of a ring, R$_t^2$. The density $\rho$ is often modeled as a power law scaling with R$^{-p}$, with p depending on assumptions such as the disk viscosity prescription and thermal structure. Under this approach, the ratio of the accretion rates shown by the same star during the Herbig and the IMTT phases is  $\dot{\rm M}_{\rm acc}$ (Herbig)/$\dot{\rm M}_{\rm acc}$ (IMTT) = (R$_{t, IMTT}$/R$_{t, Herbig}$)$^{p-1.5}$. Thus, to increase the mass accretion rate as the magnetosphere shrinks, it is required that p $>$ 1.5, which is consistent with essentially all models of gas density distributions in protoplanetary disks \citep[e.g.,][]{DAlessio98}. In particular, for the canonical value p = 2.25, the increase by a factor of $\sim$ 4 in the mass acccretion rate that is typically observed during the evolution from the IMTT to the Herbig regime can be recovered when the magnetosphere size is reduced by a factor of $\sim$ 6, which agrees in general with the observations in the top panel of Fig. \ref{fig:disk_locking}. This view, which links accretion with the size of the magnetosphere, does not contradict the potential role that the increase of the FUV luminosity might also play in boosting accretion rates \citep{Ronco24,Brittain25}. 

It is relevant for the broad picture outlined below to note that the previous scenario also predicts that the IMTTs that rotate more slowly, that is, that are disk-locked at larger distances from the star, are expected to show lower accretion rates than faster IMTTs whose truncation radii were already reduced during their evolution toward the MS. This trend is observed in Fig. \ref{fig:disk_locking}, where the slowest IMTTs (large red symbols) also show the lowest accretion rates. In turn, Fig. \ref{fig:disk_locking} also shows that during the Herbig stage, the accretion rates and rotational periods tend to decrease, which agrees in general with canonical models of evolution based on disk dissipation, and with a lower disk-locking efficiency. 

\begin{figure}
   \centering
   \includegraphics[width=9cm]{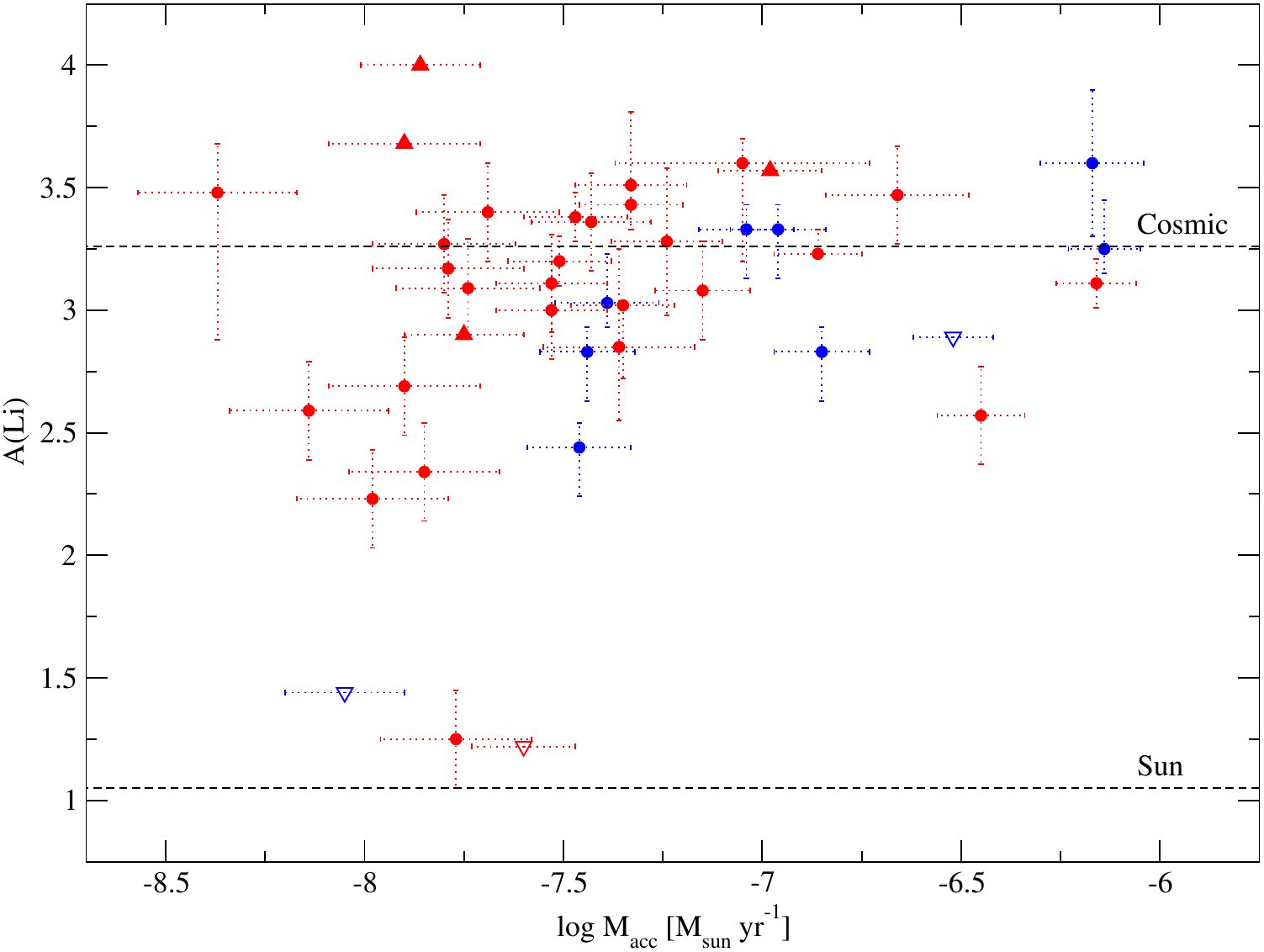}
      \caption{Li abundance vs. mass accretion rate for IMTTs (red) and Herbigs (blue). The solid and open symbols, triangles, and dashed lines are the same as in Fig. \ref{fig:Li_age}.}  
         \label{fig:Li_accretion}
   \end{figure}

With the previous discussion in mind, we show in Fig. \ref{fig:Li_accretion} the potential relation of the Li content and accretion. In the whole sample, a weak correlation links higher abundances with higher accretion rates (p $\sim$ 13 $\%$), but this is stronger for the Herbig stars alone (p $\sim$ 3 $\%$) than for the IMTTs (p $\sim$ 28 $\%$). The emerging picture based on all previous analyses is summarized below. 

The initial accretion rates of IMTTs increase because the size of the magnetosphere decreases (and possibly, because the FUV luminosity increases). The lack of a clear A(Li) -- $\dot{\rm M}_{\rm acc}$ trend for IMTTs may be explained by the balance between the initial abundance, Li burning when the stellar rotation is disk locked, and Li enhancement due to freshly accreted material. During the Herbig phase, the accretion rates and rotational periods tend to decrease, in line with canonical evolution. Because standard Li depletion observed in low-mass TTs is not expected for more massive sources, the accreted material might even produce a stronger Li enrichment, which would explain the A(Li) -- $\dot{\rm M}_{\rm acc}$ correlation in Herbig stars. In turn, the A(Li) -- P correlation observed during the Herbig phase and later on during the MS might be caused by the stars that remain disk-locked for a comparatively longer time. These stars would incorporate fresh Li from the circumstellar medium while keeping relatively low rotational velocities. The presence of planets might also affect A(Li), and we explore this next.

\subsection{Li abundance and planets}
\label{Sect:Li_SEDs}

The direct detection of exoplanets in protoplanetary disks around pre-MS stars remains a very challenging task \citep[with very few confirmed candidates; e.g.,][]{Keppler18,Capelleveen25}, and indirect methods are still necessary to infer these exoplanets \citep[e.g.,][]{vioque26}. The underabundance of metals in the stellar photospheres of young intermediate-mass stars with group I SEDs has been related with the presence of forming planets in their protoplanetary disks. Theoretical and observational works support that giant planets capable of opening a gap in the disk trap refractory elements and prevent their transfer onto the central stars, which accrete metal-poor material and thus show lower metallicities than young stars without such planets \citep{Kama15,Jermyn18,GuzmanDiaz23}. The well-known correlation between the presence of giant planets and higher stellar metallicities of MS stars \citep[e.g.,][and references therein]{Gonzalez97,Adibekyan19} does not contradict the previous lines of evidence. The planet-metallicity correlation is also observed in sufficiently evolved MS and post-MS intermediate-mass stars that show primordial metal abundances, when the mixing processes with the stellar interiors become efficient \citep{Maldonado25}. In turn, the presence of Li on stellar surfaces has been associated with the engulfment of giant planets \citep[e.g.,][]{Israelian01,Spina21}, which is expected to occur with a greater likelihood as the stellar mass increases and their magnetospheres become too weak to halt inward planet migration \citep{Mendigutia24}. On the other hand, \citet{Israelian04,Israelian09} found that Sun-like MS stars hosting planets exhibit significantly lower A(Li) than those without detected planets. The origin of this relation and the type of stars for which it applies remains the subject of debate \citep[e.g., the recent works][and references therein]{Rathsam23,Jinxiao25,Carlos25}.

\begin{figure}
   \centering
   \includegraphics[width=9cm]{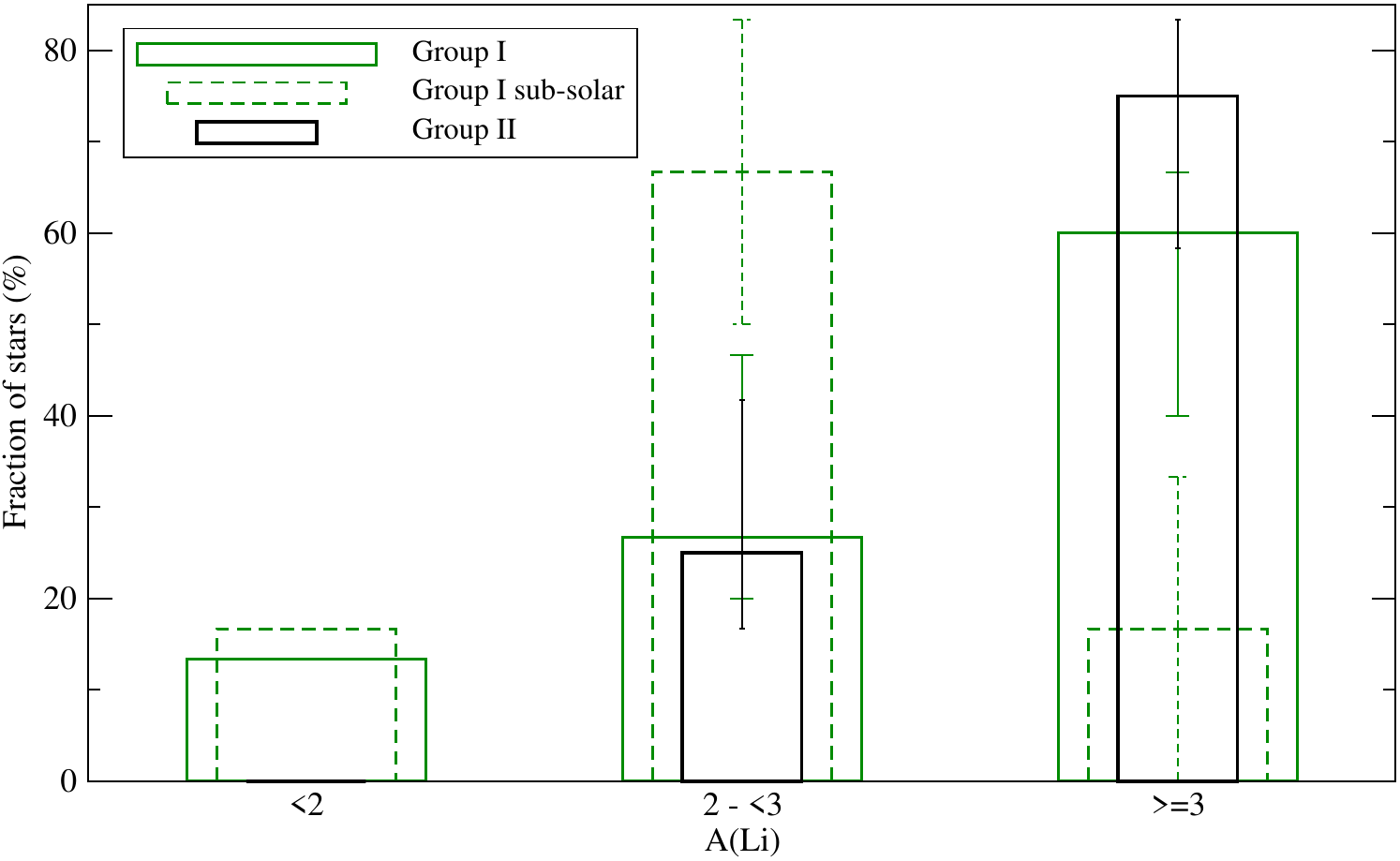}
      \caption{Distribution of A(Li) for 15 group I (solid green), 6 group I with subsolar metallicity (dashed green), and 12 group II (black) young intermediate-mass stars.} 
         \label{fig:Li_hist_GI_vs_GII}
   \end{figure}

Figure \ref{fig:Li_hist_GI_vs_GII} shows the distributions of A(Li) for SED group I sources, group I sources with subsolar metallicity (above the error bars), and group II sources. Following previous works \citep[e.g.,][and references therein]{Kama15,Jermyn18,GuzmanDiaz23,Maldonado25}, group I was used as a probe of protoplanetary disks with holes presumably caused by giant planets, group I  with subsolar metallicity served as a proxy of the best giant-planet host candidates, and group II represented the stars without holes in their protoplanetary disks that are thus the less likely giant-planet hosts. For A(Li) $\geq$ 2, the distributions of groups I and II coincide within the error bars. Based on our sample, we therefore see no trend that would link the potential presence of giant planets with higher values of A(Li). The reason might be the absence of planet engulfment in our sample or that planet engulfment does not result in significant differences in A(Li) \citep[as proposed, e.g., by][and references therein]{Sun25}. In contrast, the only stars with A(Li) $<$ 2 are group I sources, which suggests that giant planet formation might be linked to Li depletion. The inclusion of group I sources with subsolar metallicity reinforces the previous conclusions. Four out of six of these sources have already been proposed as planet hosts based on high-resolution imaging of their disks: HD 169142 \citep[e.g.,][]{Pohl17,Bertrang18,Toci20,Hammond23}, V599 Ori \citep{Valegard24}, HD 139614 \citep{Muro-Arena20}, and HD 143006 \citep{Perez18,Benisty18,Ballabio21}. No such observations are available for the remaining two sources (HBC 217 and PDS 2). In either case, the A(Li) values of all group I sources with subsolar metallicity except for HBC 217 are well below the cosmic abundance. 

Although the previous analysis is based on a small number of pre-MS stars with not yet confirmed exoplanets, it agrees with the finding by \citet{Israelian09} and suggests that the possible link between the presence of exoplanets and Li depletion might begin during the pre-MS \citep{Bouvier08} and extend to intermediate-mass stellar hosts. The latter possibility is difficult to test on the basis of current data because it is difficult to observe Li and confirm the presence of exoplanets for stars more massive than the Sun \citep[e.g.,][and references therein]{Mendigutia24,Maldonado25}.  

\section{Summary and conclusions}
\label{Sect:summary_conclusions}
We conducted a comprehensive analysis of the early evolution of intermediate-mass stars considering their Li content. This analysis was mainly based on well-characterized IMTT and Herbig stars not surrounded by close companions that might contaminate their spectra and affect their evolution. Our main results and conclusions are listed below.

\begin{itemize}
    \item New estimates for the stellar metallicities, rotational velocities and periods, and accretion rates were provided. The curves of growth for stars hotter than 8000 K, also necessary to infer A(Li), are made available to the community.
    \item Li depletion is generally less significant for intermediate-mass stars than for their lower-mass analogs, as expected from standard evolution models of stellar interiors. However, significant Li depletion is present in $\sim$ 25 -- 30 $\%$ of intermediate-mass stars. Although standard models predict that Li depletion is more effective in stars with masses 1.5 -- 2.5 M$_{\odot}$ than in those within the mass range 2.5 -- 3.5 M$_{\odot}$, we found no significant difference. In contrast, we found hints suggesting that Li depletion might even be slightly more effective in the high-mass subsample, which agrees with previous findings \citep{Maldonado25}. 

    \item We provided evidence to support that disk-locking, and thus, the magnetospheric accretion paradigm, also applies to young intermediate-mass stars. Nevertheless, the timescale on which it operates is shorter by a factor of $\sim$ 2 than for lower-mass stars. In particular, we showed that disk-locking is efficient mainly during the IMTT phase, before the magnetosphere shrinks when the star enters the Herbig phase.

    \item We confirmed the recent result by \citet{Brittain25}, which indicated that Herbig stars show typically higher mass accretion rates than IMTTs. However, we found an increase by a factor of $\sim$ 4, which is smaller than the order of magnitude reported in that work. We explained this inconsistency mainly by different estimates of the accretion rates, ours based on an empirical calibration valid for IMTT and Herbig stars. We showed that the increase in the rotation by a factor of 3 and the increase in accretion by a factor of 4 that is observed during the transition from the IMTT to the Herbig regime can be explained by the above-mentioned reduction of the magnetosphere size. 
    
    \item While for IMTTs we found a trend relating lower A(Li) with slow rotators that is similar to that reported for lower-mass young stars, the trend reverses for Herbig stars and intermediate-mass stars in the MS. In addition, higher accretion rates and Li abundances are correlated in Herbig stars, but only weakly in IMTTs. We proposed that Li-rich accreted material plays a major role in explaining the previous trends.    
 
    \item Our data tentatively suggest that the possible link between the presence of exoplanets and Li depletion \citep{Israelian09} might extend to the intermediate-mass regime and might begin during the pre-MS evolution. 

\end{itemize}

Our results for the Li abundances are affected by the lack  of comprehensive spectroscopic studies focused on the thousands of young intermediate-mass stars that are currently identified \citep[e.g.,][]{Vioque20}. A detailed analysis of the early evolution of intermediate-mass stars also requires the development of specific models dealing with nonstandard processes of Li depletion, similar to those that are currently available for lower-mass sources.    

\section*{Data availability}
Tables \ref{table:sample}, \ref{table:observations}, \ref{table:results} and \ref{table:cog} are available in electronic form at the CDS via anonymous ftp to cdsarc.u-strasbg.fr (130.79.128.5) or via http://cdsweb.u-strasbg.fr/cgi-bin/qcat?J/A+A/.

\begin{acknowledgements}
The authors acknowledge the anonymous referee, whose suggestions have served to improve the manuscript. Based on data obtained from the ESO Science Archive Facility with DOIs: https://doi.org/10.18727/archive/27, https://doi.org/10.18727/archive/33, https://doi.org/10.18727/archive/50, and https://doi.org/10.18727/archive/71. I.M., L.F.G and G.M. are funded by grant PID2022-138366NA-I00, by the Spanish Ministry of Science and Innovation/State Agency of Research MCIN/AEI/10.13039/501100011033 and by the European Union. IM was also funded by a "Ramón y Cajal" fellowship RyC2019-026992-I. J.C.-W. acknowledges funding by the European Union under the Horizon Europe Research \& Innovation Programme 101039452 (WANDA). Views and opinions expressed are, however, those of the author(s) only and do not necessarily reflect those of the European Union or the European Research Council. Neither the European Union nor the granting authority can be held responsible for them. G.M.M acknowledges financial support from Junta de Andalucía through the program Emergia (EMEC$\_$2023$\_$00533)
\end{acknowledgements}

%
\bibliographystyle{aa}
\bibliography{myrefs}

\begin{appendix}




\section{Notes on specific stars}
\label{appendix:stars}
BD-07 1129, PDS 2, and HD 100453: These are the stars with the smallest A(Li) in the sample, with 1.25 $\pm$ 0.2, $<$ 1.2, and $<$ 1.4, respectively. The stellar parameters and Li EW of BD-07 1129 were taken from the spectroscopic analysis in \citet{King93}. The Li EW was corrected and transformed into abundance following the procedures described in Sect. \ref{Sect:abundances}. The A(Li) obtained coincides within errorbars with that provided by \citet{King93} (1.2 $\pm$ 0.2). There are no reports indicating that BD-07 1129 could be a binary/multiple system, to our knowledge. The surface temperature spectroscopically derived in the APOGEE survey looking for close companions in young stars \citep{Kounkel19} coincides within error bars with the one listed here, but the log g is significantly larger (4.29 $\pm$ 0.07, vs. 3.5 $\pm$ 0.2 reported here). However, the use of the log g value in \citet{Kounkel19} leads to the same A(Li) within error bars. Concerning PDS 2 and HD 100453, their upper limits on A(Li) are based on the non-detections of the \ion{Li}{i} feature in the corresponding ESO spectra. This lack of detections can hardly be explained by Li ionization or rotational dilution of the spectral line, based on the relatively low stellar temperatures and projected rotational velocities derived by \citet{GuzmanDiaz23}.

HD 101412: This is the star with the smallest $v \sin i$ in the sample, 3 $\pm$ 1 km s$^{-1}$. This value has been adopted from the spectroscopic study in \citet{GuzmanDiaz23}, which reports consistency with previous works. There is disagreement on whether its inclination is closer to pole-on \citep[$\sim$ 30\degr; e.g.,][]{vanderplas08} or to edge-on \citep[$\sim$ 80\degr; e.g.,][]{Fedele08,Ilee14}, for which an intermediate value expected from  random orientations ($\sim$ 52\degr) is adopted here. HD 101412 also shows one of the strongest magnetic fields ever measured in a Herbig star \citep[][]{Hubrig10}

Brun 555 and HBC 415: These are the only super-Li rich stars in the sample, with A(Li) $>$ 3.9 and $>$ 4.0, respectively. The Li EW and stellar parameters for Brun 555 were taken from \citet{King93}, and those for HBC 415 were measured in the corresponding ESO spectrum and taken from \citet{Valegard21}. Li EWs were corrected and transformed into abundances following the procedures described in Sect. \ref{Sect:abundances}, leading to lower limits due to the limitations of the computed COG described in that section. The A(Li) obtained by \citet{King93} is consistently larger than our lower limit (4.4 $\pm$ 0.2). None of the stars has reported companions, to our knowledge.

\section{Overall lithium evolution}
\label{appendix:evolution}

\begin{figure}
   \centering
   \includegraphics[width=9cm]{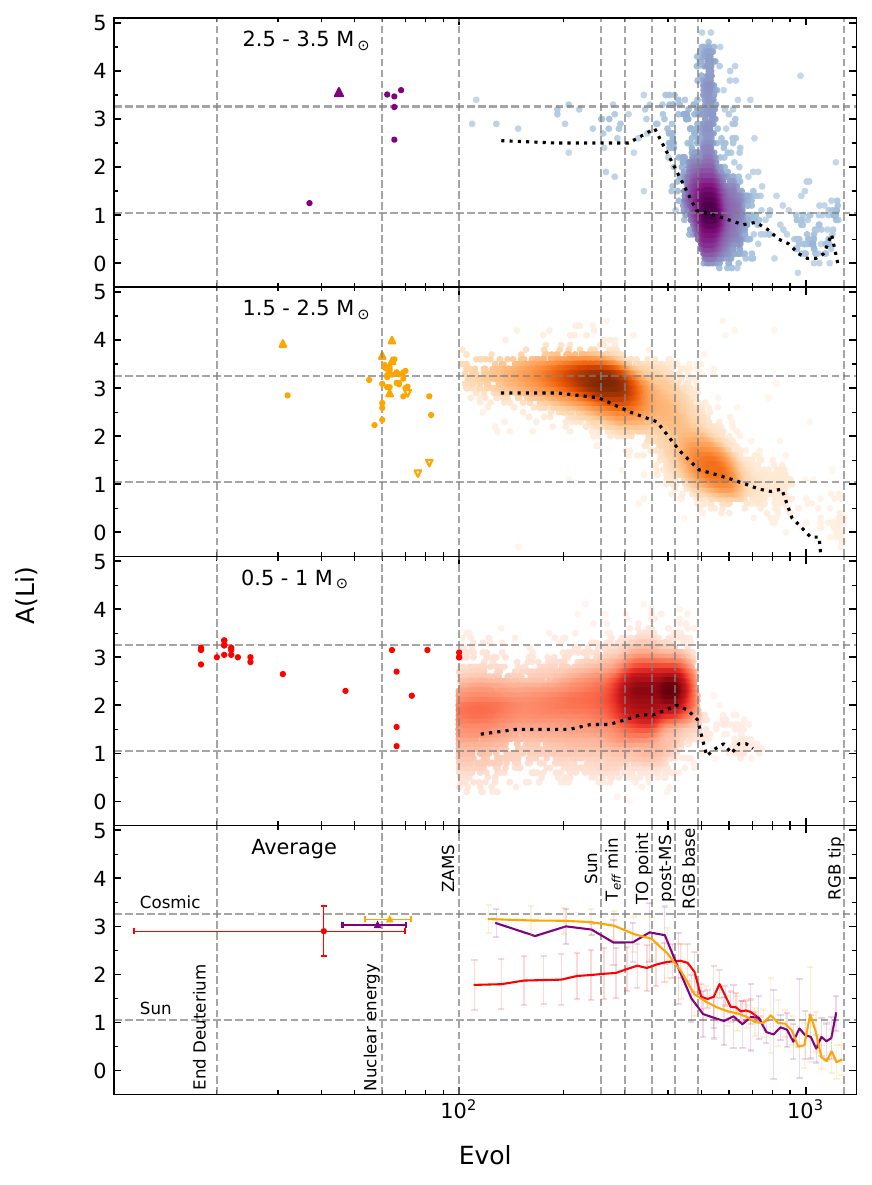}
      \caption{Expanded version of Fig. \ref{fig:Li_evol_pre_MS}.}  
         \label{fig:Li_evol}
   \end{figure}

Figure \ref{fig:Li_evol} shows the evolution of A(Li) for different stellar masses and phases including the pre-MS, the MS, and the post-MS. The pre-MS data come from this work for the intermediate-mass stars, and from \citet{Martin94b} for the (sub-)solar mass stars. For more evolved sources, data from the LAMOST survey \citep{Cui12} were filtered using the prescriptions in \citet{Ding24}. Thus, stars with low S/N spectra $\leq$ 20 and maximum difference and dispersion of A(Li) from multiple exposures $\geq$ 1.5 and $\geq$ 0.5, were discarded. The reported uncertainty for A(Li) is $\sim$ 0.1 dex for the filtered sources, which were crossmatched with the \textit{Gaia} DR3 catalog of astrophysical parameters produced by the Apsis processing \citep{Creevey23,Fouesneau23}. Then, we selected the sources with stellar masses within the corresponding ranges and with low probability of being a binary/multiple system ($<$ 0.01). This procedure leads to nearly 50,000 stars with A(Li) detections for the \textit{Gaia} DR3 evolutionary parameter spanning from the ZAMS to the red giant branch (RGB) tip. Different stages are indicated with vertical dashed lines, following the nomenclature in \citet{Hidalgo18} .

\label{appendix:whole_sample}

Non-detections are more abundant than detections, and both show a similar pattern spanning over similar ranges on the y-axes of the first three panels. Non detections are not plotted to avoid confusion and because it is not clear whether they actually indicate upper limits for the Li content or result from other causes (see, e.g., Sect. \ref{Sect:preliminary}). Instead, the medians of the non-detections are indicated with dotted lines, for reference. 

Two intriguing features in Fig. \ref{fig:Li_evol} are pointed out. First, based on the median values in the bottom panel, A(Li) appears to increase during the MS evolution of (sub-)solar type stars. Although such an increment is smaller than the error bars provided by the standard deviations, A(Li) is expected to decrease during the MS evolution of such stars \citep[e.g.,][]{Sestito05}. Second, although the median post-MS evolution is similar for all stellar mass ranges explored, only the higher mass stars show a clear spike of over-abundant stars. This is observed right after Evol $\sim$ 490, which corresponds to the base of the RGB for (sub-)solar mass models or to the minimum in log (L$_*$/L${_\odot}$) for intermediate-mass models. Post-MS lithium enrichment is known \citep[e.g.,][]{Charbonnel20,Martell21}, but the feature identified here only refers to relatively massive stars during a very specific stage, with no previous reports of a similar, well-defined behavior to our knowledge. It is out of the scope of this paper to analyze the origin of the two previous features, which may be the subject of future analyses.

\onecolumn
\section{Li spectra}
\label{appendix:lithium}
\begin{figure} [h]
 \centering
    \includegraphics[width=0.24\textwidth, trim=4cm 1cm 4cm 0, clip]{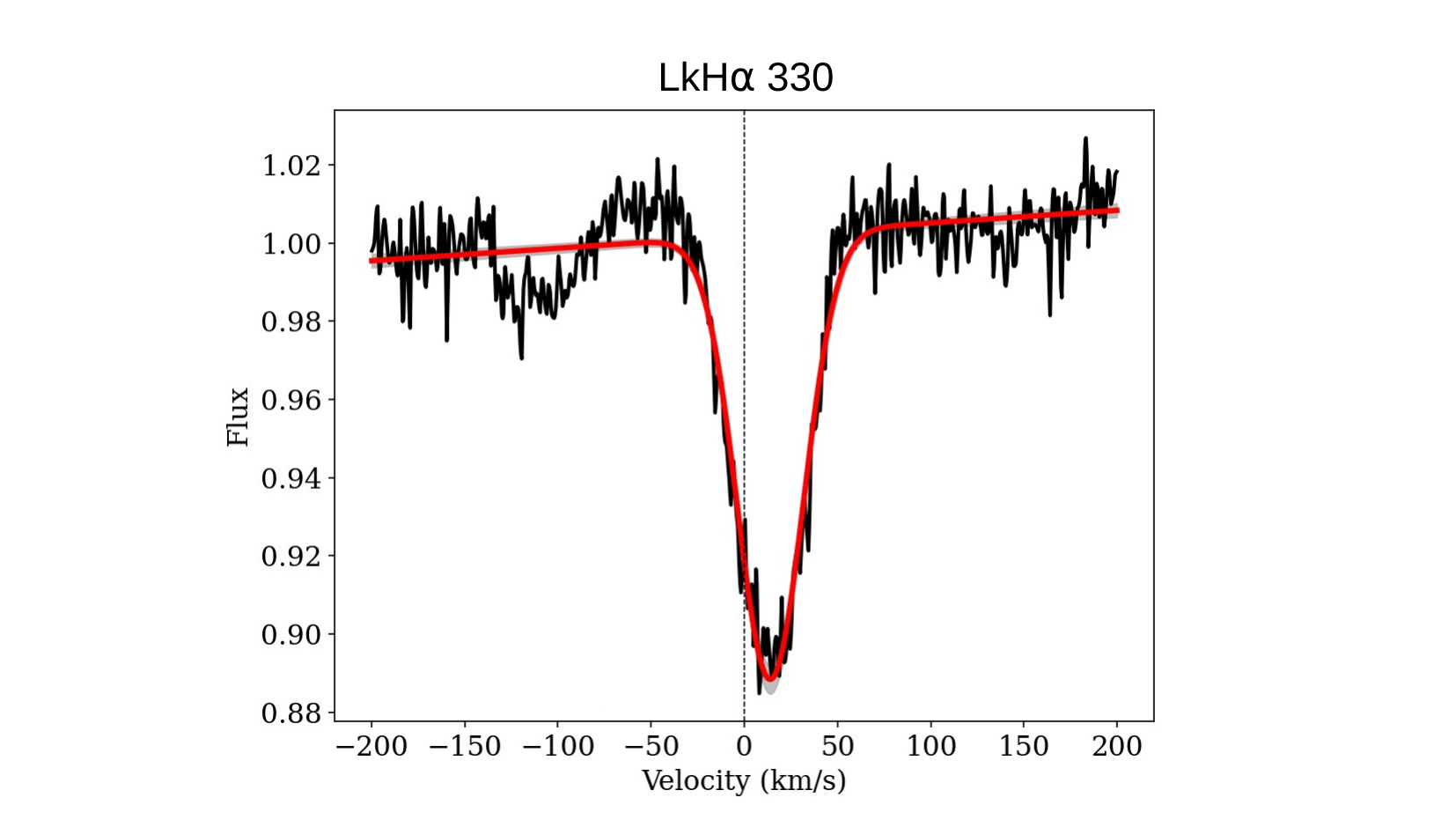}
    \includegraphics[width=0.24\textwidth, trim=4cm 1cm 4cm 0, clip]{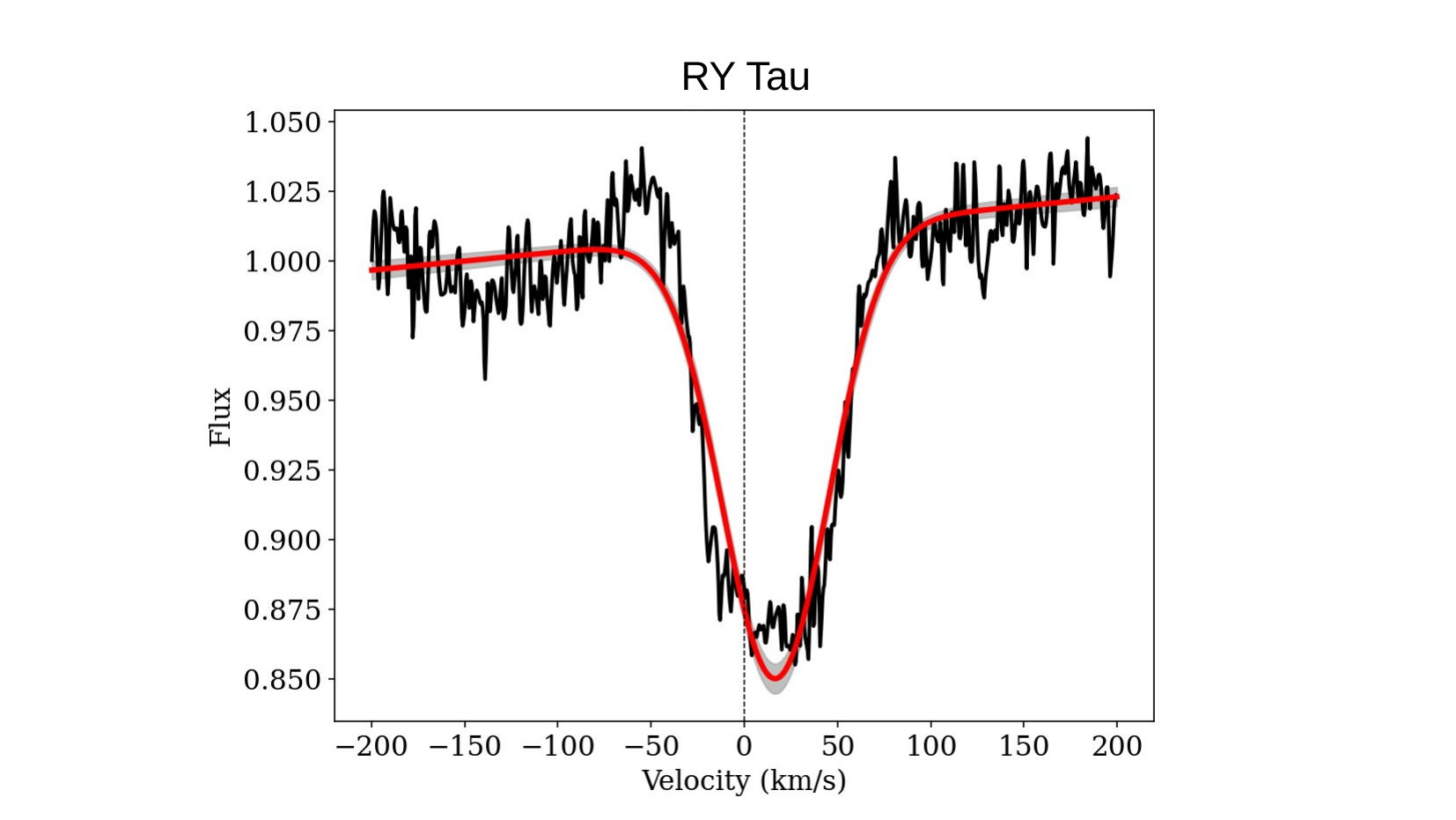}
    \includegraphics[width=0.24\textwidth, trim=4cm 1cm 4cm 0, clip]{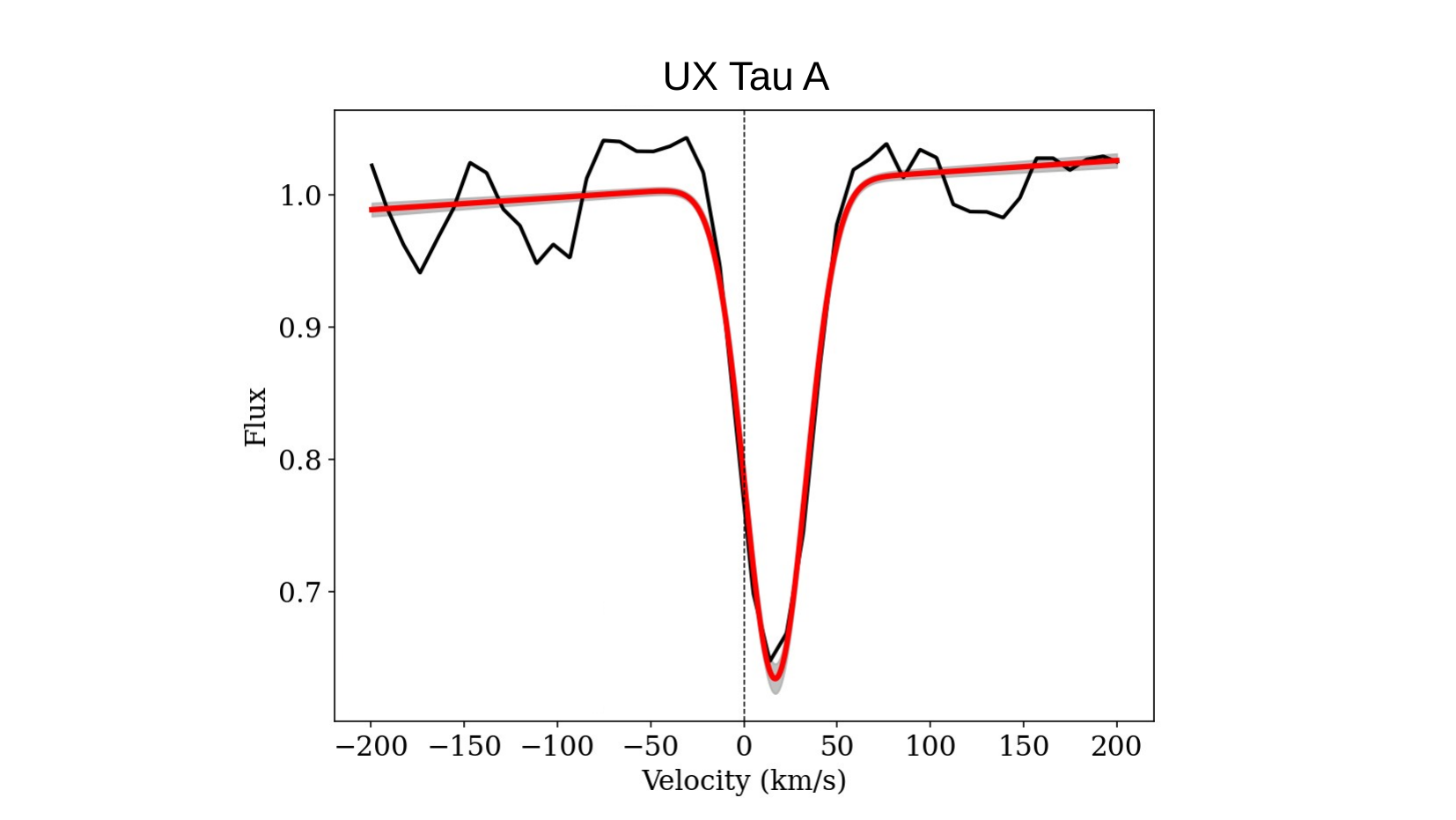}
    \includegraphics[width=0.24\textwidth, trim=4cm 1cm 4cm 0, clip]{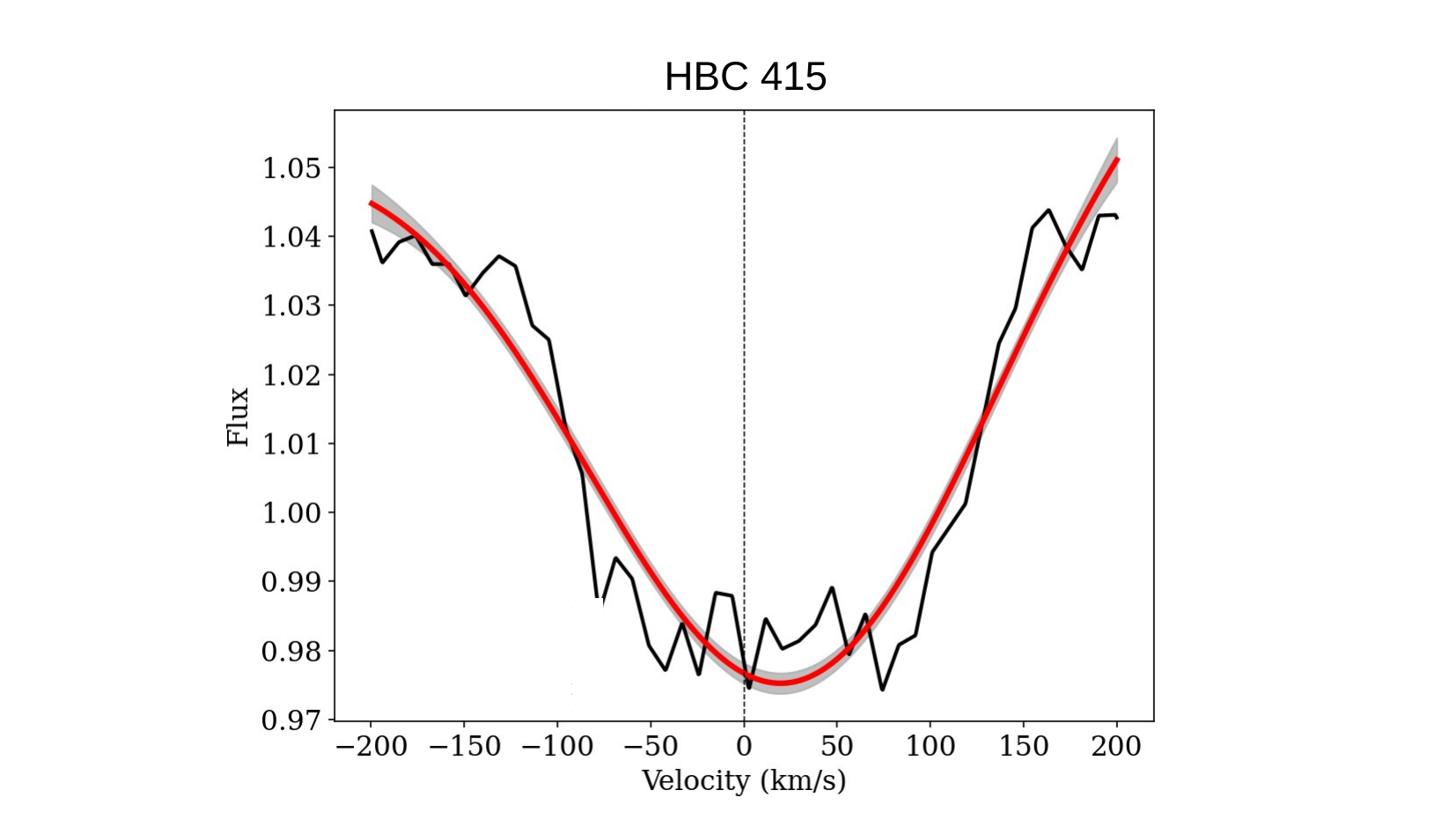}
\end{figure}
\begin{figure} [h]
 \centering
    \includegraphics[width=0.24\textwidth, trim=4cm 1cm 4cm 0, clip]{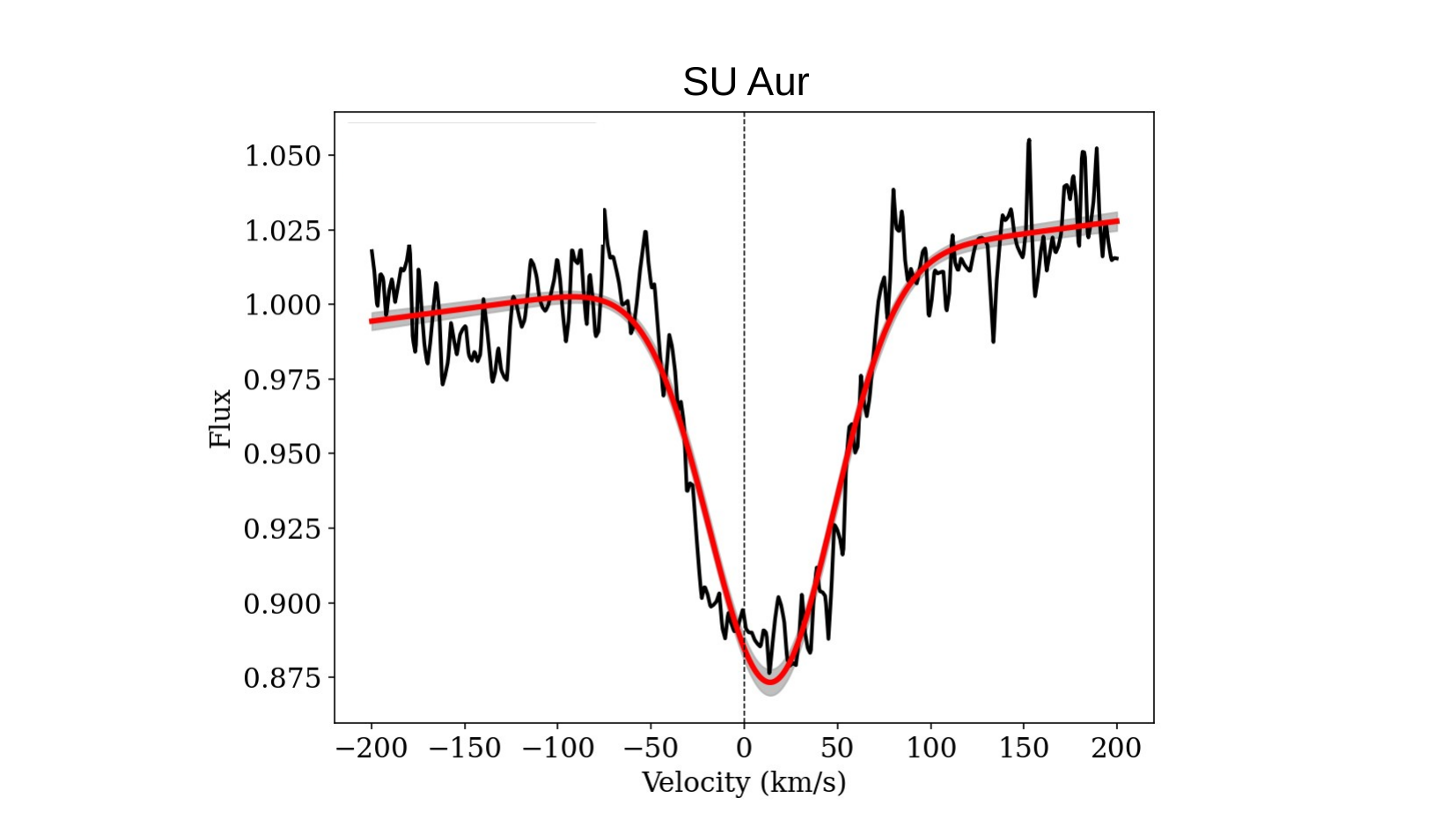}
    \includegraphics[width=0.24\textwidth, trim=4cm 1cm 4cm 0, clip]{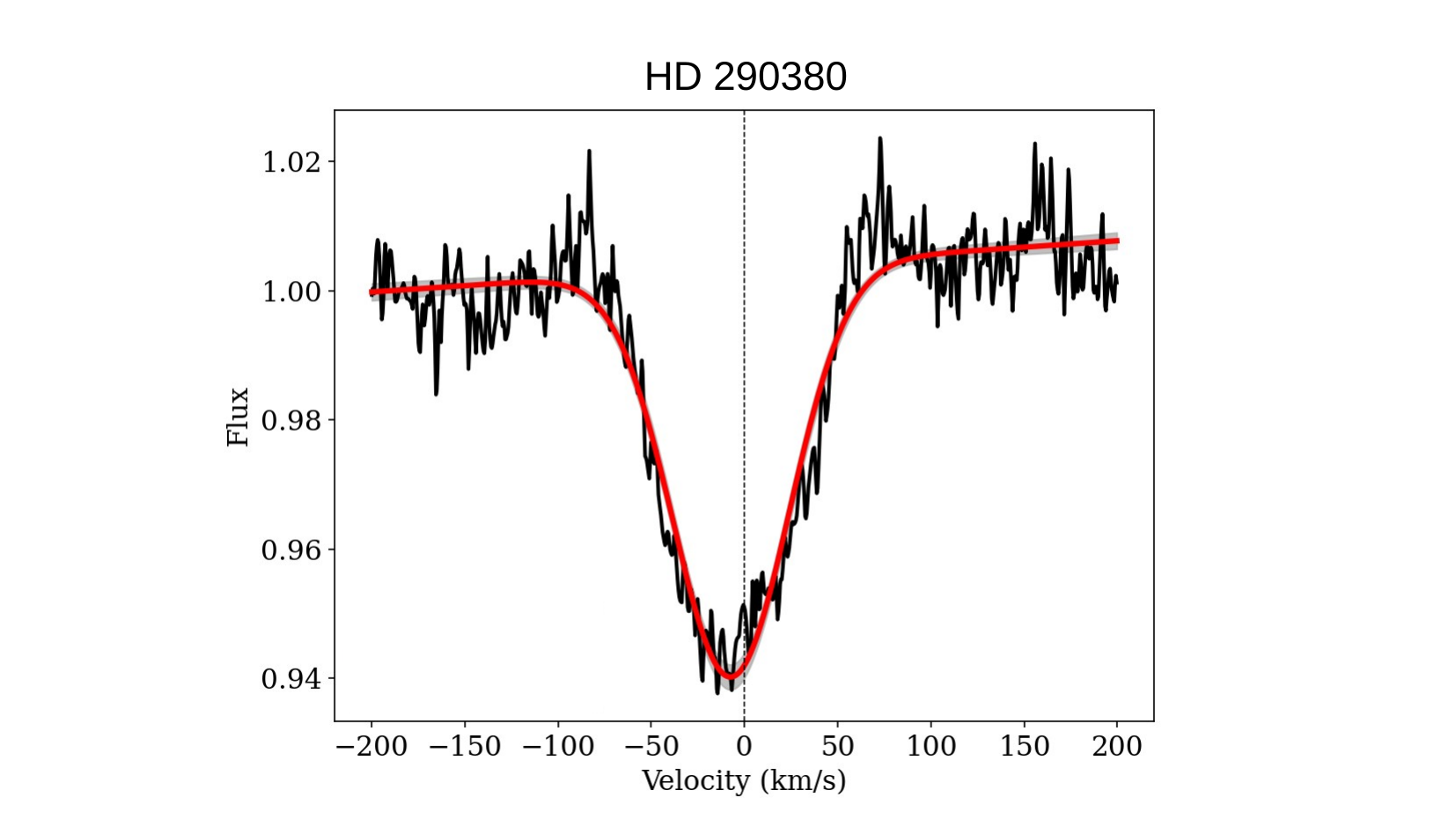}   
    \includegraphics[width=0.24\textwidth, trim=4cm 1cm 4cm 0, clip]{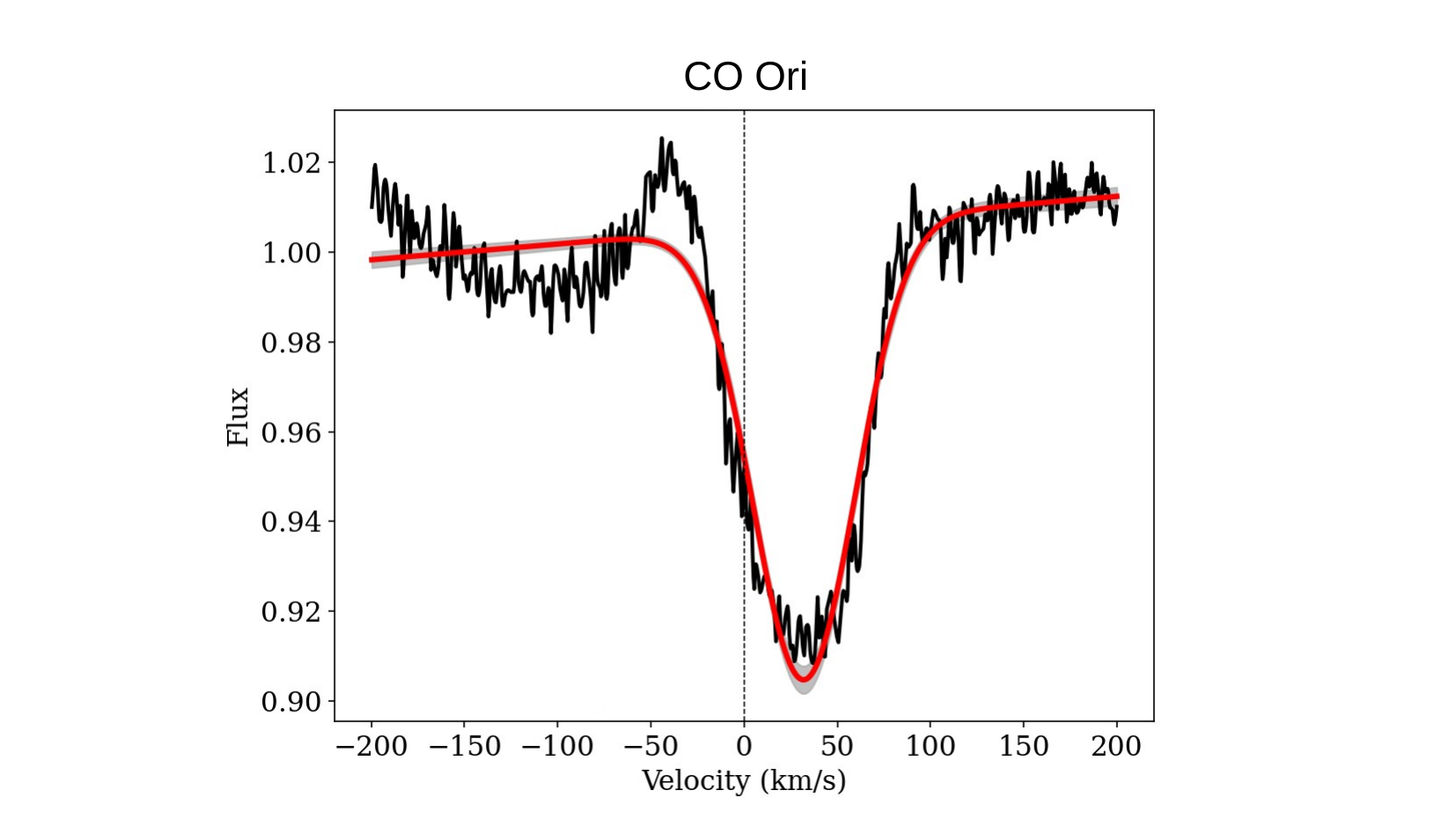}
    \includegraphics[width=0.24\textwidth, trim=4cm 1cm 4cm 0, clip]{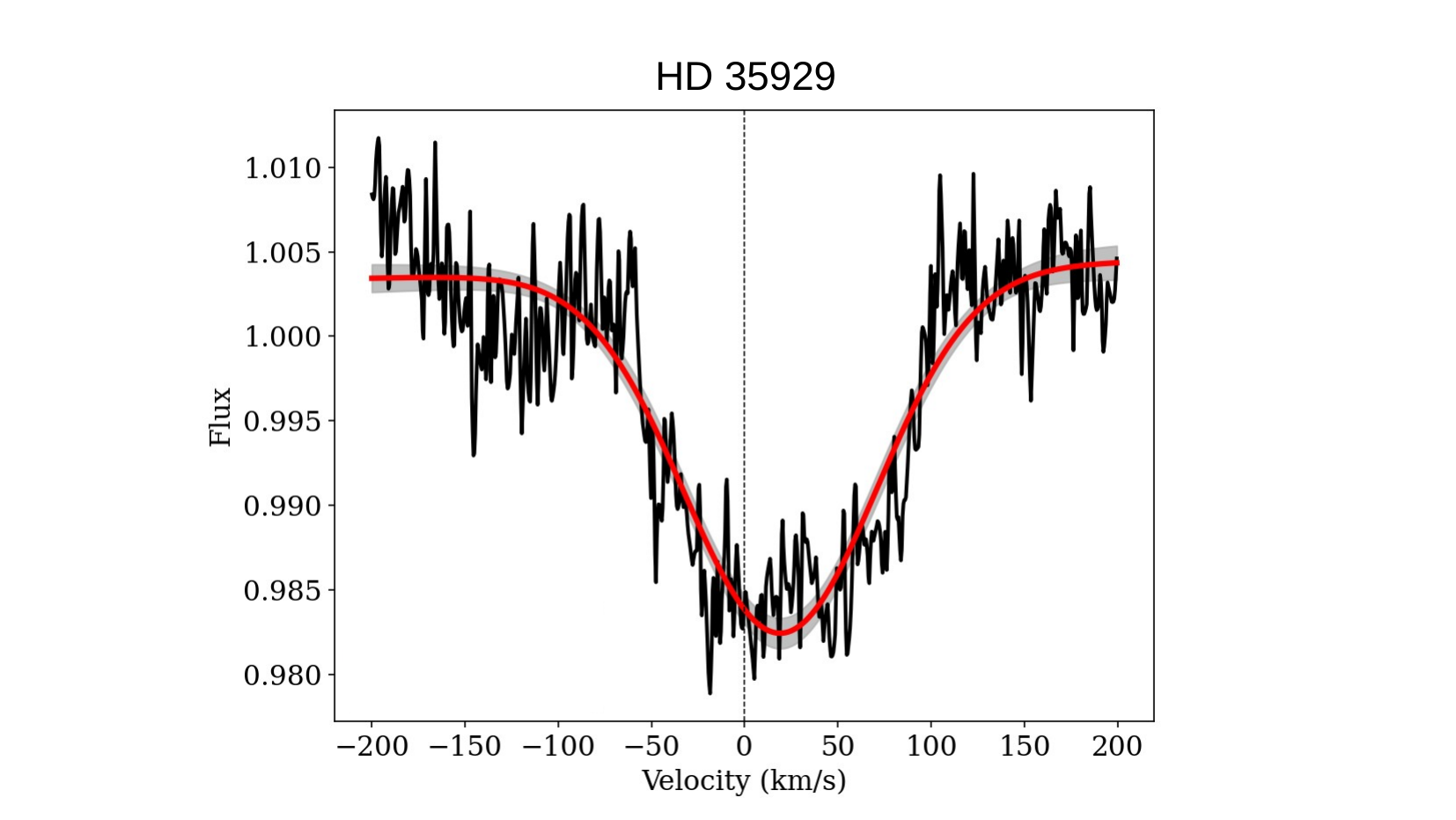}
\end{figure}
\begin{figure} [h]
 \centering
    \includegraphics[width=0.24\textwidth, trim=4cm 1cm 4cm 0, clip]{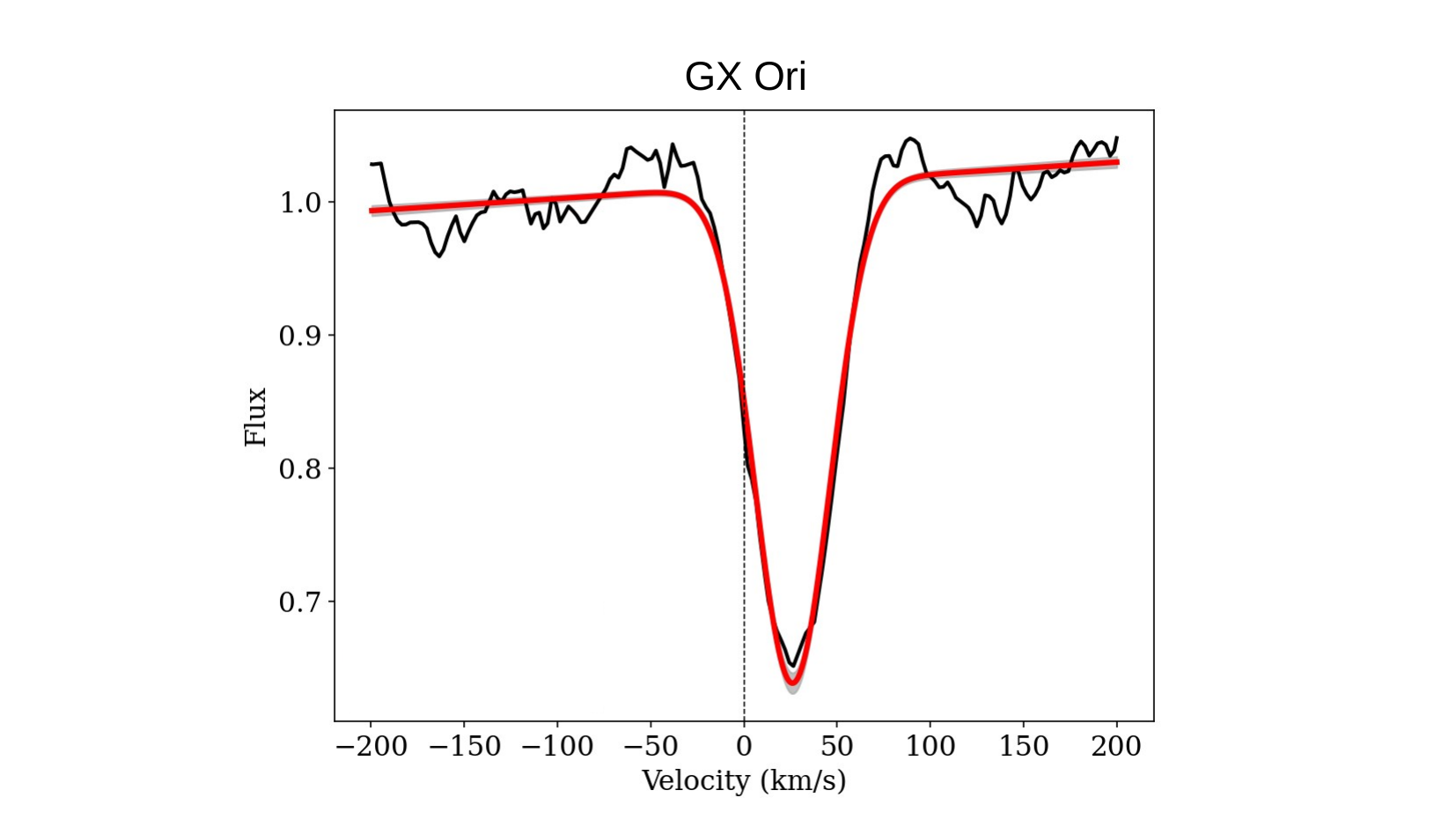}
    \includegraphics[width=0.24\textwidth, trim=4cm 1cm 4cm 0, clip]{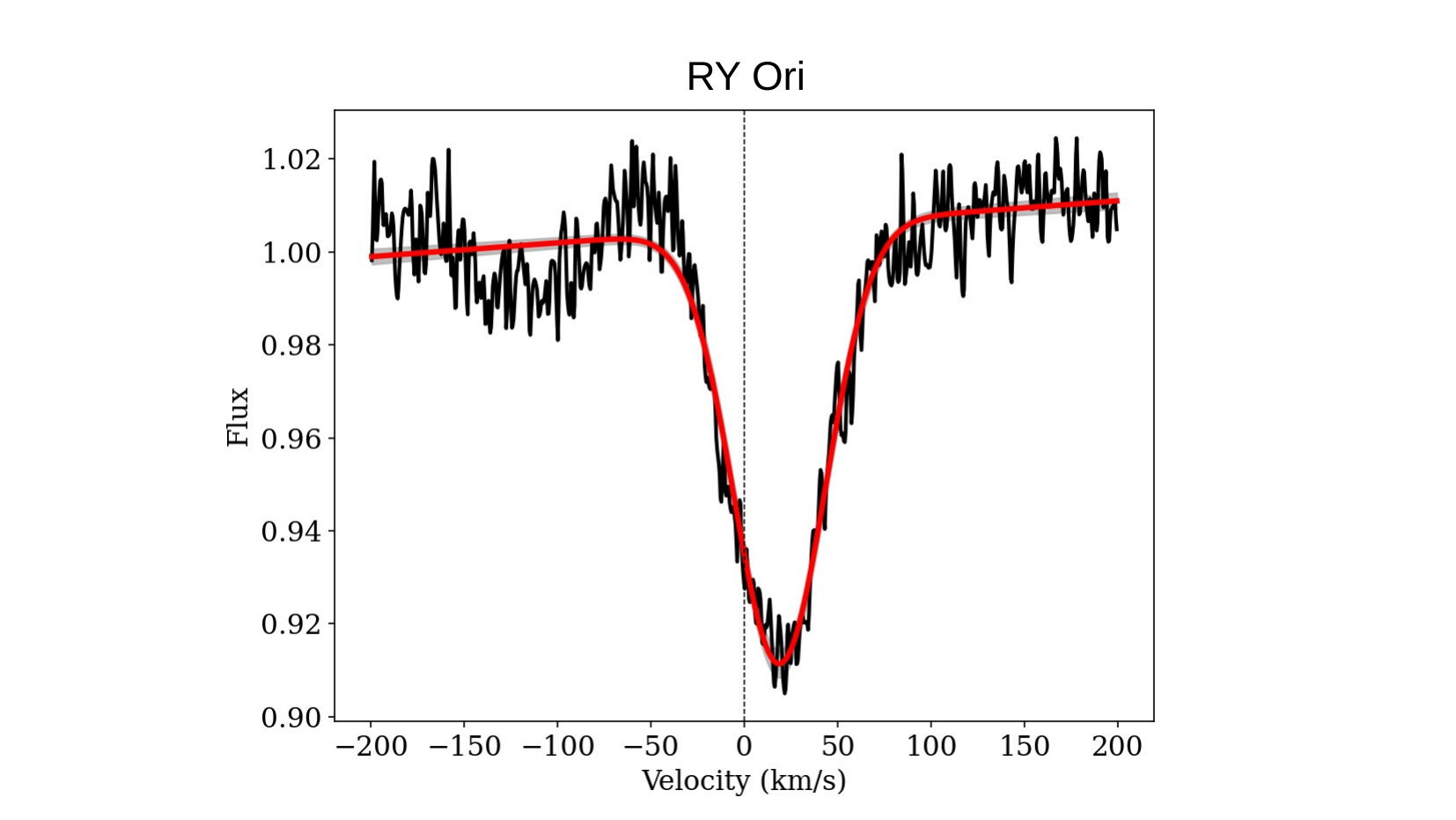}
    \includegraphics[width=0.24\textwidth, trim=4cm 1cm 4cm 0, clip]{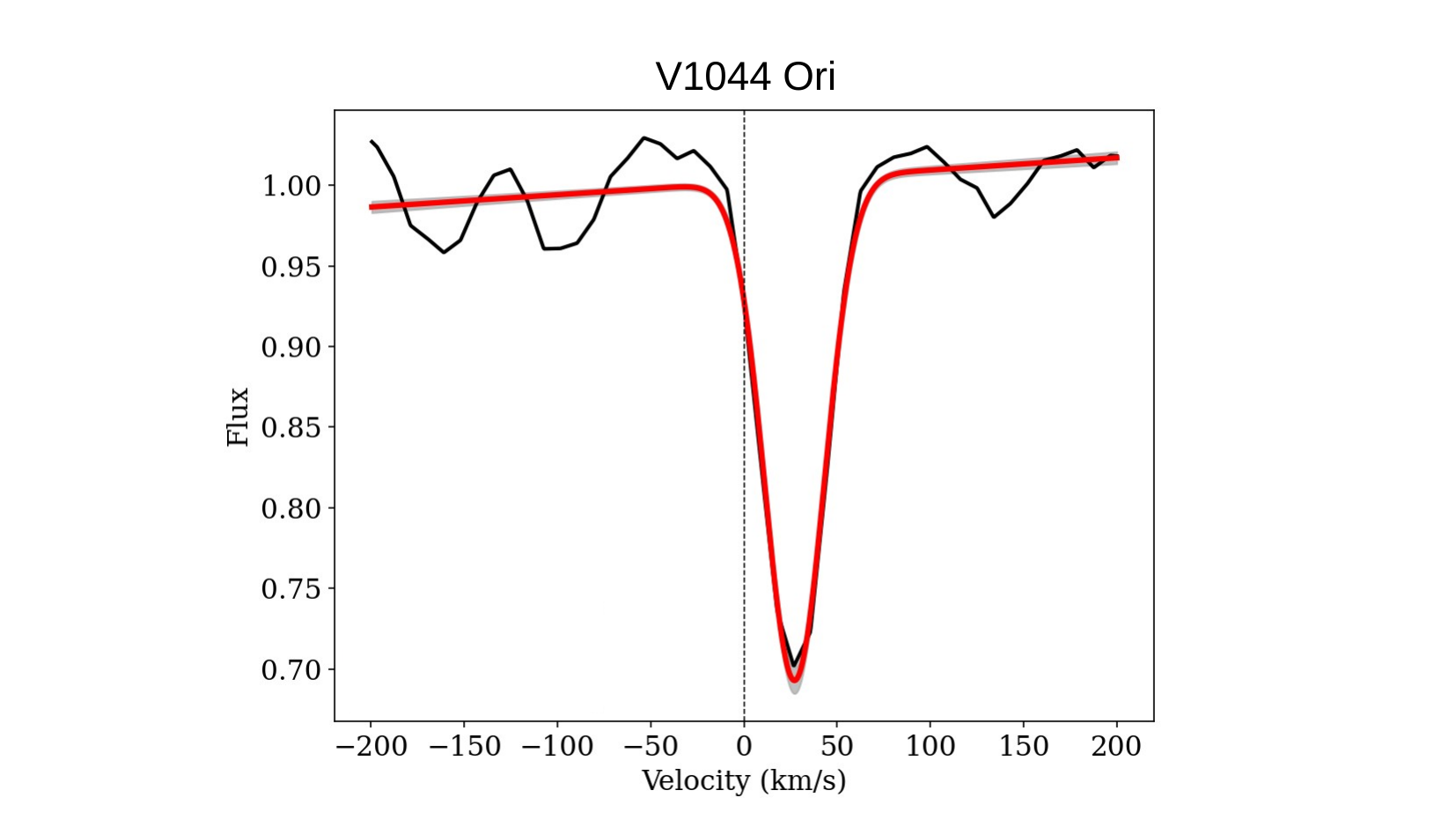} 
    \includegraphics[width=0.24\textwidth, trim=4cm 1cm 4cm 0, clip]{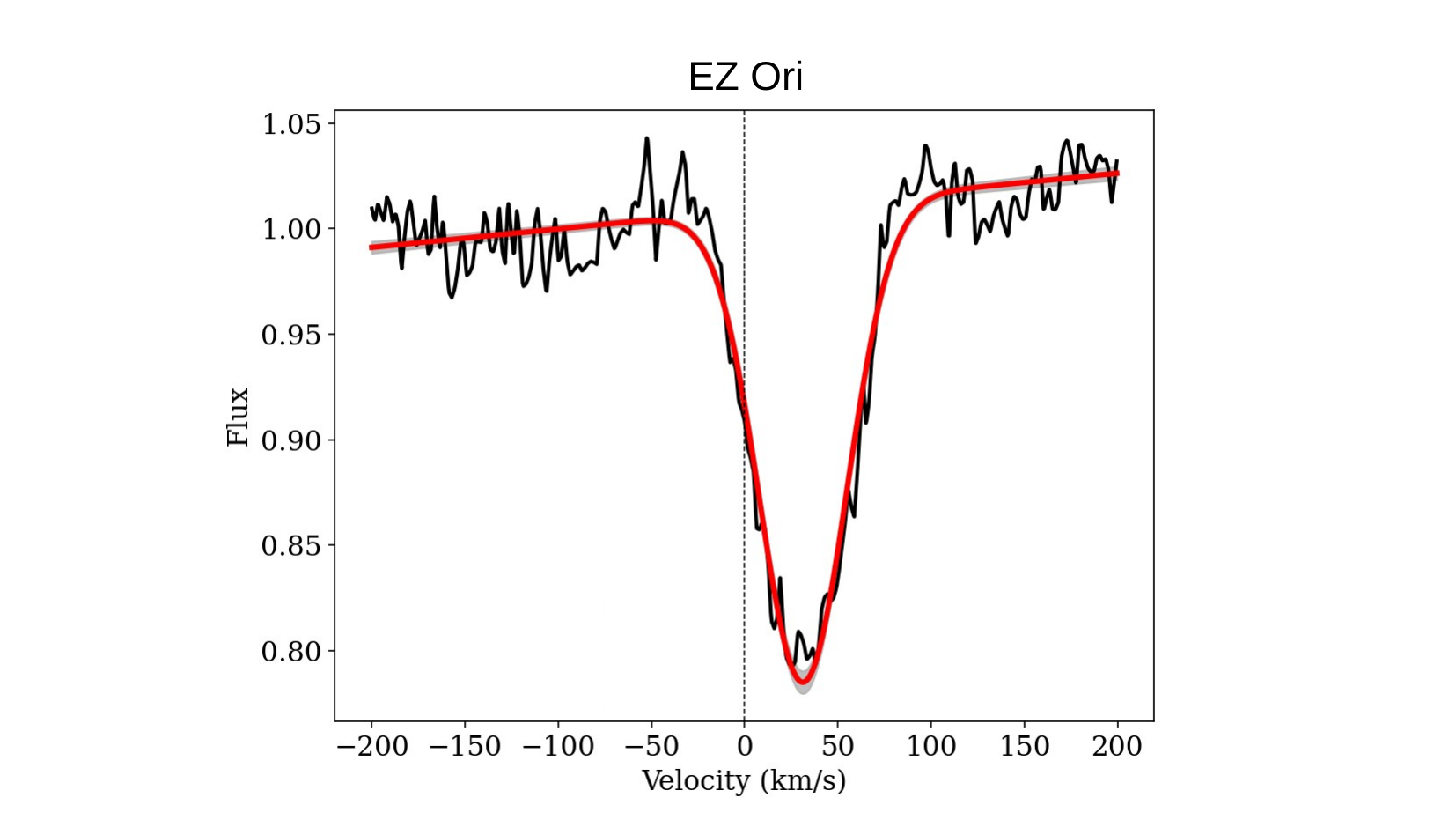} 
\end{figure}
\begin{figure} [h]
 \centering
    \includegraphics[width=0.24\textwidth, trim=4cm 1cm 4cm 0, clip]{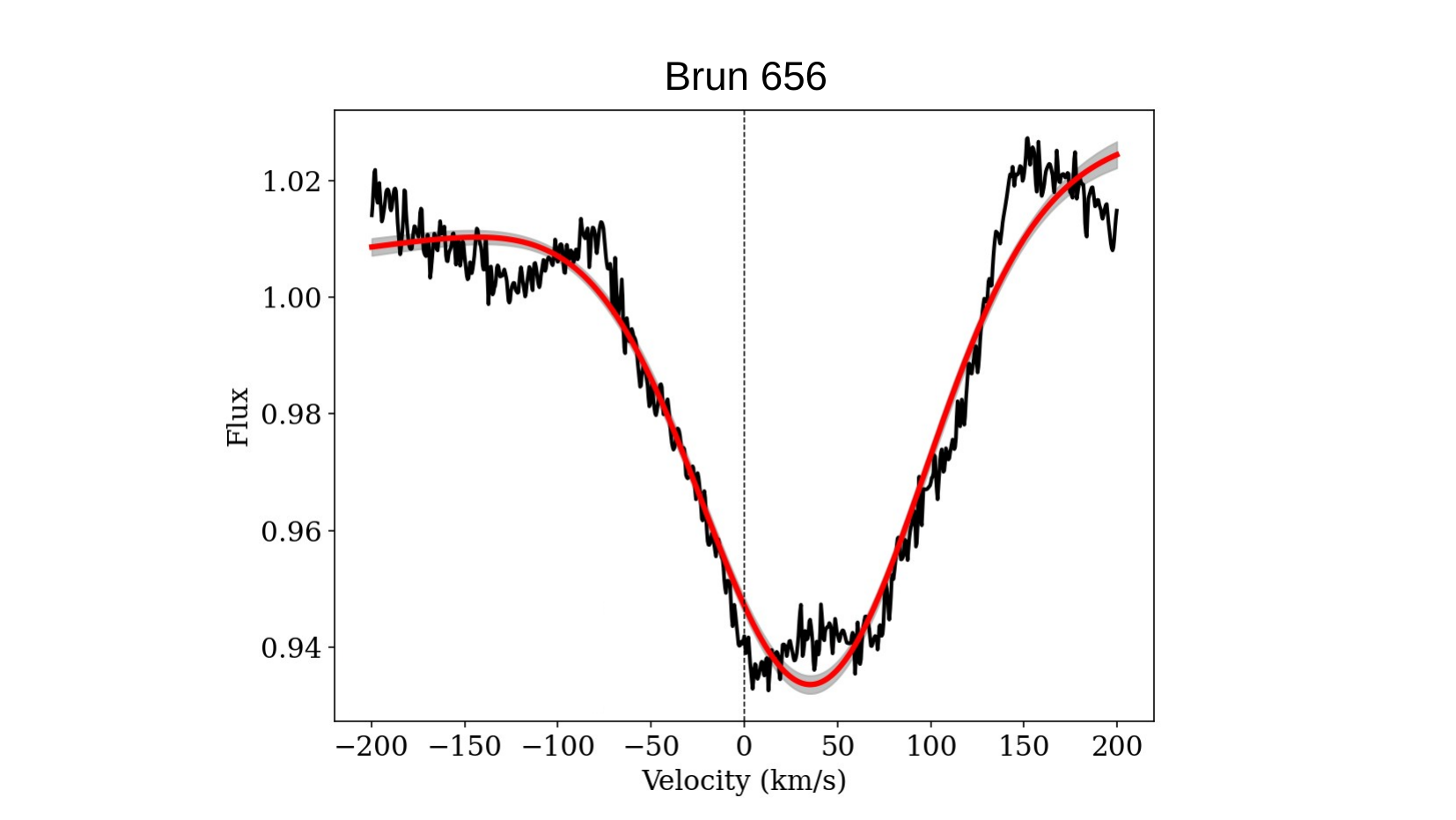}
    \includegraphics[width=0.24\textwidth, trim=4cm 1cm 4cm 0, clip]{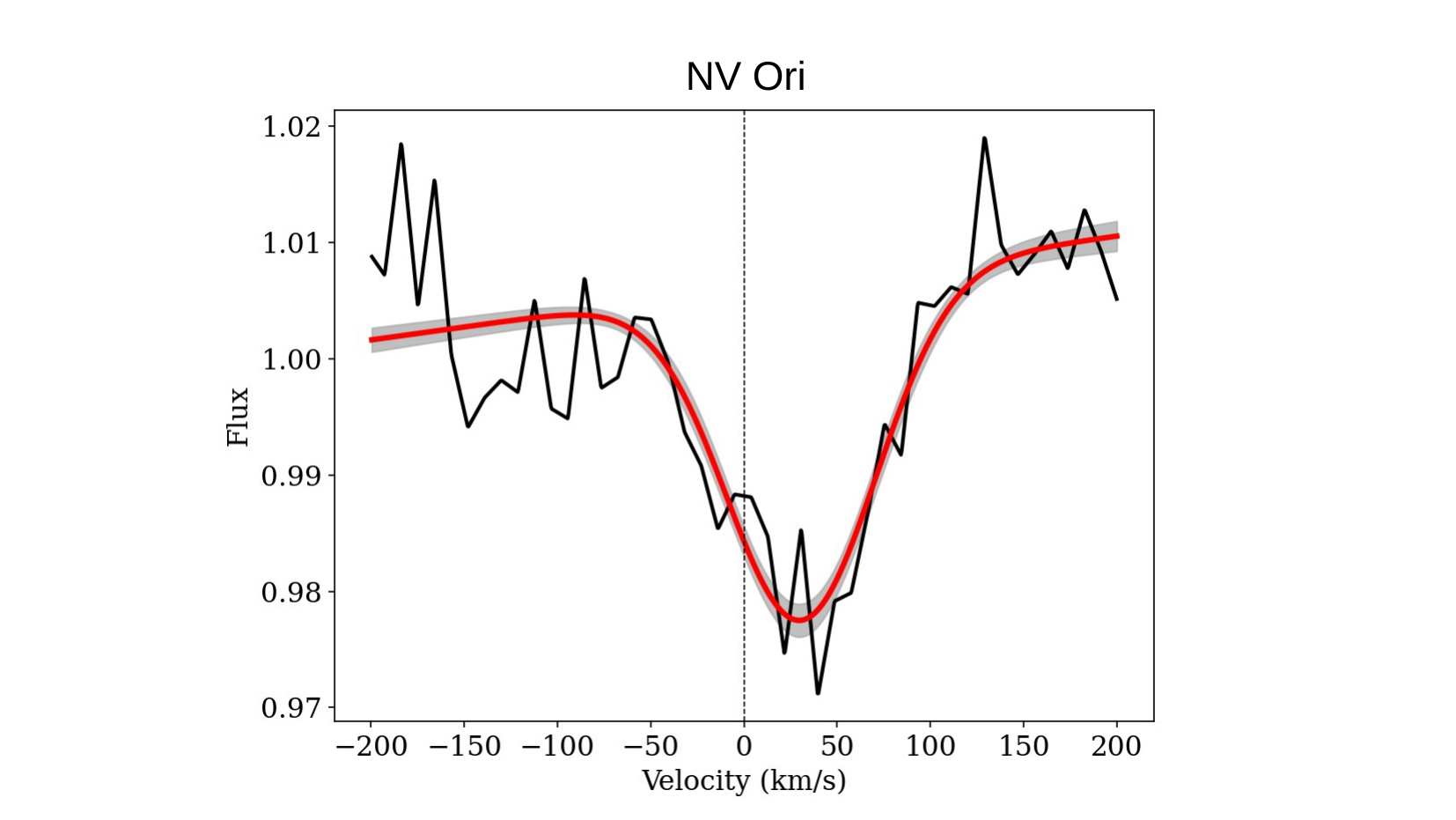}
    \includegraphics[width=0.24\textwidth, trim=4cm 1cm 4cm 0, clip]{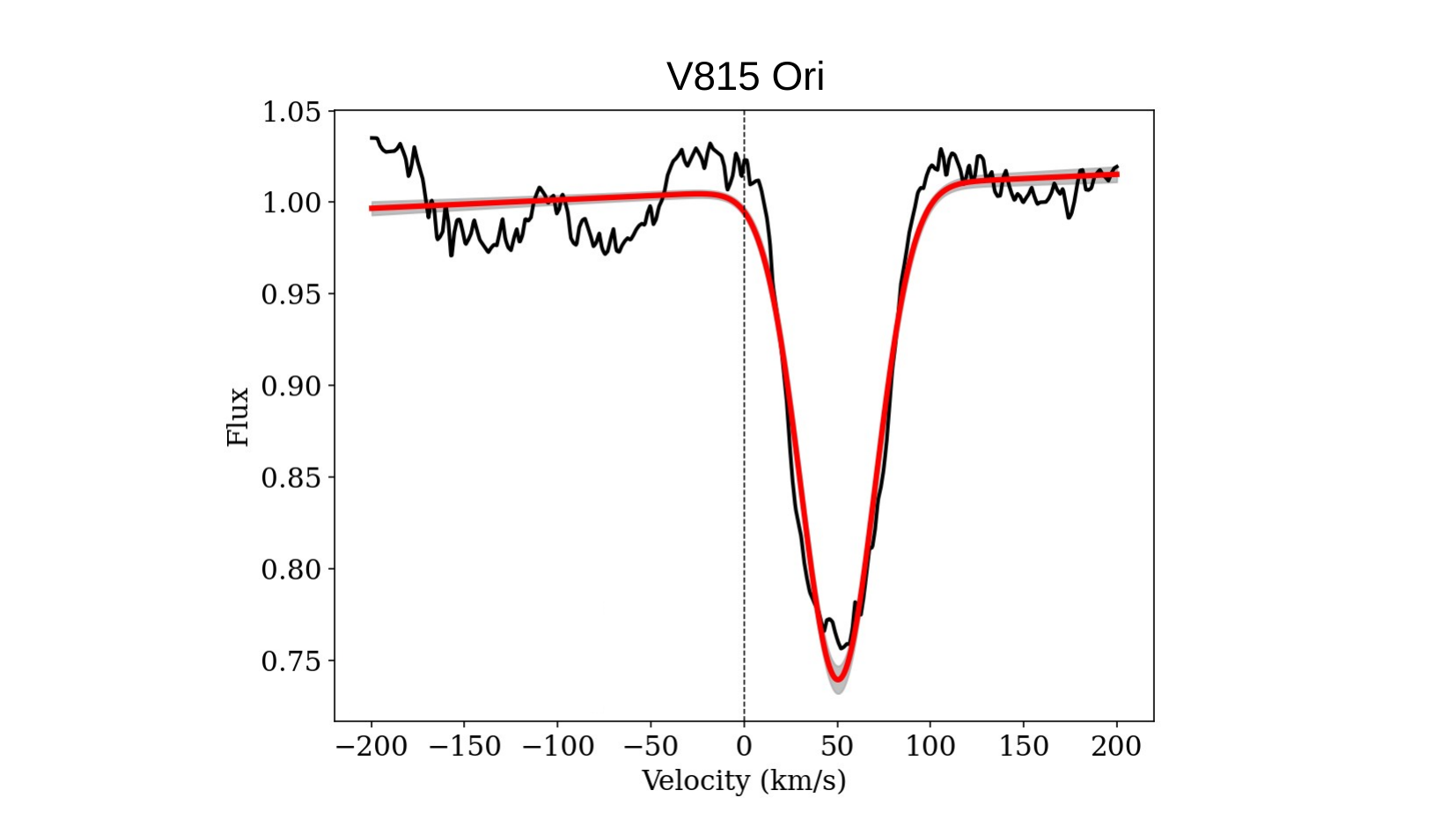} 
    \includegraphics[width=0.24\textwidth, trim=4cm 1cm 4cm 0, clip]{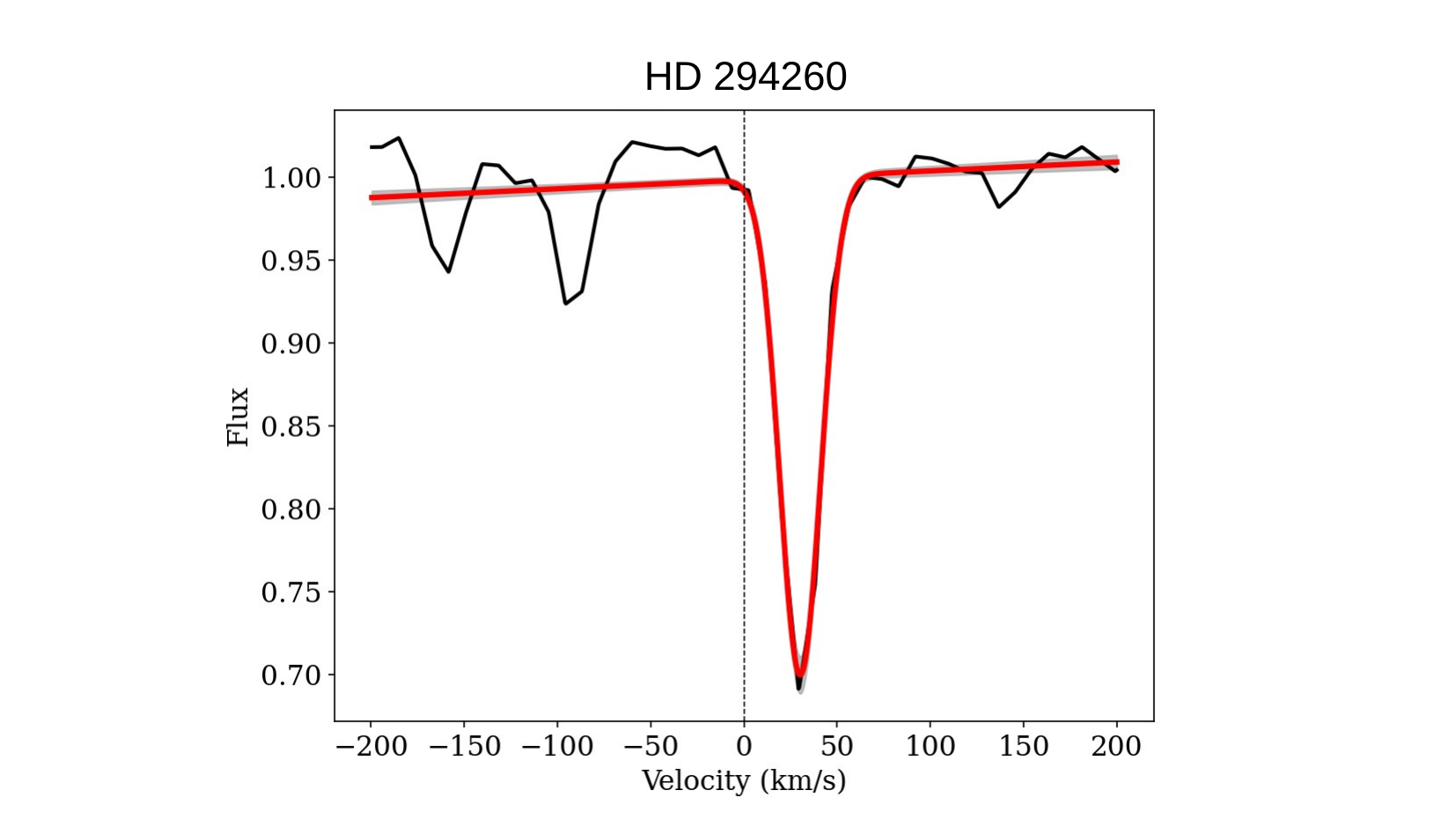}
\end{figure}
\begin{figure} [h]
 \centering
    \includegraphics[width=0.24\textwidth, trim=4cm 1cm 4cm 0, clip]{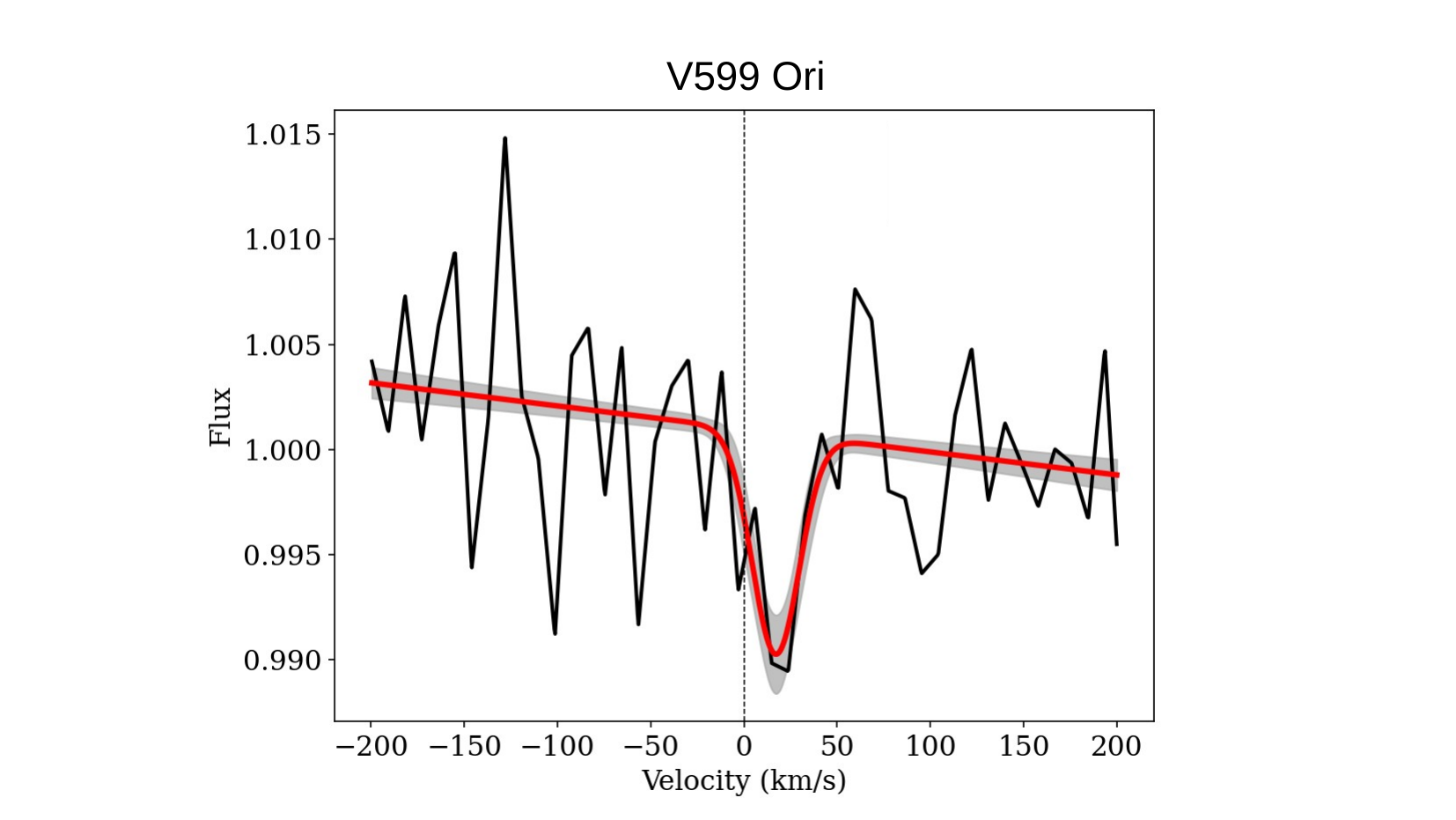} 
    \includegraphics[width=0.24\textwidth, trim=4cm 1cm 4cm 0, clip]{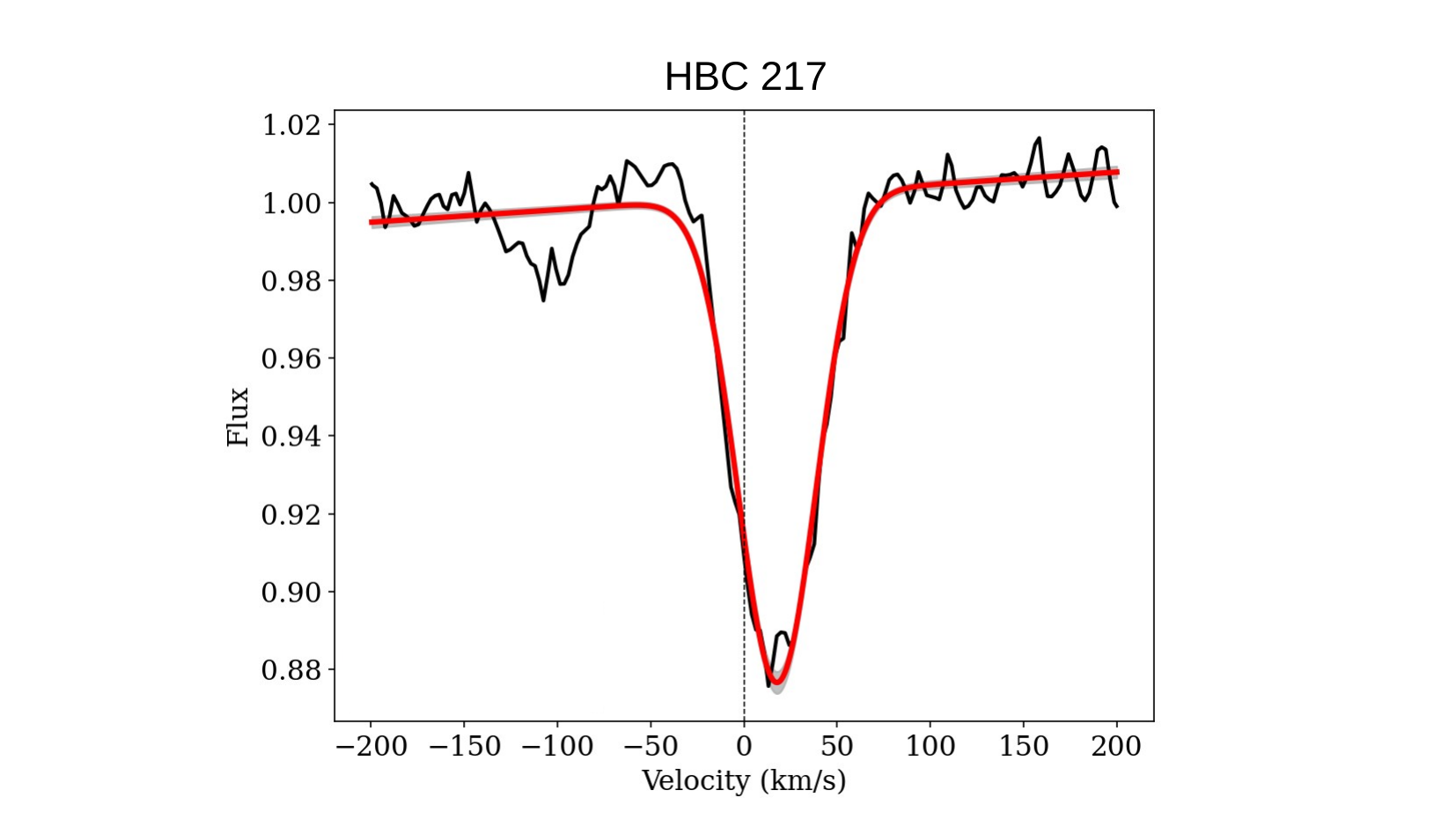}  
    \includegraphics[width=0.24\textwidth, trim=4cm 1cm 4cm 0, clip]{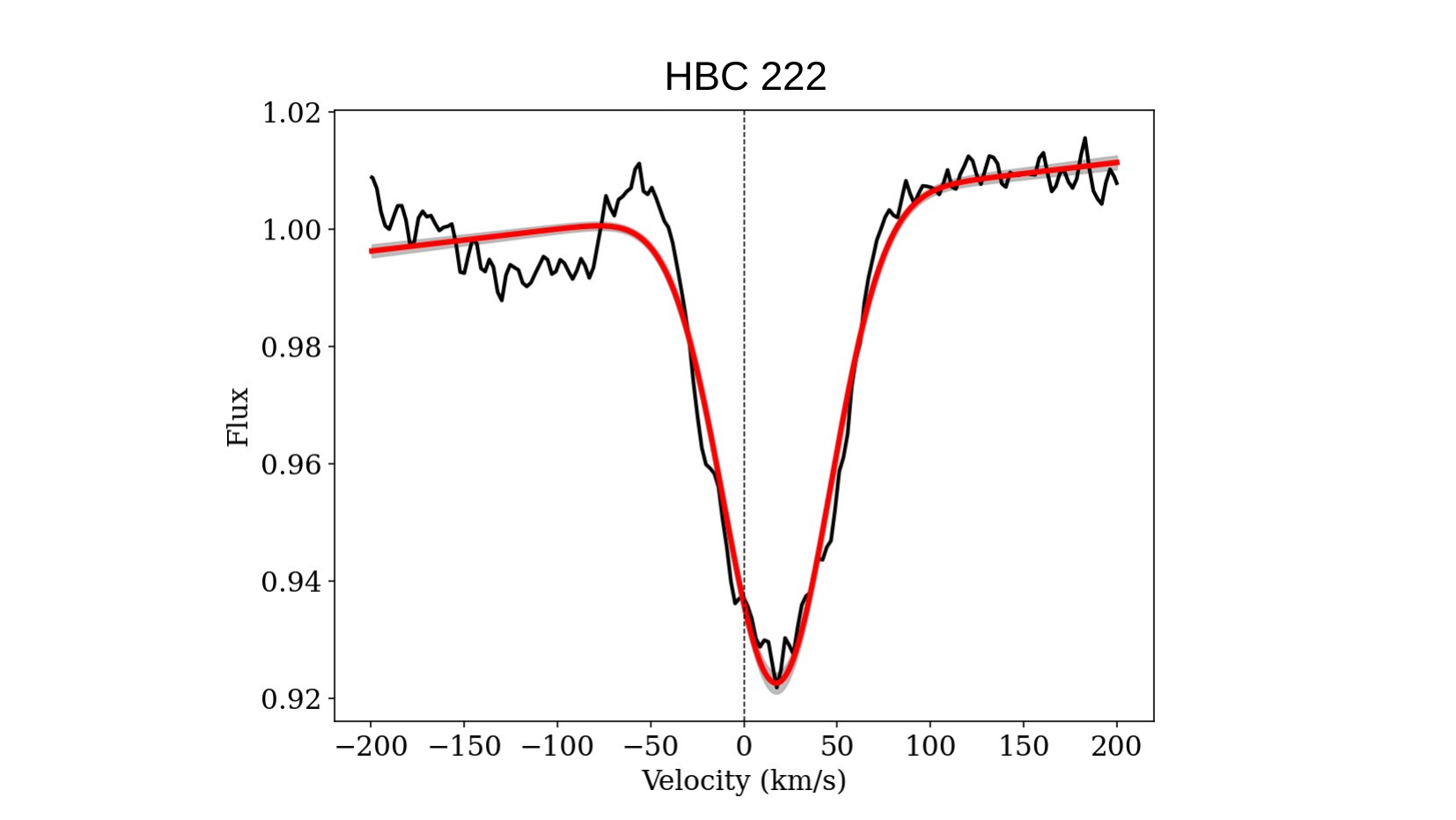}
    \includegraphics[width=0.24\textwidth, trim=4cm 1cm 4cm 0, clip]{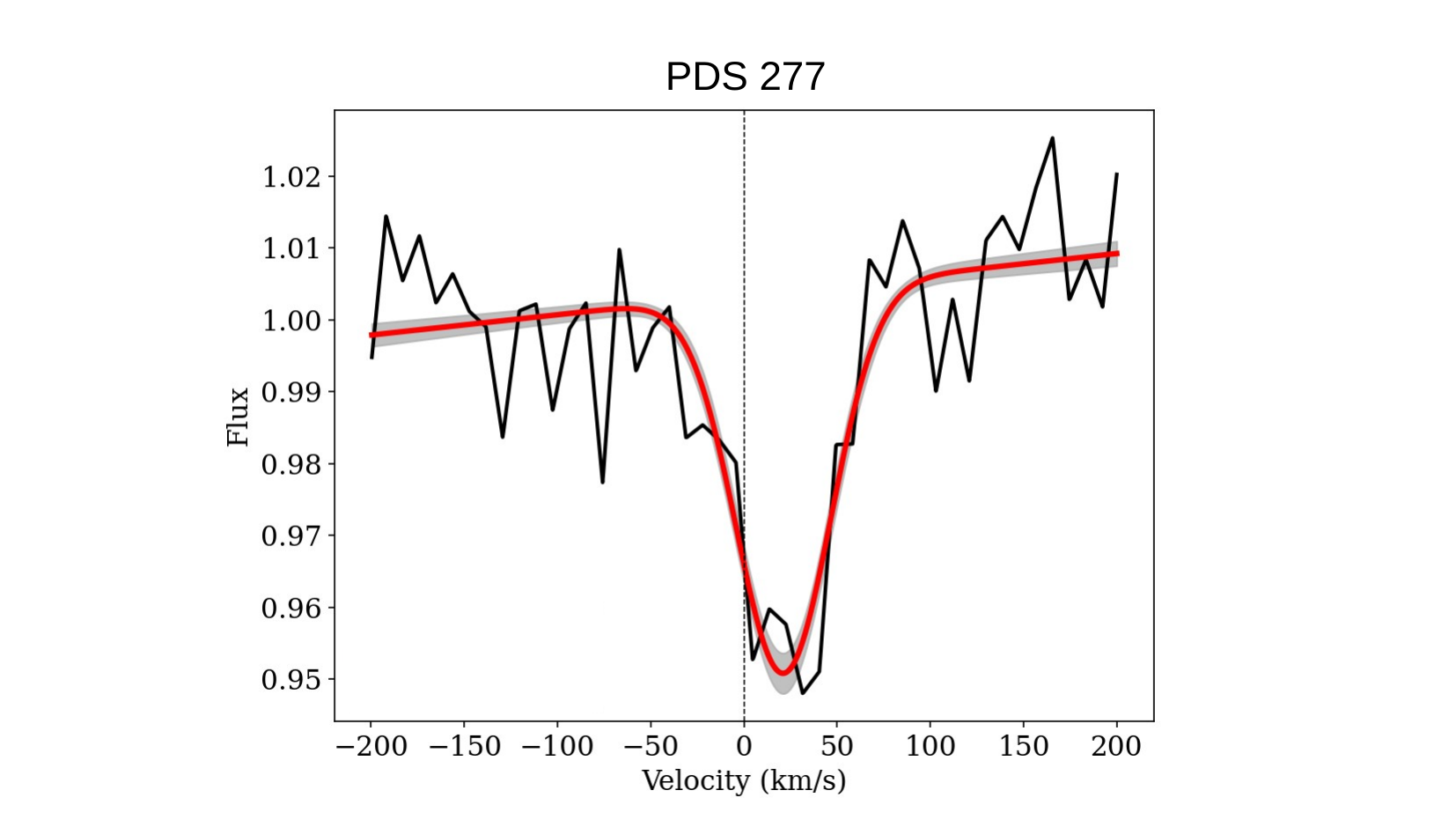} 
\caption{Observed \ion{Li}{i} spectra centered at 6707.856 {\AA} (vertical dashed line) are indicated in black, the fits necessary to derive EWs in red, and the 3$\sigma$ uncertainties in gray. See \citet{King93} for the corresponding sources in Table \ref{table:sample}.} 
\label{Fig:lithium} 
\end{figure}
\begin{figure} [h]
 \centering
 \caption*{Fig.~\ref{Fig:lithium}. continued.}
    \includegraphics[width=0.24\textwidth, trim=4cm 1cm 4cm 0, clip]{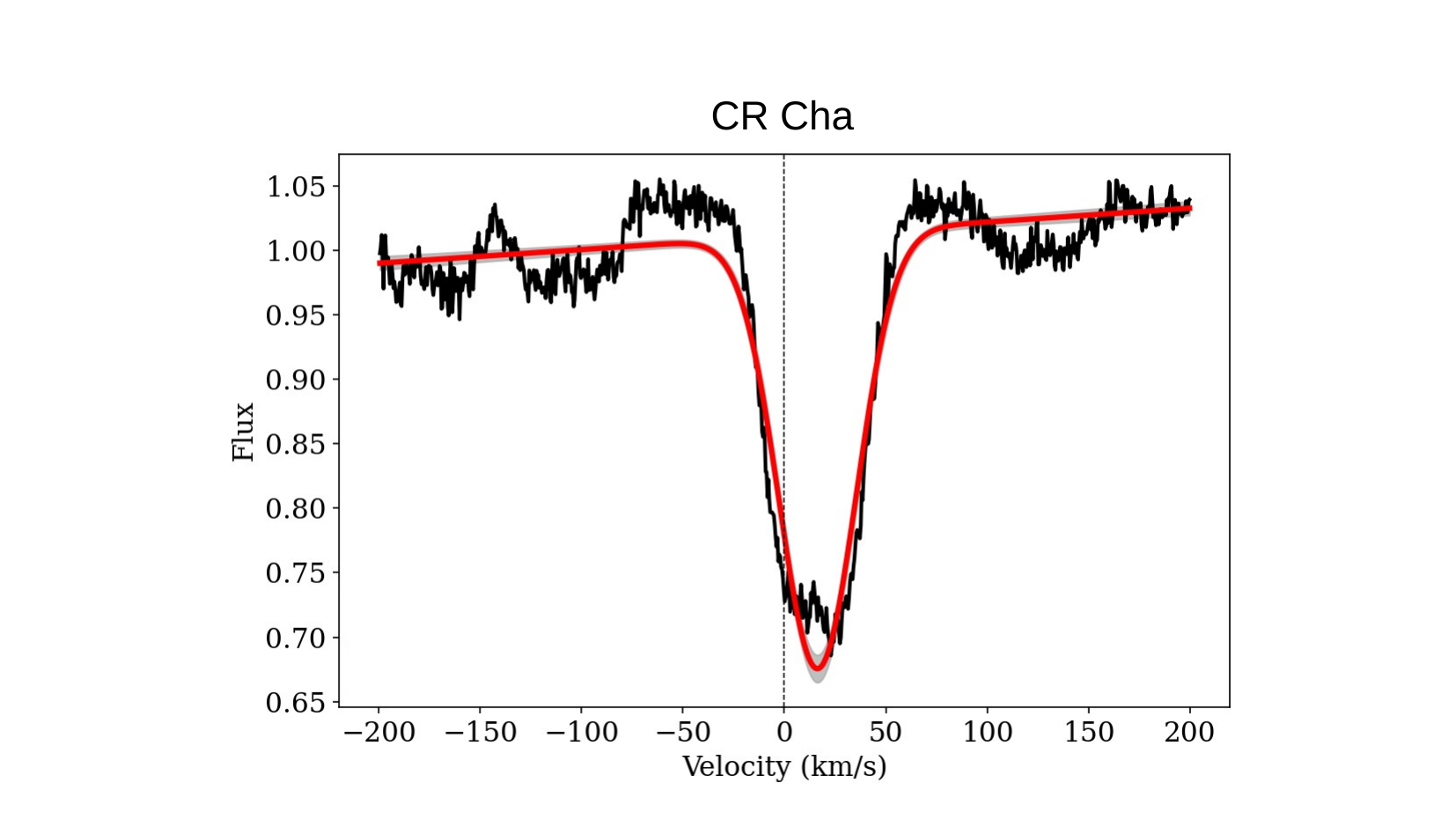}   
    \includegraphics[width=0.24\textwidth, trim=4cm 1cm 4cm 0, clip]{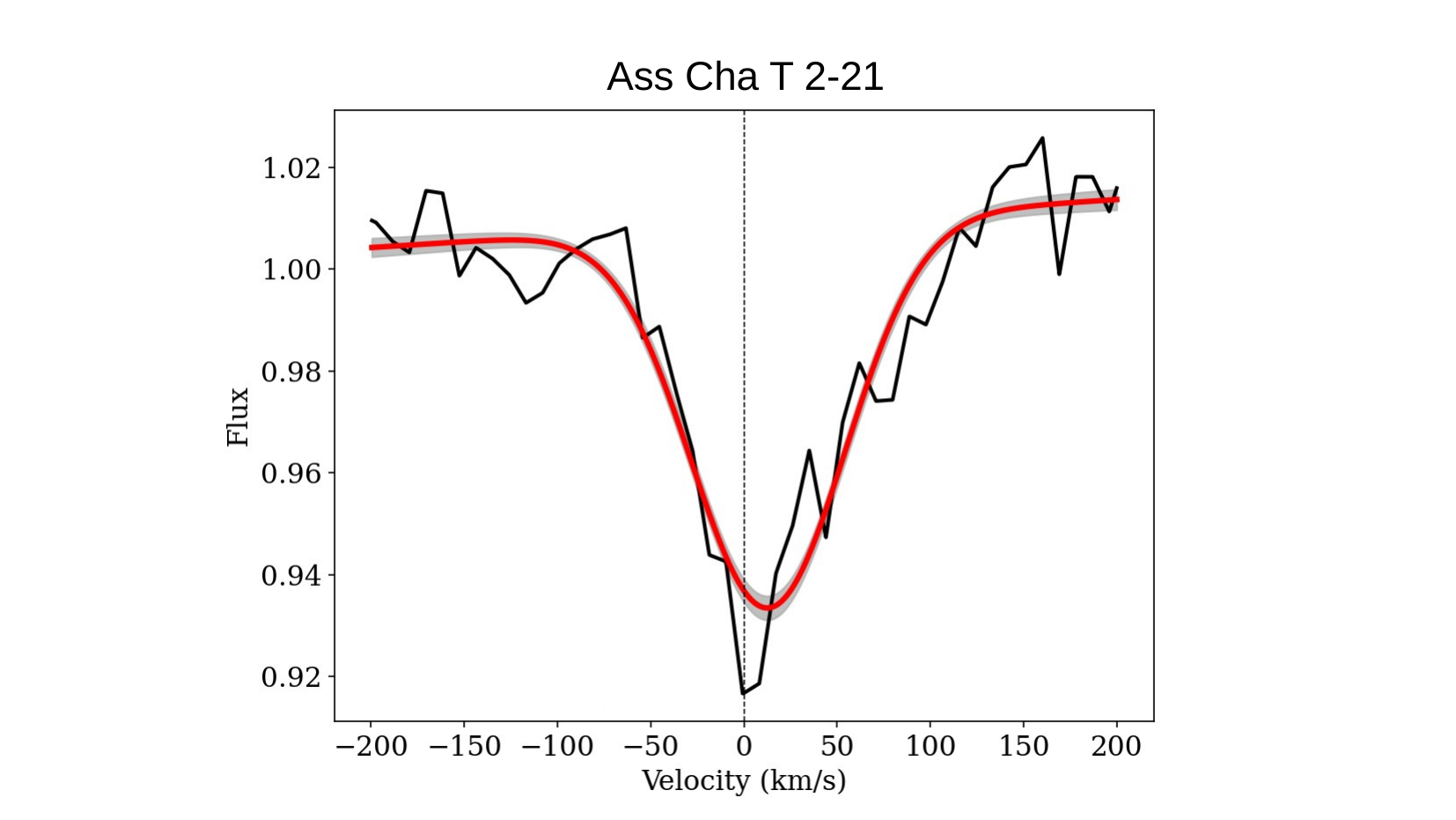}
    \includegraphics[width=0.24\textwidth, trim=4cm 1cm 4cm 0, clip]{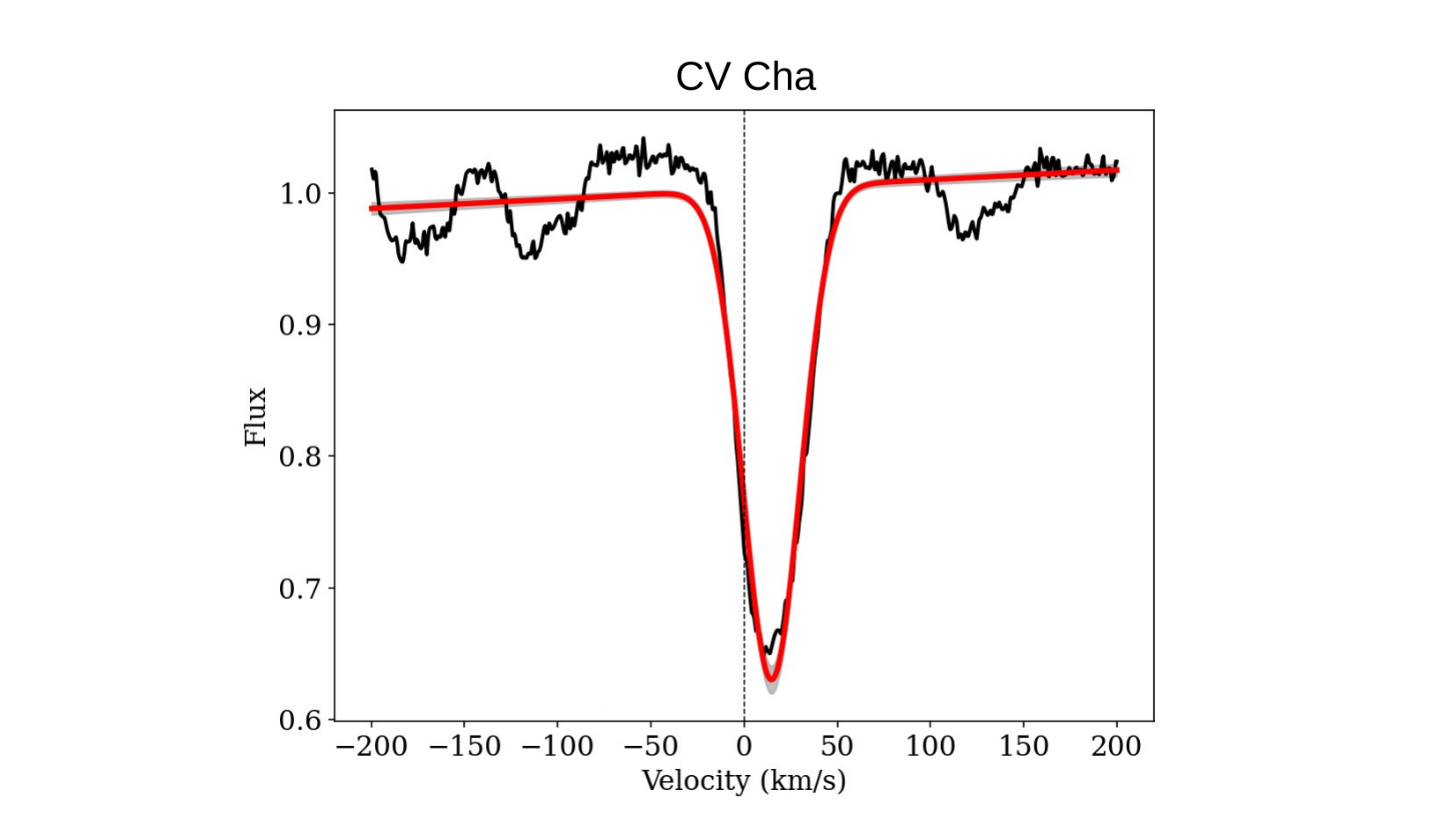} 
    \includegraphics[width=0.24\textwidth, trim=4cm 1cm 4cm 0, clip]{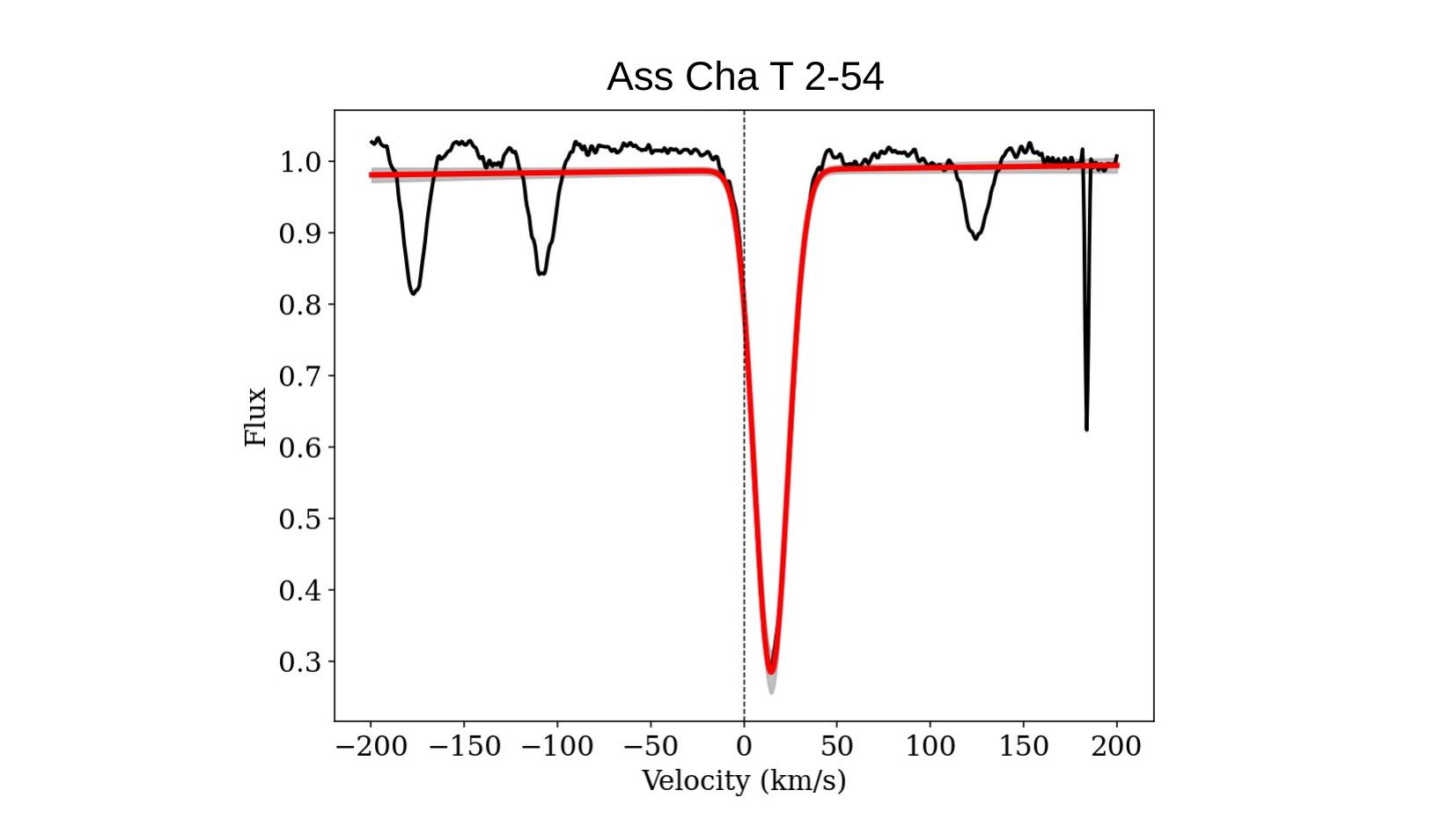}   
\end{figure}
\begin{figure} [h]
 \centering
    \includegraphics[width=0.24\textwidth, trim=4cm 1cm 4cm 0, clip]{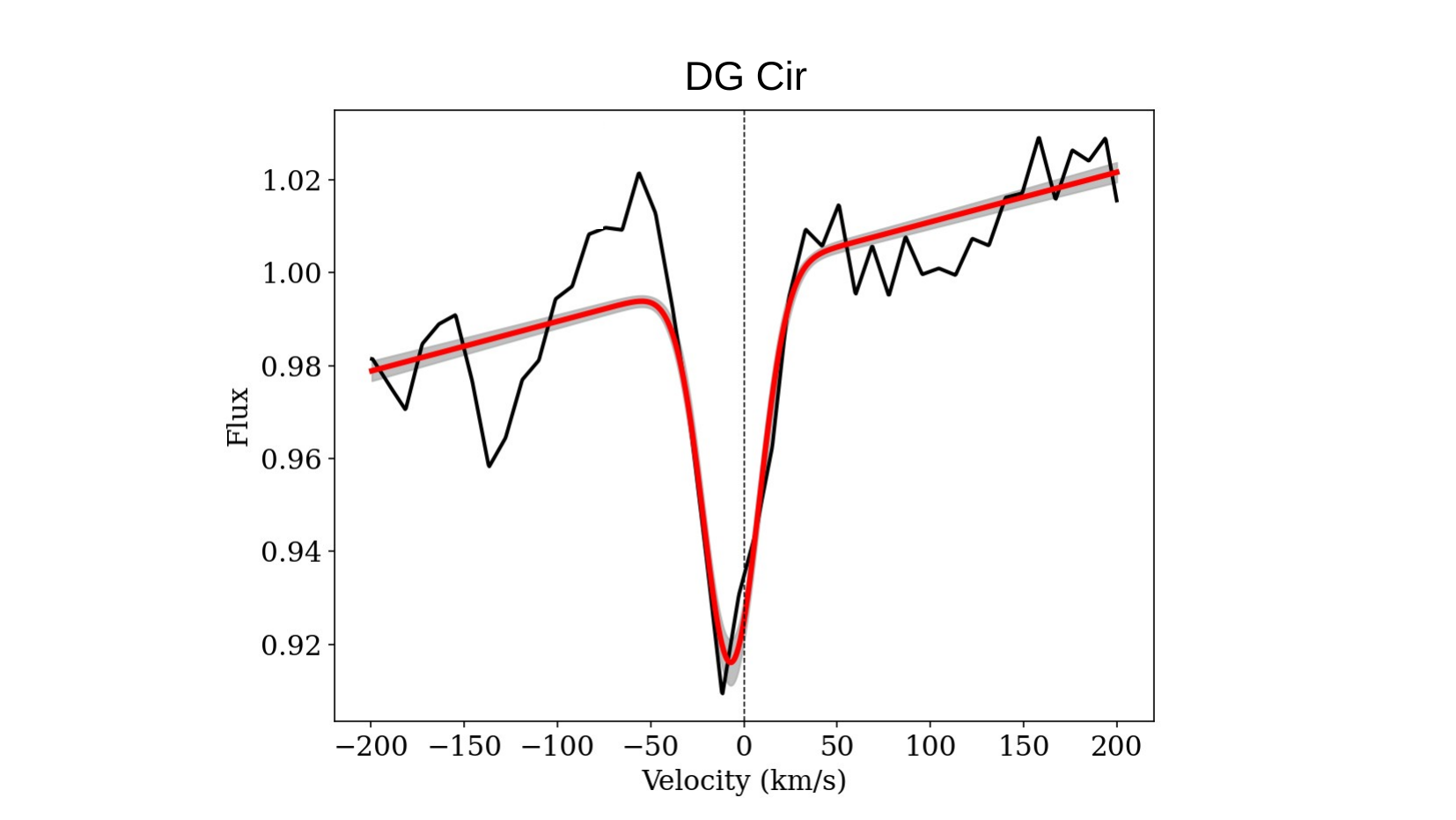}
   \includegraphics[width=0.24\textwidth, trim=4cm 1cm 4cm 0, clip]{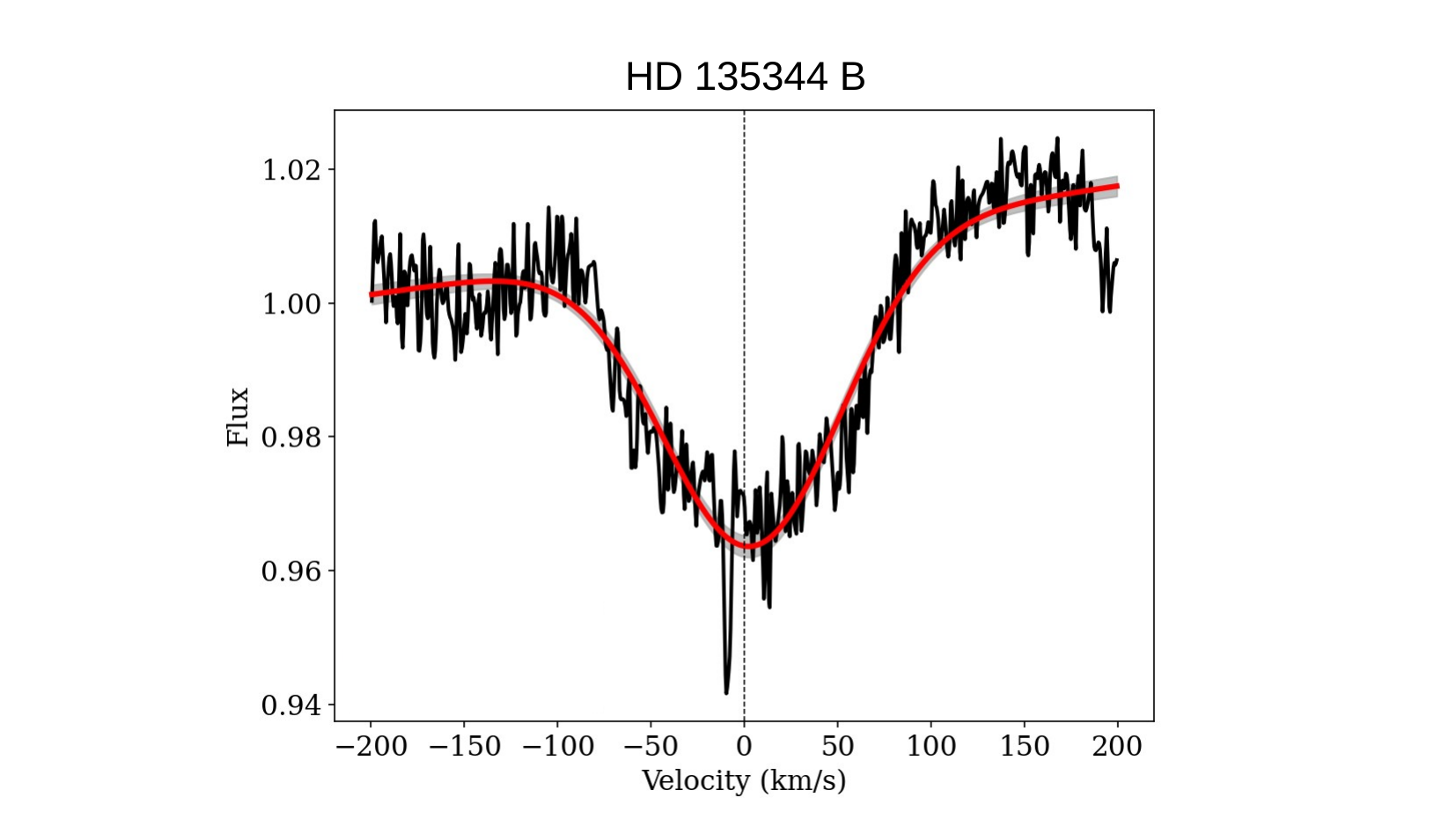} 
   \includegraphics[width=0.24\textwidth, trim=4cm 1cm 4cm 0, clip]{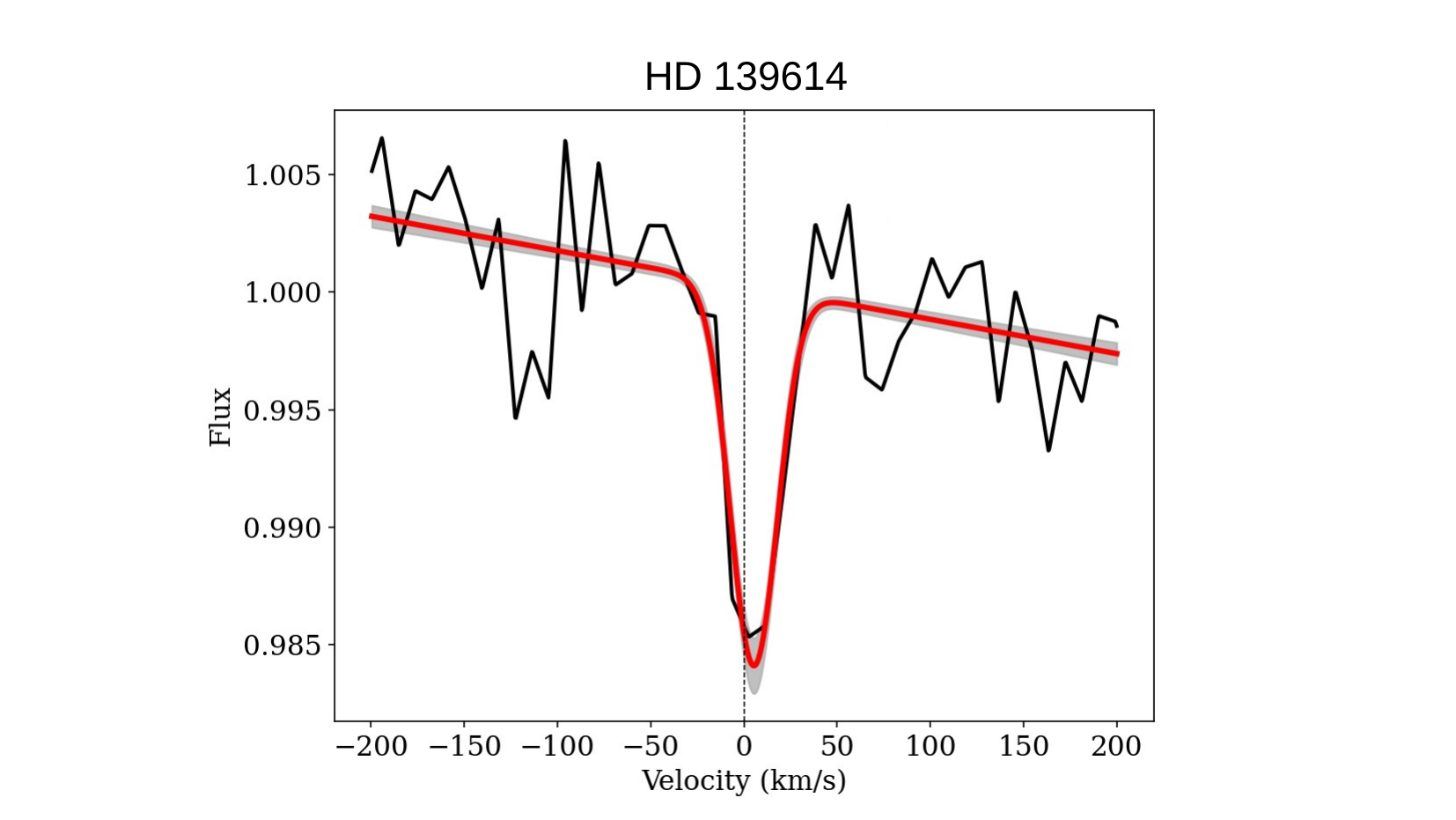} 
   \includegraphics[width=0.24\textwidth, trim=4cm 1cm 4cm 0, clip]{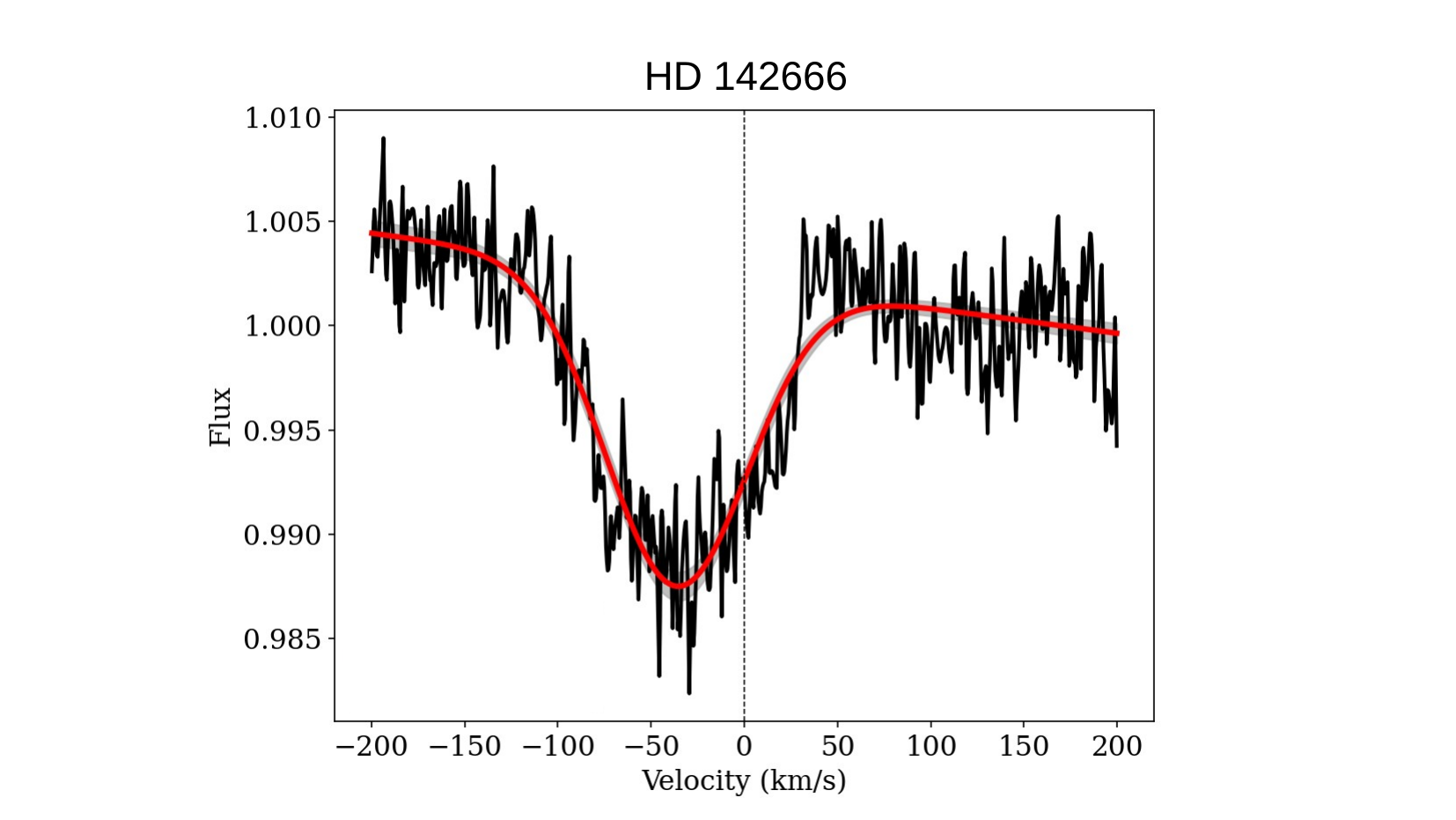} 
\end{figure}
\begin{figure} [h]
 \centering
   \includegraphics[width=0.24\textwidth, trim=4cm 1cm 4cm 0, clip]{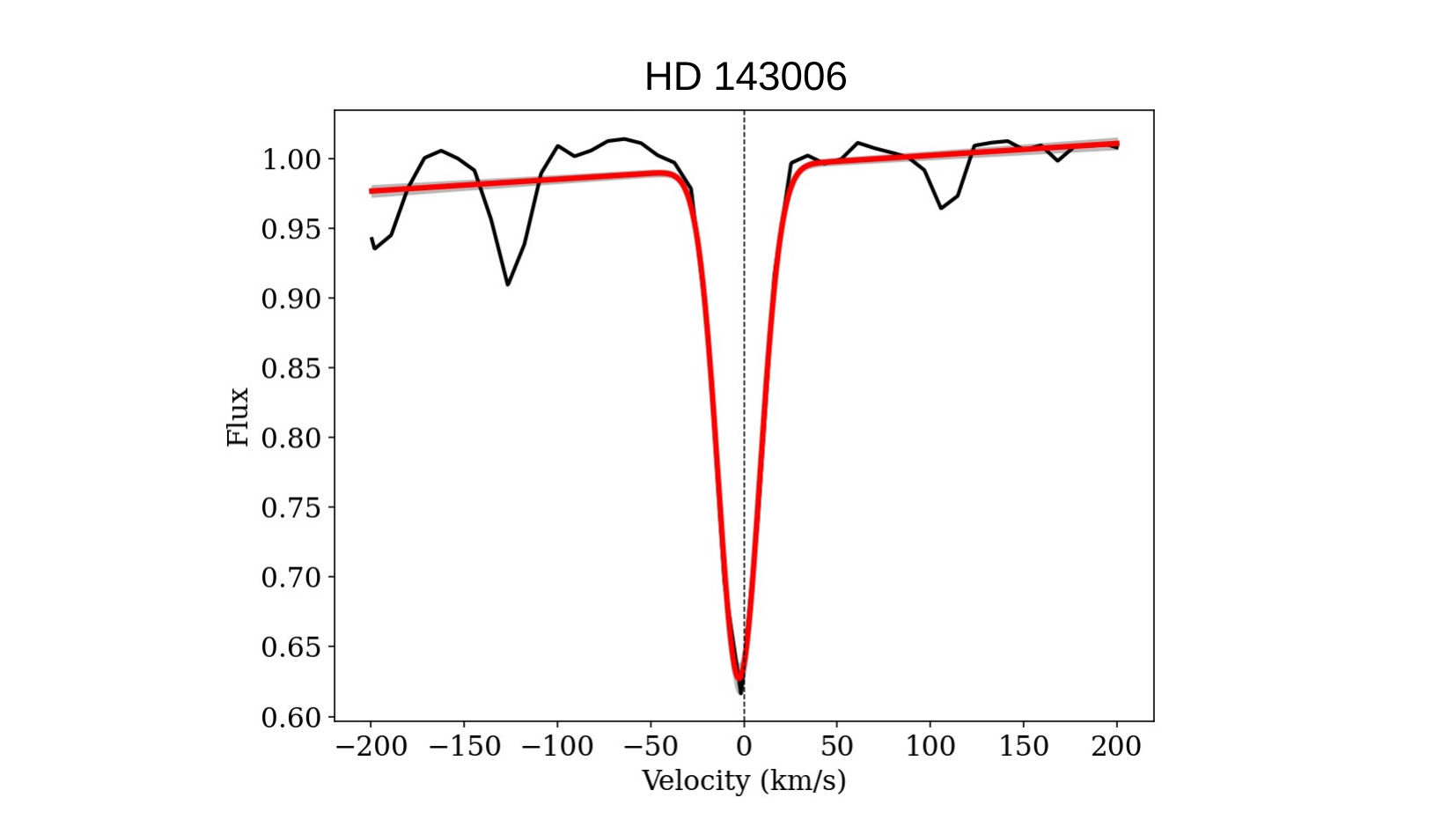} 
   \includegraphics[width=0.24\textwidth, trim=4cm 1cm 4cm 0, clip]{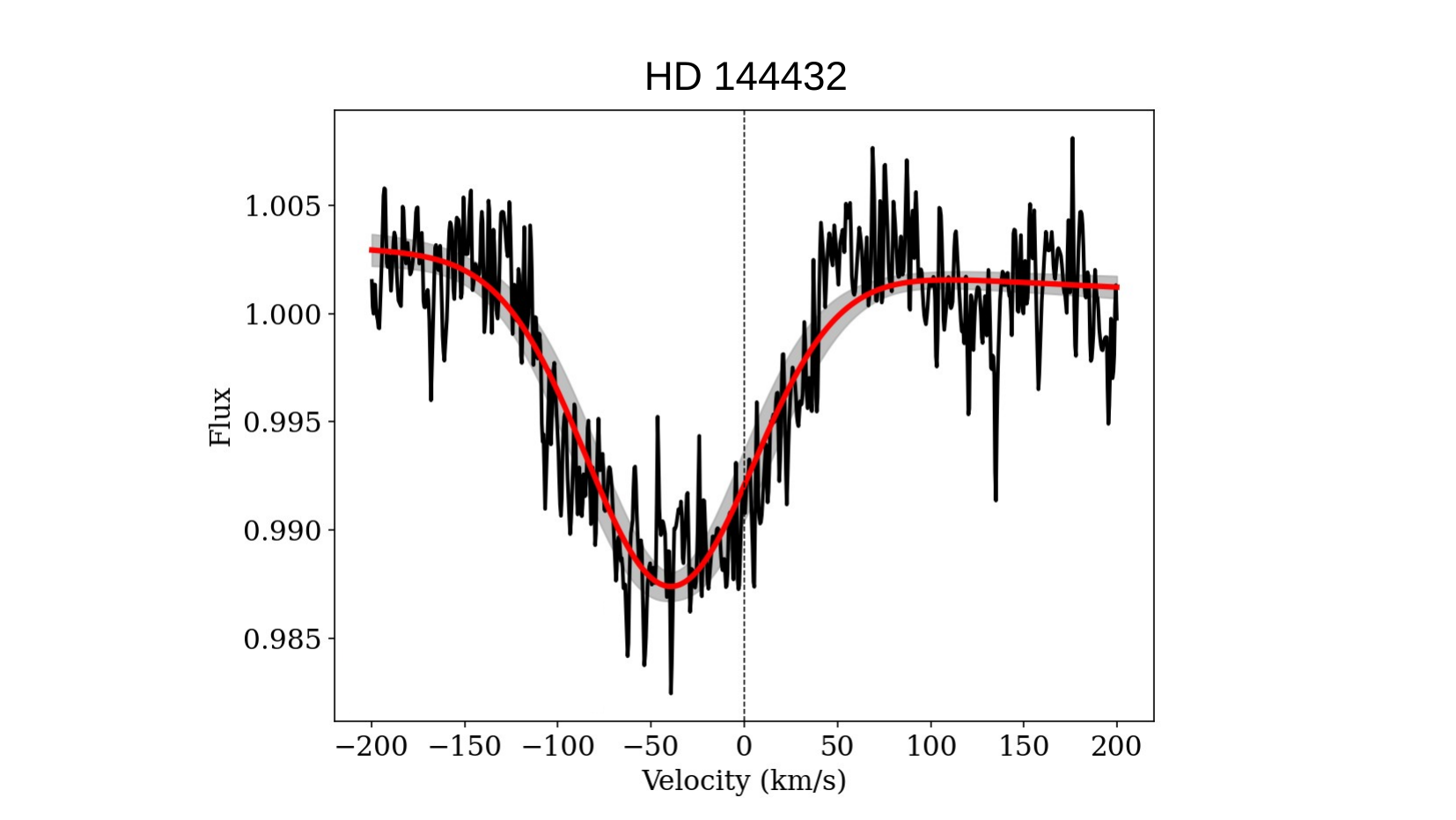}
   \includegraphics[width=0.24\textwidth, trim=4cm 1cm 4cm 0, clip]{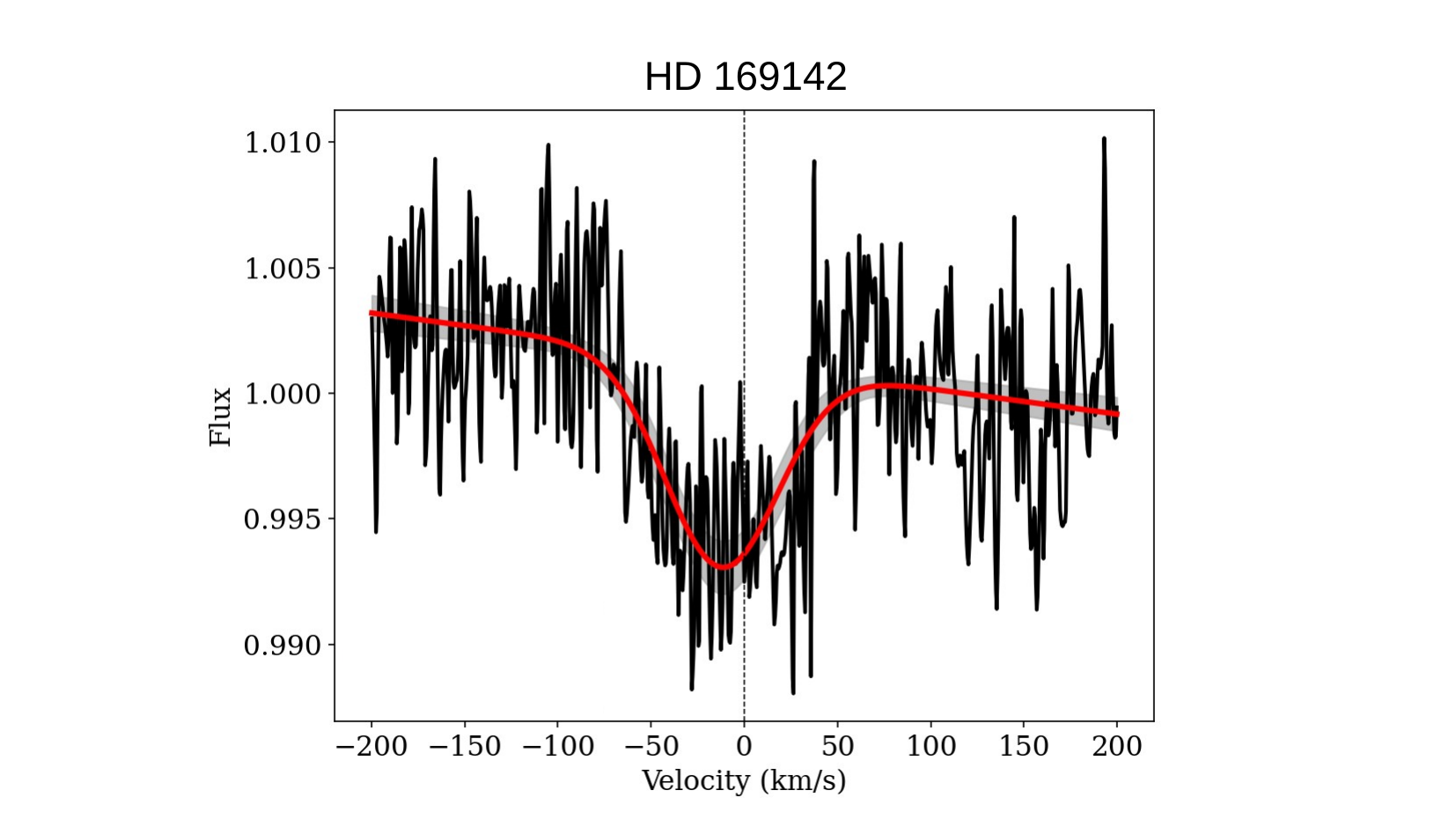}
   \includegraphics[width=0.24\textwidth, trim=4cm 1cm 4cm 0, clip]{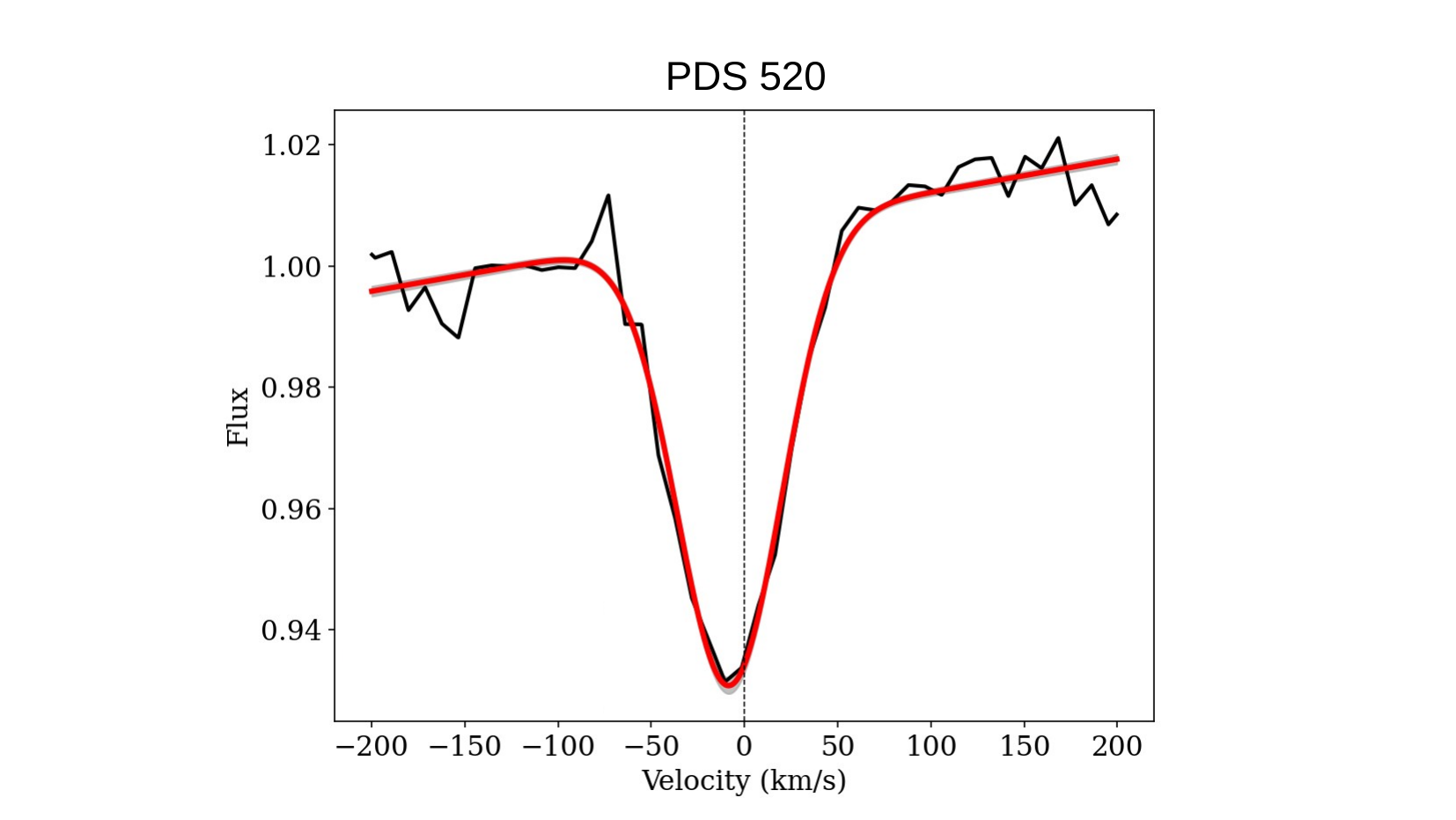}    
\end{figure}
\begin{figure} [h]
 \centering
   \includegraphics[width=0.24\textwidth, trim=4cm 1cm 4cm 0, clip]{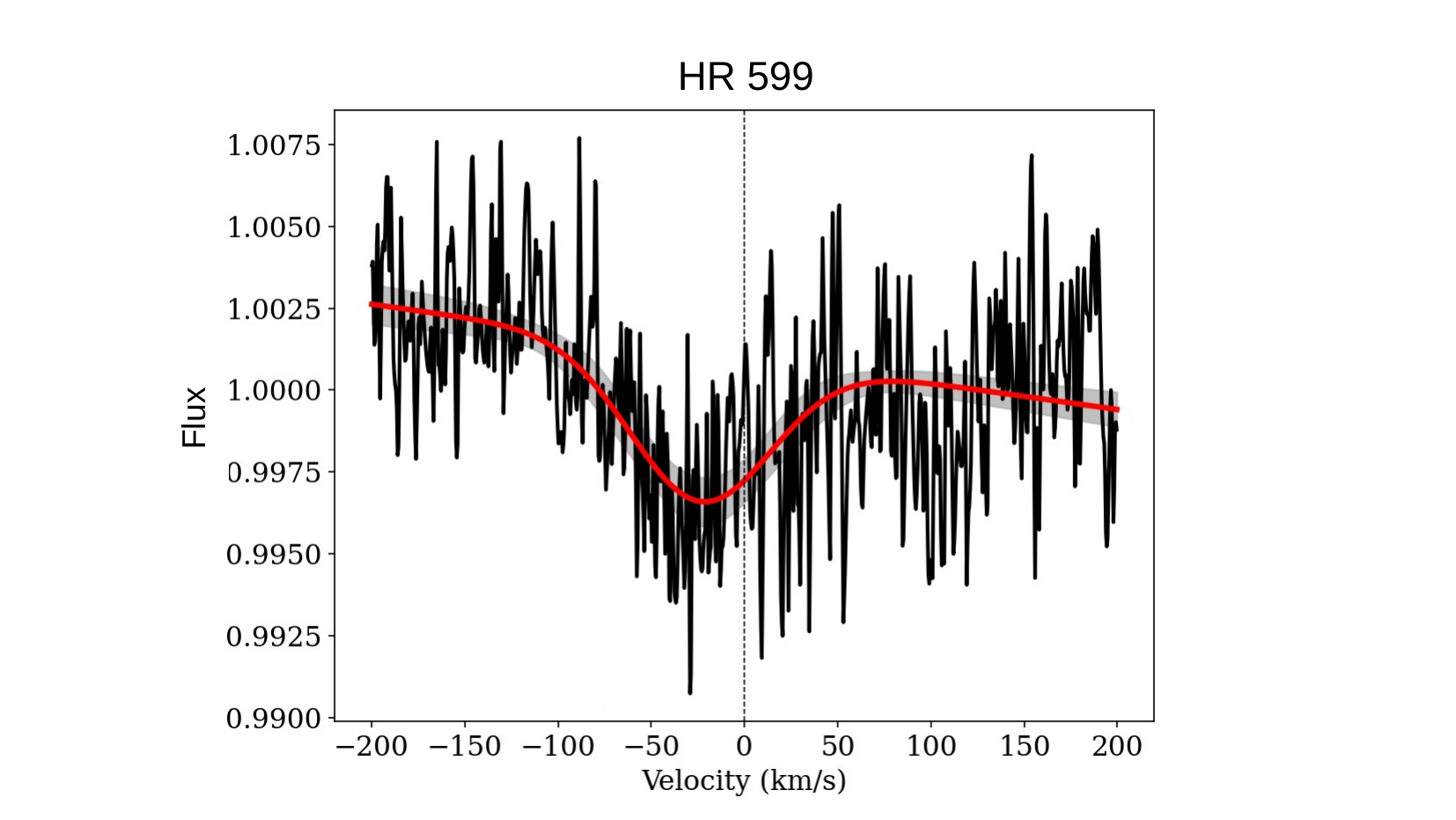} 
   \includegraphics[width=0.24\textwidth, trim=4cm 1cm 4cm 0, clip]{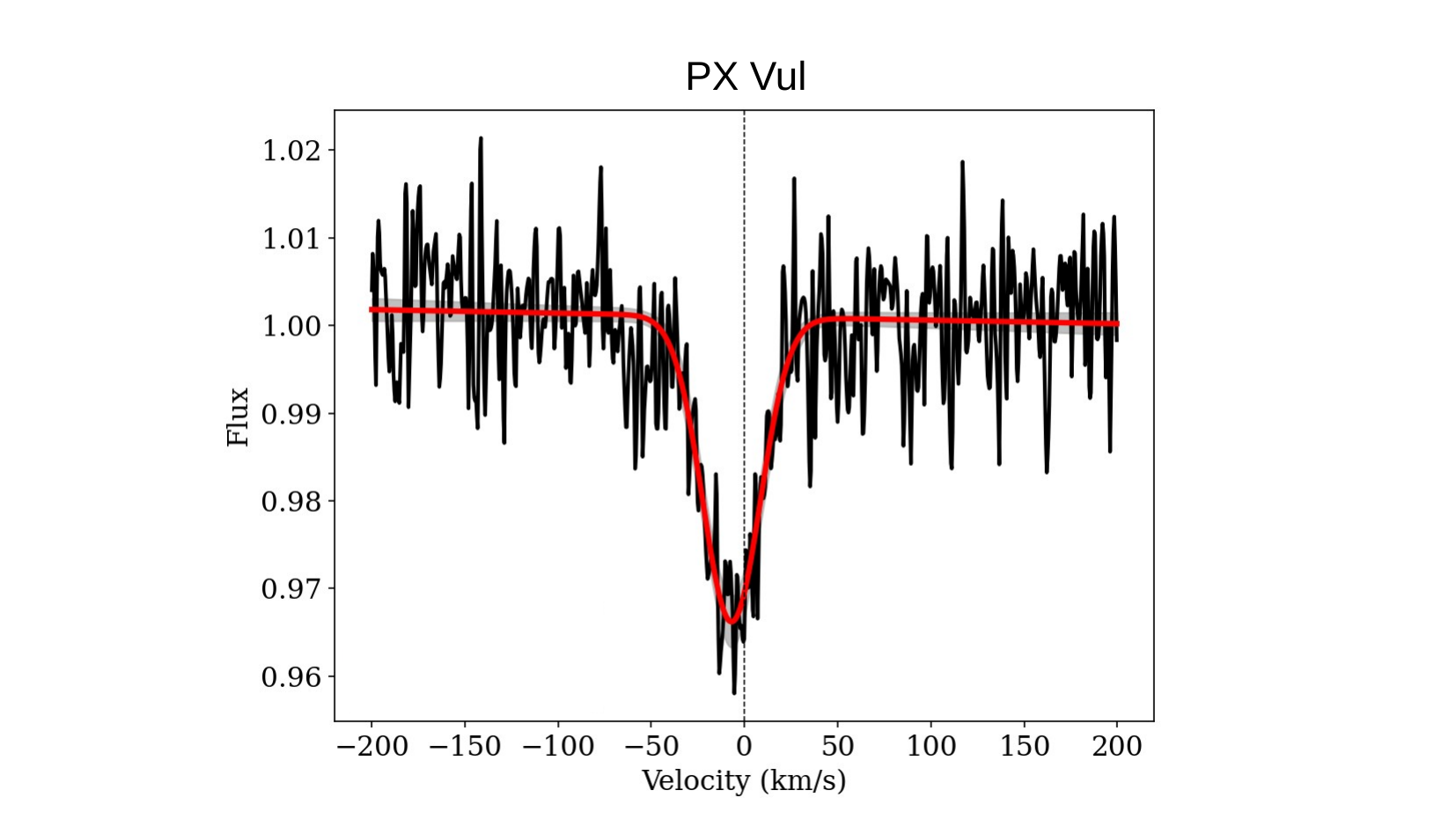} 
\end{figure}

\twocolumn

\end{appendix}
\end{document}